\documentclass[twoside,11pt]{article}
\usepackage{jmlr2e}

\usepackage{enumerate}
\usepackage{amsmath}
\usepackage{amsfonts}
\usepackage{amssymb}
\usepackage{amsbsy}
\usepackage{isomath}
\usepackage{dsfont}
\usepackage{algorithm}
\usepackage{algorithmic}
\usepackage{mathrsfs}
\usepackage{paralist}
\usepackage{epsfig}
\usepackage{caption}
\usepackage{subcaption}
\usepackage{makeidx} 
\usepackage{color}
\usepackage{hyperref}
\usepackage{tikz}
\usepackage{textcomp}
\usepackage{float}
\usepackage{pgfplots}
\usetikzlibrary{pgfplots.groupplots}
\usetikzlibrary{positioning}
\usepackage{booktabs}
\usepackage{hhline}

\usepgfplotslibrary{patchplots,colormaps}

\numberwithin{equation}{section} 

\newcommand\Reals {{\mathds{R}}}

\newcommand \CA {{\mathcal{A}}}

\newcommand \CF {{\mathcal{F}}}

\newcommand \BFA {\mathbf{A}}

\newcommand \BFD {\mathbf{D}}
\newcommand \BFE {\mathbf{E}}
\newcommand \BFL {\mathbf{L}}
\newcommand \BFM {\mathbf{M}}
\newcommand \BFP {\mathbf{P}}
\newcommand \BFR {\mathbf{R}}
\newcommand \BFU {\mathbf{U}}
\newcommand \BFX {\mathbf{X}}
\newcommand \BFZ {\mathbf{Z}}
\newcommand \BFW {\mathbf{W}}

\newcommand \spn {{\bf span}}
\newcommand \core {{\bf core}}
\newcommand \degree {{\bf deg}}

\DeclareMathAlphabet{\mathpzc}{OT1}{pzc}{m}{it}

\newcommand \node[1] {\nu_{#1}}

\def\clap#1{\hbox to 0pt{\hss#1\hss}}

\usetikzlibrary{automata}
\usetikzlibrary{topaths}

\tikzstyle{select}=[rectangle,draw=black,fill=red!20,inner sep=0mm, minimum size=9mm]
\tikzstyle{hidden}=[circle,dashed,draw=black,fill=green!20,inner sep=0mm, minimum size=9mm]
\tikzstyle{observed}=[circle,draw=black,fill=blue!10,inner sep=0mm, minimum size=9mm]

\graphicspath{{figures/}}

\usepackage{epstopdf}

\jmlrheading{}{}{}{}{}{}

\ShortHeadings{CoreCluster: A core decomposition framework for graph clustering}{Giatsidis, Malliaros, Tziortziotis, Dhanjal, Kiagias, Thilikos and Vazirgiannis}
\firstpageno{1}

\begin{document}

\title{A $k$-core Decomposition Framework for Graph Clustering}

\author{\name Christos Giatsidis \email giatsidis@lix.polytechnique.fr \\
       \addr LIX, \'{E}cole Polytechnique\\
       Palaiseau, 91120, France
       \AND
       \name Fragkiskos D. Malliaros \email fmalliaros@lix.polytechnique.fr \\
       \addr  LIX, \'{E}cole Polytechnique\\
       Palaiseau, 91120, France
       \AND
       \name Nikolaos Tziortziotis \email ntziortziotis@lix.polytechnique.fr\\
       \addr  LIX, \'{E}cole Polytechnique\\
       Palaiseau, 91120, France
       \AND
       \name Charanpal Dhanjal \email charanpal@gmail.com\\
       \addr TSI, T\'{e}l\'{e}com ParisTech\\
       Paris, 75013, France
       \AND
       \name Emmanouil Kiagias \email e.kiagias@gmail.com\\
       \addr  LIX, \'{E}cole Polytechnique\\
       Palaiseau, 91120, France
       \AND
       \name Dimitrios M. Thilikos \email sedthilk@thilikos.info\\
       \addr CNRS, LIRMM, University of Athens\\
       Montpellier, 34095, France and\\
       Athens, 15784, Greece
       \AND
       \name Michalis Vazirgiannis \email mvazirg@lix.polytechnique.fr\\
       \addr LIX, \'{E}cole Polytechnique\\
       Palaiseau, 91120, France
}

\editor{-}

\maketitle

\begin{abstract}
  Graph clustering or community detection constitutes an important task for investigating the internal structure of graphs, with a plethora of applications in several domains.
  Traditional techniques for graph clustering, such as spectral methods, typically suffer from high time and space complexity.
  In this article, we present \textsc{CoreCluster}, an efficient  graph clustering framework based on the concept of graph degeneracy, that can be used  along with any known graph clustering algorithm.
  Our approach capitalizes on processing the graph in an hierarchical manner provided by its core expansion sequence, an ordered partition of the graph into different levels according to the $k$-core decomposition.
 Such a partition provides an efficient way to process the graph in an incremental manner that preserves its clustering structure, while making the execution of the chosen clustering algorithm much faster due to the smaller size of the graph's partitions onto which the algorithm operates. 
  An experimental analysis on a multitude of real and synthetic data demonstrates that our approach can be applied to any clustering algorithm accelerating the clustering process, while the quality of the clustering structure is preserved or even improved.
\end{abstract}


\section{Introduction} \label{sec:intro}
Detecting clusters or communities in graphs constitutes a cornerstone problem with many applications in several disciplines.
Characteristic application domains include social and information network analysis, biological networks, recommendation systems and image segmentation. Due to its importance and multidisciplinary nature, the problem of graph clustering has received great attention from the research community and numerous algorithms have been proposed \citep[see][for a survey in the area]{fortunato-survey}.

Spectral clustering \citep[e.g.][]{NgJW01onsp, Luxburg07} is one of the most sophisticated methods for capturing and analyzing the inherent structure of data and can have highly accurate results on different data types such as data points, images, and graphs.
Nevertheless, spectral methods impose a high cost of computing resources both in time and space regardless of the data on which it is going to be applied \citep{fortunato-survey}.
Other well-known approaches for community detection are the ones based on modularity optimization \citep{NewGirv,PhysRevE.70.066111}, stochastic flow simulation \citep{satuluri-mcl-kdd09} and local partitioning methods \citep{fortunato-survey}. In any case, scalability is still a major challenge in the graph clustering task, especially nowadays with the significant increase of the graphs' size.

Typically, the methodologies for scaling up a graph clustering method can be divided into two main categories: (i) algorithm-oriented and (ii) data-oriented.
The former considers the algorithm of interest and appropriately optimizes, whenever is possible, the ``parts'' of the algorithm responsible for scalability issues.
Prominent examples here are the fast modularity optimization method \citep{PhysRevE.70.066111} and the scalable flow-based Markov clustering algorithm \citep{satuluri-mcl-kdd09}.
The latter constitutes a widely used class of methodologies relied on sampling/sparsification techniques.
In this case, the size of the graph onto which the algorithm will operate is reduced, by disregarding nodes/edges.
However, possible useful structural information of the graph (i.e., nodes/edges) is ignored in this approach.

In this paper, we propose \textsc{CoreCluster}, a graph clustering framework that capitalizes on the notion of graph degeneracy, also known as $k$-core decomposition \citep{seidman1983}.
The main idea behind our approach is to combine any known graph clustering algorithm with an \textit{easy-to-compute}, \textit{clustering-preserving} hierarchical representation of the graph -- as produced by the $k$-core decomposition -- towards a scalable graph clustering tool.
The $k$-core of a graph is a maximal size subgraph where each node has at least $k$ neighbors in the subgraph (we say that $k$ is the {\em rank} of such a core). 
The maximum $k$ for which a graph contains a $k$-core is known as its {\em degeneracy}.
We refer to this core as ``the densest core''.
Intuitively, the $k$-core of such a graph is located in its ``densest territories''.
Based on this idea, we show that the densest cores of a graph are roughly maintaining its clustering structure and thus constitute \textit{good starting points} (seed subgraphs) for computing it.
Given the fact that the size of the densest core of a graph is orders of magnitude smaller than that of the original graph, we apply a clustering algorithm starting from its densest core and then, on the resulting structure, we incrementally cluster the rest of the nodes in the lower rank cores in decreasing order, following the hierarchy produced by the $k$-core decomposition. 

At a high level, the main contributions of this paper are three-fold:
\begin{itemize}
\item \textit{Clustering Framework}: 
  We introduce \textsc{CoreCluster}, a scalable degeneracy-based graph clustering framework, that can be used along with any known graph clustering algorithm.
  We show how \textsc{CoreCluster} utilizes the $k$-core decomposition of a graph in order to  (i) select seed subgraphs for starting the clustering process and (ii) expand the already formed clusters or create new ones.

\item \textit{Scalability and Accuracy Analysis}: We discuss analytically the ability of \textsc{CoreCluster} to scale-up, describing its expected running time.
  We also justify why the $k$-core decomposition provides the direction under which to perform clustering incrementally.
  More specifically, we demonstrate that the $k$-core structure captures the clustering properties of a graph, thus being able to indicate good seed subgraphs for a clustering algorithm. Furthermore, we derive upper bounds about the difference between eigenspaces of successive cores. Finally, we show that the cluster expansion process of the \textsc{CoreCluster} framework is closely related to the minimization of the graph cut criterion.

\item \textit{Experimental Analysis}:  We perform an extensive experimental evaluation regarding  the efficiency and accuracy of the \textsc{CoreCluster} framework.
  A large set of experiments were conducted both on synthetic and real-world graphs. This is to evaluate on ground truth information from the synthetic datasets and on the quality of the  clusters from the real-world graphs.
  The empirical results show that the time complexity is improved by 3-4 orders of magnitude (compared to a baseline algorithm), especially for large graphs.
Moreover, in the case of graphs with inherent community structure, the quality of the results is maintained or even improved.
In additionally, the initial experimentation of \citet{giatsidis2014corecluster} is extended here to a wide range of algorithms in order to identify the properties of the graphs and algorithms that optimize the performance our framework. The \textsc{CoreCluster} framework was introduced on the basis of studies of degeneracy in real social graphs \citep{giatsidis2011d}. Given the interdisciplinary application of graphs, we do not hold claims of universal applicability (i.e., not all types of graphs have the same properties). We conduct here an extensive evaluation of \textsc{CoreCluster} in order to outline detectable properties that indicate the conditions under which the proposed framework is useful or meaningless.
\end{itemize}

The remainder of this paper is organized as follows. Section~\ref{sec:related-work} discusses related work and Section~\ref{PRELIMINARIES} introduces some preliminary notions. Section~\ref{sec:method} formally introduces the \textsc{CoreCluster} framework. The theoretical analysis and the computational complexity of the proposed framework are presented in Sections~\ref{sec:TheoreticalAnalysis} and~\ref{sec:timecomplexity}, respectively.  Finally, our extended empirical analysis is presented in Section~\ref{sec:experiments} and we conclude with a discussion about the advantages of the \textsc{CoreCluster} framework along with suggestions for future work in Section~\ref{sec:conclusions}. 

\section{Related Work} \label{sec:related-work}

In this section, we review the related work regarding the graph clustering problem, approaches for scaling-up graph clustering and applications of the k-core decomposition.

\noindent \textbf{Graph clustering.} The problem of community detection and graph clustering has been extensively studied from several points of view.
Some well-known approaches include spectral clustering \citep[e.g.,][]{NgJW01onsp,normalized-cuts,rohtua2}, modularity optimization \citep[e.g.,][]{PhysRevE.69.066133,PhysRevE.70.066111}, multilevel
graph partitioning \citep[e.g., Metis,][]{metis}, flow-based methods \citep{satuluri-mcl-kdd09}, hierarchical methods \citep{NewGirv} and many more. 
A very informative and comprehensive review over the different approaches can be found in \citet{fortunato-survey}.
Also, \citet{rohtua} have conducted a comparative analysis on the performance of some of the most recent algorithms, in artificial data produced by their parameterized generator of benchmark graphs.
In our work, we use the same graph generator as in~\citet{rohtua} to evaluate our framework.
Another recent empirical comparison of community detection algorithms has been performed by \citet{leskovec-www10}.
There, due to lack of ground-truth data, the evaluation of the produced clusters is achieved applying quality measures, such as conductance.
As we will present during our experimental analysis (see Section \ref{sec:experiments}), we follow a similar practice to evaluate our framework in the case of real-world graph data.

\noindent \textbf{Scaling-up graph clustering.}
The efficiency of graph clustering can be improved in various ways.
Two well-known approaches are the ones of \textit{sampling} and \textit{sparsification}.
In the case of spectral clustering, sampling-based approaches include the Nystr\"{o}m method \citep{kumar:on} and randomized SVD algorithm \citep{drineas}.
The approach of  \citet{kumar:on} capitalizes on the Nystr\"{o}m column-sampling method \citep{NIPS2000_1866} which is an efficient technique to generate low-rank matrix approximations.
More specifically, a novel technique has been proposed that follows a non uniform sampling of columns of the affinity matrix.
Although, Nystr\"{o}m method suffers from high time and memory complexity \citep{Yan:2009}.
\citet{drineas} proposed the randomized SVD algorithm that essentially samples a number of columns of the Gram matrix with probability proportional to their norms, and performs Principal Component Analysis on the selected features.
Nevertheless, the randomized SVM algorithm may need to sample a large number of columns in order to get a sufficient small appproximation error in some cases. 
In \citep{Yan:2009} a fast approximate algorithm for spectral clustering has been presented where two different preprocessors have been used in order two reduce the size of the initial data structure. The first one is the classical k-means algorithm while the other one is the Random Projection tree. In a nutshell, neighbouring data points correspond into a set of local representative points and then the spectral algorithm is executed only on the reduced set of the representative points.

Concerning graph sampling, the goal is to produce a graph of smaller size (nodes and edges), preserving a set of desired graph properties (e.g., degree distribution, clustering coefficient) \citep{leskovec-sampling-kdd06}.
The work by \citet*{maiya-www10}, presents a method, based on the notion of expansion properties, to sample a subgraph that preserves the community structure, i.e., contains representative nodes of the communities.
Then, the community membership of the nodes that do not belong to the sample can be expressed as an inference problem. Unlike the previous methods that sample both nodes and edges, the graph sparsification algorithm presented in \citet{Satuluri:2011:LGS:1989323.1989399} reduces only the number of edges (focusing on inter-community edges) in order improve the running time of a
clustering algorithm.
In contrast to the aforementioned methodologies, our approach keeps the structure of the graph intact, without excluding any structural information from the clustering process.

\noindent  \textbf{$k$-core decomposition.}
The $k$-core decomposition constitutes a well-established approach for identifying particular subsets of a graph, called $k$-cores, in an hierarchical way, starting from external vertices and moving to more central ones.
The $k$-cores are important structures in graph theory and their study goes back to the 60’s \citep{Erdos63onth}.
\citet{seidman1983} first applied the $k$-core decomposition to study the cohesion of social networks.
Since then, $k$-core decomposition has been applied in several graph-related tasks, such as graph visualization \citep{vespignani,parthasarathy-icde12}, dense subgraph discovery \citep{andersen-waw-2009} and as an edge ordering criterion for graph coarsening \citep{karypis-ipdps}.
Some very recent studies include the extension of the notion of $k$-cores to directed graphs \citep{d-cores-icdm11} and core decomposition in massive graphs \citep{core-icde11}.

\section{Preliminaries}
\label{PRELIMINARIES}

Given an undirected graph $G$, we denote by $V(G)$ and $E(G) \subset V(G) \times V(G)$ the sets of its vertices and edges, respectively.
Given a set $S\subseteq V(G)$, we denote  by $G[S]$ the induced subgraph of $G$ that is obtained  if we remove from it all vertices that do not belong in $S$.
We also define as $n$ and $m$ the number of vertices and edges of graph $G$, respectively, i.e., $n=|V(G)|$ and $m=|E(G)|$.
The neighbourhood of a vertex $v\in V(G)$ is denoted by $N_{G}(v)$ and contains all vertices of $G$ that are adjacent to $v$.
The degree $\degree_{G}(v)$ of a vertex $v$ in $G$, is the number of edges incident to the vertex and is equal to $|N_{G}(v)|$. 
The {\em minimum degree} of graph $G$, denoted by $\delta(G)$, is the minimum degree of the vertices in $G$, i.e.,
\begin{equation*}
  \delta(G)=\min\{\degree_{G}(v)\mid v\in V(G)\}.
\end{equation*}

Given a non-negative integer $k$, we define the {\em $k$-core} of a graph $G$, denoted by $G_k  \triangleq \core_{k}(G)$, as the maximum size subgraph of $G$ with minimum degree $k$, where $k$ is the {\em rank} of the core.
In a nutshell, in the {\em $k$-core} of a graph $G$ every vertice is adjacent with at least $k$ vertices.
The {\em degeneracy} of a graph $G$ is the maximum $k$ for which $G$ contains a non-empty core.
In a more formal way, the degeneracy of a graph is denoted by $\delta^{*}(G)$ and is given as:
\begin{equation}
  \delta^{*}(G)=\max\{\delta(H)\mid H\subseteq G\}.
\end{equation}
Intuitively, the dense cores of a graph, i.e., those whose ranks are close to $\delta^{*}(G)$, may serve as starting seeds for any clustering algorithm as they are expected to preserve the clustering structure of the original graph.
Later in this paper, we make this statement more precise, providing the necessary theoretical as well as experimental justification.

\begin{figure}[t]
  \centering
  \scalebox{0.5}{\input{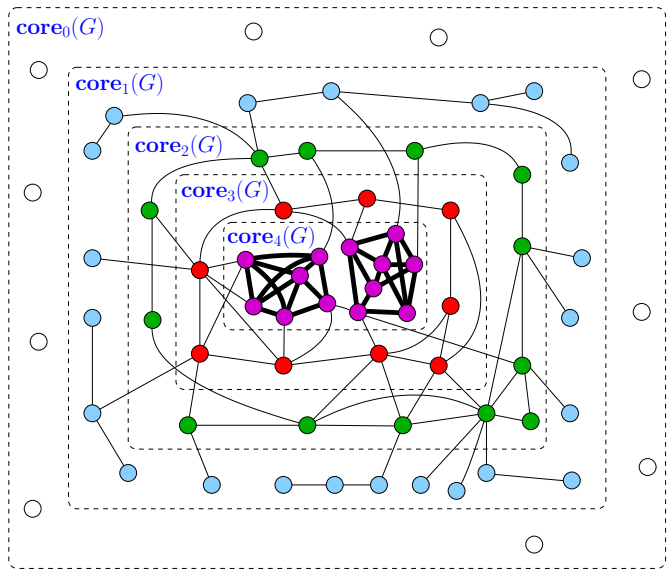}}
  \caption{A graph $G$ of degeneracy $4$ and its cores. The different colors express the  partition of the vertices of the graph to layers $V_{4},V_{3},V_{2},V_{1}$, and $V_{0}$. Fat-edges indicate parts of a clustering of the graph.}
  \label{fig-gamma-k}
\end{figure}

For a given graph $G$ with $\delta^{*}(G)=k$, we define its \textit{core expansion sequence} as a sequence of vertex sets  $\{V_{k},V_{k-1},...,V_{0}\}$, computed in a recursive top-down way as follows:
\begin{equation}
  V_{i} = \left\{
  \begin{array}{ll}
    V(\core_{k}(G)), & \text{if } i = k, \\
    V(\core_{i}(G))\setminus V_{i+1}, & \text{if } i=k-1,\ldots,0.
  \end{array} \right .
\end{equation}
We refer to the sets of a core expansion sequence as {\em layers}, with set $V_{i}$ being its $i^{th}$ layer (see Fig. \ref{fig-gamma-k} for an illustrative example).

In general terms, detecting the $i$-core of a graph $G$ is an easy process \citep{batagelj-2003}: just remove vertices, and edges incident to them, of degree less than $i$ until this is not possible any more.
It can be easily verified that $O(k \cdot n)$ steps are required for the computation of the $k$-core decomposition.
Typically, the degeneracy of real world graphs is small, making the computation of the  {\em core expansion sequence} an easy computational task.

\section{Graph Clustering with Degeneracy} \label{sec:method}
In this section, we describe in detail the proposed graph clustering framework.
\textsc{CoreCluster} capitalizes on the concept of degeneracy to improve the efficiency of graph clustering.
The main idea behind our approach is that the $k$-core decomposition preserves the clustering structure of a graph and therefore the ``best'' $k$-core subgraph can be used as good starting point for a clustering method.
At the same time, the decomposition provides an hierarchical organization of the nodes in the graph, that can serve as a ``guide'' for the clustering process.

\subsection{The \textsc{CoreCluster} Framework}
One of the main attributes of the proposed framework, which constitutes our initial motivation, is its ability to accelerate the application of computationally complex clustering algorithms without any significant loss in their accuracy.
Suppose we have a clustering algorithm that takes as input a graph $G$ and outputs a partition of $V(G)$ into a number of sets that form a clustering of $G$.
As our method could be applied at any clustering algorithm, we do not further specify the attributes of such an algorithm and we, abstractly, name it {\bf Cluster}.
Furthermore, we assume that {\bf Cluster} algorithm runs in $O(n^{3})$ steps.
This assumption coincides with the computational complexity of the  spectral clusetering algorithm proposed by \cite{NgJW01onsp} and utilized in our initial experiments as well as during the theoretical analysis of our methodology.
In the totality of our experiments (see Section \ref{sec:experiments}), we study more than the computational complexity and we utilize various algorithms in the place of {\bf Cluster} including  ``fast'' algorithms.  

In a nutshell, the \textsc{CoreCluster} framework applies the {\bf Cluster} algorithm at the highest $k$-core of the graph and then it iterates from the highest to the lowest core applying the following strategy:
\begin{enumerate}[i.]
\item assign based on a simple criterion all the nodes that can be assigned to the already existing clusters, and
\item apply the {\bf Cluster} algorithm  to the remaining nodes (in order to create new clusters).
\end{enumerate}

\begin{algorithm}[t!]
  \begin{algorithmic}[1]          
    \REQUIRE A graph $G$.
    \ENSURE A partition of $V(G)$ into clusters.
    \STATE $k:=\delta^{*}(G)$
    \STATE $q:=0$
    \STATE Let $V_{k},\ldots,V_{0}$ be the core expansion sequence of $G$
    \STATE Let $G_{i}$ be the $i$-core of $G$, for $i=0,\ldots,k$
    \STATE $S_{k} = V_{k}$
    \STATE $\CA_{k}=\{C_{1}^{k},\ldots,C_{\rho_{k}}^{k}\}={\bf Cluster}(G[S_{k}])$
    \FOR {$i=k-1$ {\bf to} 0}
    \STATE $S_{i}=$\mbox{\em Select}$(G_{i},\CA_{k}\cup \ldots \cup\CA_{i+1},V_{i})$
    \STATE $\CA_{i}=(C_{1}^{i},\ldots, C^{i}_{\rho_{i}})={\bf Cluster}(G[S_{i}])$
    \ENDFOR
    \STATE {\bf Return} $\CA_{k}\cup \cdots\cup \CA_{0}$
  \end{algorithmic}
  \caption{\textsc{CoreCluster}$(G)$} \label{alg:CoreClusterFramework}
\end{algorithm}

A sketch of \textsc{CoreCluster} framework is given in pseudocode in Alg.~\ref{alg:CoreClusterFramework}.
Initially, \textsc{CoreCluster} performs  $k$-core decomposition to obtain the core expansion sequence of the graph. Then, algorithm \textbf{Cluster} is applied to the $k$-core subgraph, creating the first set of $\rho_k$ clusters.
The procedure $\mbox{\em Select}$ (see Section~\ref{sec:selection}, for a detailed description) takes as input the, so far, created clusters, i.e., the sets in  $\CF_{i+1}=\CA_{k}\cup\cdots\cup \CA_{i+1}$ and the $i^{th}$ layer $V_{i}$, and tries to assign each of the vertices of $V_{i}$  in some cluster in $\CA_{k}\cup\cdots\cup \CA_{i+1}$.
After this update, the procedure $\mbox{\em Select}$ returns the unassigned vertices.
The choice of the selection procedure considers the way under which the vertices of $V_{i}$ are adjacent with the vertices of the clusters in $\CA_{k}\cup\cdots\cup \CA_{i+1}$.
This selection can be done by using several heuristic approaches, and next we describe such a procedure. 
Afterwards, the \textbf{Cluster} is applied to the unassigned vertices returned from the $\mbox{\em Select}$ procedure.
The whole procedure is implemented for each layer ($V_{i}$) in a recursive manner, starting from the highest level.

\subsection{Spectral Clustering}
\label{sec:spectal_alg}

It should be stressed that the \textsc{CoreCluster} can be essentially seen as a ``meta-algorithmic procedure'' that can be applied to any graph clustering algorithm, accelerating the execution and preserving the clustering structure of the latter.
The discussion that follows in the experimental analysis (Section~\ref{sec:experiments}) of the paper argues that this indeed can accelerate a high-demanding clustering algorithm without any significant expected loss in its performance.

The spectral clustering algorithm, proposed by \cite{NgJW01onsp}, is one of the most important and well-known algorithms in data mining and constitutes the initial motivation of the proposed framework, since its computational complexity restrict its applicability to large graph networks.
Under this prism, we use spectral clustering as the baseline and the basis for the theoretical analysis of the \textsc{CoreCluster} framework (algorithm   \textbf{Cluster}).
The central idea of the spectral clustering algorithm is to keep the top $\rho$ eigenvectors of the normalized Laplacian matrix, defined as:
\begin{equation}
  \BFL = \BFD^{-1/2} (\BFD - \BFW) \BFD^{-1/2} = \textbf{I} - \BFD^{-1/2} \BFW \BFD^{-1/2},
  \label{eq:LaplacianMatrix}
\end{equation}
where  $\BFW \in \Reals^{n \times n}$ and $\BFD \in \Reals^{n \times n}$ are the adjacency and the degree matrix, respectively.
The adjacency matrix gives the similarity between pairs of graph vertices, such that $\BFW_{ij} > 0$ if there is an edge from vertex $i$ to vertex $j$ and zero ($\BFW_{ij} = 0$) otherwise. If the graph is undirected, as happens in our case, the adjacency matrix is symmetric ($\BFW_{ij} = \BFW_{ji}$).
The degree matrix is a diagonal matrix containing the degrees of the nodes on its diagonal, e.g.,  $\BFD_{ii} = \sum_j \BFW_{ij}$.
Afterward, assign the corresponding eigenvectors as columns to a matrix $Y \in \Reals^{n \times \rho}$ and apply $k$-means clustering on the rows of $Y$.
Thereafter, each row of the new formed matrix corresponds to a data point, while the parameter $\rho$ identifies the number of clusters that we are looking for.

It has been shown that the performance of the $k$-means is strongly affected by the initialization scheme \citep{BubeckMv2012}.
Until now, various approaches have been proposed for the initialization of k-means algorithm \citep{1283494, Redmond:2007, Lu2008787}.
During our empirical analysis (Section~\ref{sec:experiments}), we are using the $k$-means$++$ algorithm proposed by \citet{1283494}, instead of the classical $k$-means algorithm, taking advantage of performing better seeding during the initialization process. Moreover, since we desire to have an automatic choice of $\rho$, we define it by noticing the ``sudden drop'' in the eigenvalues as it is suggested by \citet{DBLP:conf/nips/PolitoP01}.

\subsection{Selection procedure}
\label{sec:selection}

In this section, we give a detailed description of the selection procedure (\textit{Select}) presented in Line 8 of Alg. \ref{alg:CoreClusterFramework}.
The procedure takes as input the, so far, created clustering $\CF_{i+1}=\CA_{k}\cup\cdots\cup \CA_{i+1}$ and the vertex set $V_{i}$.
Then, this procedure assigns some of the vertices of $V_{i}$ to the clusters in $\CF_{i+1}$ and outputs the remaining ones.

For simplicity purposes, we will call the tuple $(G,\CF,V)$ as a {\em candidate triple}, where $G$ is a graph and $\CF\cup\{V\}$ is a partition of $V(G)$.
Thus, given a {\em candidate triple} $(G,\CF,V)$,  we define the following property over the vertices of $V$:
\begin{equation}
  {\BFP}^{\alpha,\beta}(v) = \exists C \in \CF:  \frac{|N_{G}(v)\cap V(C)|}{|N_{G}(v)|} \geq \alpha \ \ \ \mbox{and} \ \ \   |N_{G}(v)|\geq  \beta,
  \label{eq:selection_criterion}
\end{equation}
where $\alpha>0.5$ and $\beta$ is a positive integer.
Notice that, as $\alpha>0.5$, the truth of ${\BFP}^{\alpha,\beta}(v)$ can be certified by a unique set $C$ in $\CF$. We call such a set the {\em certificate} of verticle $v$.

\begin{algorithm}[h]
  \caption{{\em Select}$(G,\CF,V)$}\label{alg:Select}
  \begin{algorithmic}[1]
    \REQUIRE A candidate triple $(G,\CF,V)$ 
    \ENSURE  A subset $S$ of $V$ and a partition  $\CF'$  of $V(\CF)\cup(V\setminus S)$.
    \WHILE{${\BFP}^{\alpha,\beta}(v)$ is true for some $v\in V$}
    \STATE $\CF\leftarrow (\CF\setminus\{C\})\cup\{C\cup \{v\}\}$ where $C$ is the certificate of $v$.
    \STATE $V\leftarrow V\setminus\{v\}$.
    \ENDWHILE
    \STATE $V^{1} \! \! = N_{G}(V(\CF))$
    \STATE $V^{2}=V\setminus (V^{1}\cup \CF)$
    \IF{$V^{2}$ is either empty or an independent set of $G$,}
    \STATE $\CF\leftarrow$ {\em Assign}$(G,\CF,V,V^{1})$
    \STATE $\CF\leftarrow${\em Assign}$(G,\CF,V,V^{2})$
    \STATE return $\emptyset$
    \ELSE
    \STATE return $V^{1}\cup V^{2}$.
    \ENDIF
  \end{algorithmic}
\end{algorithm}

The selection procedure first (lines 1--4) tries to assign vertices of $V$ to clusters of $\CF$ using the criterion of the property ${\BFP}^{\alpha,\beta}$ that assigns a vertex to a cluster only if the vast majority of its neighbours belong in this cluster. 
The quantification of this ``vast majority'' criterion is done by the constants $\alpha$ and $\beta$.
The vertices that cannot be assigned at an already existing cluster are partitioned into two groups:~(i) $V^{1}$ contains those that have neighbours in vertices that are already classified in the clusters of $\CF$ and (ii) $V^{2}$ contains the rest. 
As the vertices in $V^{2}$ have no neighbours in the clusters, they have at least $k$ neighbours out of them, it is most likely that they may not enter to any existing cluster in the future, unless, possibly, they are completely disjoint.  
If this is not the case, a further (milder) classification is attempted by the {\em assign} procedure  that first  classifies the vertices in $V^{1}$ in the existing clusters and then we do the same for the vertices in $V^{2}$.
It is worth noting that the last selection procedure has been found useful in our experiments, especially in the case of cores with low rank (where many independent vertices may appear).

The procedure {\em assign} is a heuristic that classifies each vertex to the cluster that has the majority of its neighbours. 
Before presenting the  {\em Assign}  routine, given by Alg. \ref{alg:Assign}, let us first give some definitions.
Given a candidate triple $(G,\CF,V)$ and a vertex $v\in V$, we define as $\spn(v)=\max\{|N_{G}(v)\cap V(C)|\mid C\in \CF\}$.
We also define ${\bf argspan}(v)$ as a minimum size $C\in \CF$ with the property that $|N_{G}(v)\cap V(C)|=\spn(v)$.
\begin{algorithm}[h]
  \caption{{\em Assign}$(G,\CF,V,S)$}\label{alg:Assign}
  \begin{algorithmic}[1]
    \REQUIRE A candidate triple $(G,\CF,V)$ and a subset $S$ of $V$
    \ENSURE A partition  $\CF$
    \WHILE{$S\neq\emptyset$}
    \STATE ${\ell} = \max\{\spn(v)\mid v \in S\}$
    \STATE $L=\{v\in S\mid \spn(v)= \ell \}$
    \FOR{each $v\in L$}
    \STATE $C'=C\cup\{v\}$, where $C={\bf argspan}(v)$\
    \STATE $S\leftarrow S\setminus\{v\}$
    \STATE $\CF \leftarrow ({\CF}\setminus\{C\})\cup\{C'\}$
    \ENDFOR
    \ENDWHILE
    \STATE return $\CF$
  \end{algorithmic}
\end{algorithm}

\begin{figure}[h]
	\centering
	\includegraphics[scale=0.8]{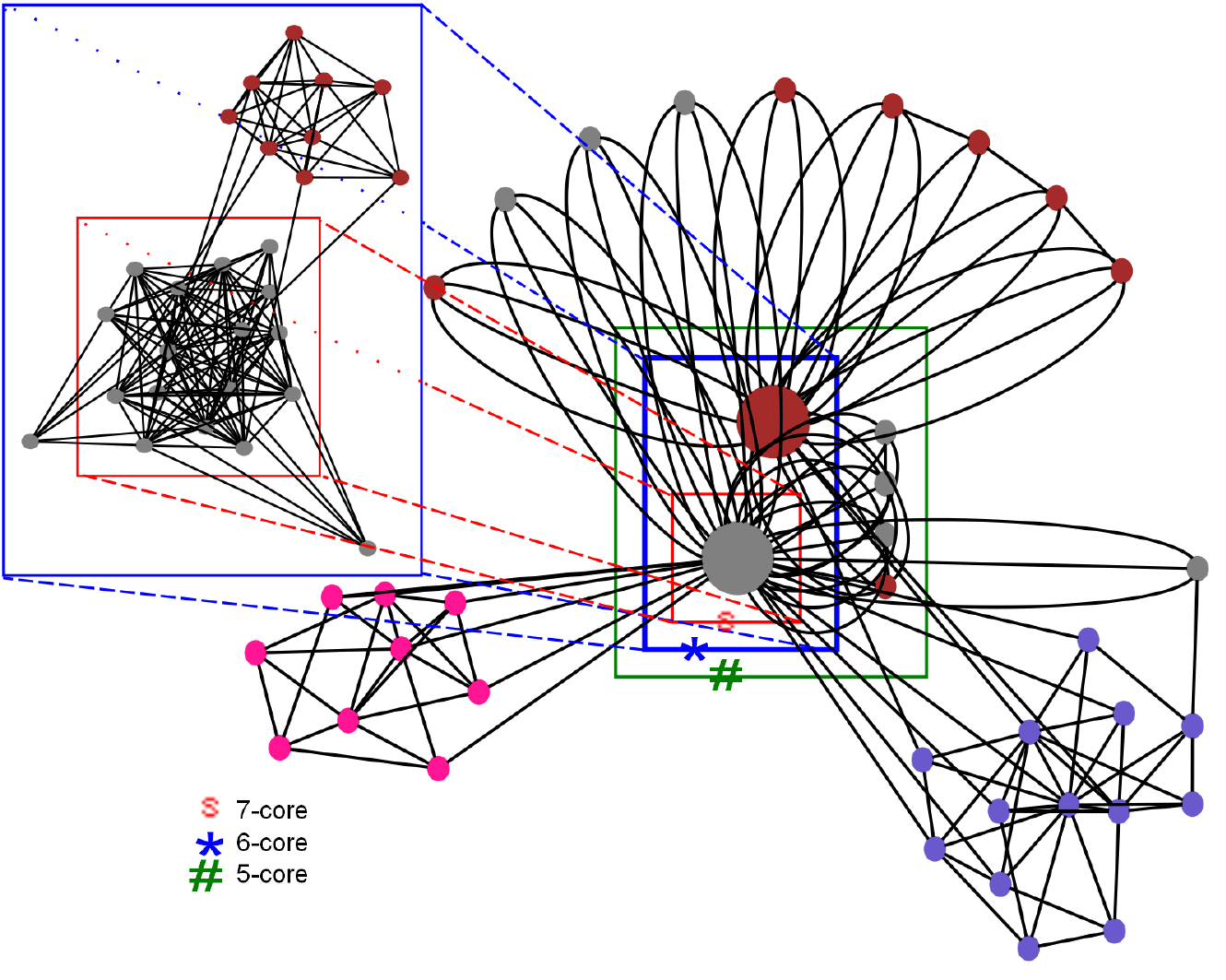}
	\caption{An example of the operation of the \textsc{CoreCluster} procedure for a portion of a graph used through our experimental study. This graph consists of the core sequence members $V_7$, $V_6$, $V_5$, and $V_4$ that are included in the black, green ($\#$),  blue ($*$), and red ($s$) squares respectively. Vertices of different colors correspond to different clusters.}
	\label{fig:coreclustering}
\end{figure}

A number of illustrative examples about the choices of the selection procedure can be extracted by Fig.~\ref{fig:coreclustering}, that is an instance of a graph derived from our experimental data (in particular from the $D_1$ dataset – see Sec. \ref{sec:artfc}). In the particular graph, the $7$-core (i.e., the graph delimited by the red square) consists exclusively of the vertices of the “grey” cluster. The set $V_6$ contains all the vertices of the $6$-core that are not in the $7$-core. All but two of these vertices are sparsely connected with the grey cluster, while they exhibit a strong interconnection between them. Therefore, they form the red cluster, while the remaining two vertices, that have all of their neighbours in the gray cluster, become a member of it. A similar assignment to the gray and red clusters is happening for the vertices in $V_5$, i.e., the vertices in the $5$-core that do not belong in the $6$-core. Finally, the remaining vertices of the graph (the vertices in $V_4$) either clearly belong to the existing clusters or are forming two new clusters, the ping and the purple ones. The experimental analysis indicates that a similar clustering behaviour characterizes the entire data set and that it is captured by the \textsc{CoreCluster} framework.

\section{Theoretical Analysis of the CoreCluster's Quality}
\label{sec:TheoreticalAnalysis}
The intuition behind the \textsc{CoreCluster} framework is that the core expansion sequence $V_{k},V_{k-1},\ldots,V_{0}$ gives a good sense of direction on how to perform clustering in an incremental way.
After that, the procedure considers $V_{k-1}$ as the remaining vertices of the  $(k-1)$-core $G_{k-1}$, and tries to assign them one by one to the already existing clusters $C_{1}^{k},\ldots,C_{\rho_{k}}^{k}$. 
The vertices  for which an assignment to an already existing cluster is not possible, form the set $S_{k-1}$  and the  {\bf Cluster} is now applied on $G[S_{k-1}]$.
As the algorithm continues, the existing clusters grow up and the vertices for which this is not possible, are grouped to new clusters.
The fact that this procedure approximates satisfactorily  the result of the application of  {\bf Cluster}  to the whole graph is justified by the observation that the early $i$-cores  (i.e., $i$-cores where $i$ is close to $k$) are already dense, and therefore \textit{sufficiently coherent}, to provide a good starting clustering that will expand well because of the selection criterion.
In fact, the subgraphs obtained by the $k$-core decomposition, provide an $(1/2)$-approximation algorithm for the \textsc{Densest-Subgraph} problem \citep{andersen-waw-2009}.

In the remaining of this section, a number of theoretical results are presented which confirms that the core expansion sequence offers an reliable way for performing any clustering methodology incrementally. Our theoretical analysis could be divided on three main axis: 
\begin{enumerate}
\item Initially, we justify why the $k$-core subgraph can be supposed as a seed subgraph for any clustering algorithm, as nodes with a high clustering coefficient are usually retained at the highest $k$-core subgraph.
\item Afterwards,  based on the perturbation theory, we demonstrate that the difference between eigenspaces of successive cores is upper bounded. Under this prism, we are able to analyze the conditions under which the eigenspaces of successive cores vary significantly with each other.
\item Last but not least, we highlight that the cluster expansion process, which assigns the vertices of $V_i$ to an already existing cluster by using the selection criterion, is closely related to minimizing the graph cut criterion. 
\end{enumerate}

\subsection{Quality of the maximal core as a starting seed}
\label{sec:theory}
We claim that the decomposition identifies subgraphs that progressively correspond to the most central regions and  connected parts of the graph.
Here we show that nodes with high clustering coefficient in $G$, are more likely to``survive" at the highest k-core subgraph by the pruning ($k$-core decomposition) procedure.
At this point, let us describe the clustering coefficient \citep{watts-strogatz-1998} measure which will be used later through the analysis of our framework. 
The clustering coefficient measures the density of triangles in the graph and indicates the degree to which nodes tend to cluster together.
More specifically, the local clustering coefficient of a vertex is defined as the ratio of the links that the vertex has with the vertices of its neighbourhood to the total number of edges that could exist among the vertices within its neighbourhood.
On the other hand, the global clustering coefficient is defined as the mean of the local clustering coefficients over all graph vertices.
Our claim is based on the following theorem.

\begin{theorem}[\cite{gleich-kdd12}] \label{thm:gleich}
  Let $G$ be a graph with heavy-tailed degree distribution, and let $C_G$ be the (global) clustering coefficient of $G$.
  Then, there exists a $k$-core in $G$ for $k \ge C_G ~ \dfrac{d_{\max}^{\varepsilon}}{2}$, where $\varepsilon < 1$ is a constant such that most edges are incident to a node with degree at least $d_{\max}^{\varepsilon}$ (typically $\varepsilon = 2/3$), where $d_{\max}$ is the maximum degree of the nodes.
\end{theorem}

The above theorem implies that graphs with heavy-tailed degree distribution and high global clustering coefficient $C_G$, have large degeneracy.
Next we present our claim for the relationship between the local clustering coefficient $C_v, ~ \forall v \in V(G)$ and the $k$-core subgraph justifying the selection of the $k$-core as good seed subgraph in the clustering procedure.

\begin{corollary}
  Let $G$ be a graph with heavy-tailed degree distribution.
  The contribution of each node $v \in V(G)$  to the $k$-core decomposition of the graph is proportional to the local clustering coefficient $C_v$.
\end{corollary}

\begin{proof}
  The global clustering coefficient $C_G$ of the entire graph is given by the average of the local clustering coefficients $C_v, ~ \forall v\in V(G)$, i.e.,  $C_G = \dfrac{1}{n} \sum_{v} C_v$, where $n= |V(G)|$.
  Then, according to the Theorem \ref{thm:gleich} we get that:
  \begin{align*}
    k  \ge C_G ~ \dfrac{d_{\max}^{\varepsilon}}{2} &= \Bigg( \dfrac{1}{n}
    \sum_{v} C_v
    \Bigg) ~ \dfrac{d_{\max}^{\varepsilon}}{2} = \underbrace{\Bigg( \dfrac{1}{n} 
      \dfrac{d_{\max}^{\varepsilon}}{2} \Bigg)}_{\gamma} \sum_{v} C_v  \\
    \Rightarrow ~~  & k \ge \gamma ~ \sum_{v} C_v,
  \end{align*}
  \noindent where parameter $\gamma$ captures global characteristics of the graph (that depend on the total number of nodes and the maximum degree).
\end{proof}

Therefore, nodes with high clustering coefficient (in the original graph) are more likely to be found in the best (max-core) $k$-core ($k=\delta^\ast{(G)}$)  subgraph, since they tend to be more robust to the degeneracy process.
Thus, the $k$-core subgraph can be used as good starting point (seed subgraph) for the clustering task.
Furthermore, we have experimentally validated the above claim (see Appendix A).

\subsection{The Eigenspaces of k-cores}

The spectral clustering algorithm of \citep{NgJW01onsp} is based on the eigenvectors of a Laplacian matrix and here we study the eigenspaces of successive cores in conjunction with the \textsc{CoreCluster} approach. The key question we are interested in is under which conditions the $\rho$ largest eigenvectors of the Laplacian $\BFL^{(i)}$ for the $i^{th}$ core are similar to the corresponding eigenvectors of $\BFL^{(i+1)}$ for $i = 0,\ldots, k-1$? If the eigenspaces are indeed similar then it follows that clustering on a core is a good basis for clustering the entire graph. 

To answer this question we defer to \emph{perturbation theory} (\cite{stewart1990matrix}) so that we can examine the change of the eigenvalues along with the eigenvectors of a matrix when a small perturbation is added, corresponding to moving from one core to the next. In the case of spectral clustering, consider that the perturbation between the Laplacian matrices ($\BFL^{(i)}$ and $\BFL^{(i+1)}$) of successive cores is given by $\Delta \BFL^{(i)} = \BFL^{(i)} - \BFL^{(i+1)}$. 

The key result that we will use, is presented by \citet{Davis70}, and bounds the difference between eigenspaces of Hermitian matrices under perturbations.

\begin{theorem}[\citet{Davis70}]
  Let $\BFA \in \Reals^{n \times n}$ be a Hermitian matrix with spectral resolution given by $[\BFX_1 \BFX_2]^{\top} \BFA [\BFX_1 \BFX_2] = diag(\BFL_1, \BFL_2)$, where $[\BFX_1, \BFX_2]$ is unitary with $\BFX_1 \in \Reals^{n \times \rho}$.
  Let $\BFZ \in \Reals^{n \times \rho}$ have orthonormal columns, $\BFM$ be any Hermitian matrix of order $\rho$ and define the residual matrix as $\BFR = \BFA\BFZ - \BFZ\BFM$. Let $\lambda(\BFA)$ represent the set of eigenvalues of $\BFA$ and suppose that $\lambda(\BFM) \subset [\alpha, \beta ]$ and for some $\delta > 0$, $\lambda(\BFL_2) \subset \mathbb{R} \backslash [\alpha - \delta, \beta + \delta]$. Then, for any unitary invariant norm, we have:
  \begin{equation}
    \| sin\Theta(\mathcal{R}(\BFX_1), \mathcal{R}(\BFZ))\| \leq \frac{\| \BFR \|}{\delta} ,
  \end{equation}
  where $\mathcal{R}(\BFX_1)$ is the column space of matrix $\BFX_1$.
  \label{the:Davis70}
\end{theorem}

The spectral resolution of $\BFA$ is a transformation of it using the unitary matrix $[\BFX_1 \BFX_2]$ such that $\BFX_1$ spans the \emph{invariant subspace} of $\BFA$. The function $sin\Theta(\cdot, \cdot)$ corresponds to a diagonal matrix of dimension $\rho$, composed of sines of the \emph{canonical angles} between two subspaces. Roughly speaking, this corresponds to the angles between the bases of the subspaces and measures how the two subspaces differ \citep[see][for details]{stewart1990matrix}. 

Then, the following lemma can be easily derived.

\begin{lemma}\label{lem:pert}
Let $\BFA$ and $\tilde{\BFA} = \BFA + \BFE$ be two $n \times n $ Hermitian matrices whose eigendecompositions are given by $\BFA  = \BFU \Lambda \BFU^{\top}$ and $\tilde{\BFA}  = \tilde{\BFU} \tilde{\Lambda} \tilde{\BFU}^{\top}$, with corresponding sets of eigenvalues $\lambda_1 \leq \lambda_2 \leq \cdots \leq \lambda_n$ and $\tilde{\lambda}_1 \leq \tilde{\lambda}_2 \leq \cdots \leq \tilde{\lambda}_n$. Then the following results hold, where $\BFU_1$ and $\tilde{\BFU}_1$ are the matrices of the first $\rho$ eigenvectors of  $\BFA$ and $\tilde{\BFA}$: 
\begin{align}
  \| sin\Theta(\mathcal{R}(\BFU_1), \mathcal{R}(\tilde{\BFU}_1))\|_F &\leq \frac{\sqrt{\sum_i \lambda^{(\rho)}_i(\BFE^{\top} \BFE)}}{\tilde{\lambda}_{\rho+1} - \lambda_{\rho}},\\
  \| sin\Theta(\mathcal{R}(\BFU_1), \mathcal{R}(\tilde{\BFU}_1))\|_2 &\leq \frac{\sigma_{max}(\BFE)}{\tilde{\lambda}_{\rho+1} - \lambda_{\rho}},
\end{align}
where $\lambda^{(\rho)}(\cdot)$  is the largest $\rho$ eigenvalues of a matrix, $\sigma_{max}(\cdot)$ is the largest singular value of a matrix and $\tilde{\lambda}_{\rho+1} > \lambda_{\rho}$.
\end{lemma}
\begin{proof}
First, note that we can express the eigendecompositions of $\BFA$ and $\tilde{\BFA}$ as:
\begin{align}
  \BFA  &= \BFU \Lambda \BFU^{\top} = [\BFU_1 \BFU_2] \left[
    \begin{array}{cc}
      \Lambda_1 & 0 \\
      0 & \Lambda_2
    \end{array} \right]
        [\BFU_1 \BFU_2]^{\top}   \\
        \tilde{\BFA}  &= \tilde{\BFU} \tilde{\Lambda} \tilde{\BFU}^{\top} = [\tilde{\BFU}_1 \tilde{\BFU}_2] \left[
          \begin{array}{cc}
            \tilde{\Lambda}_1 & 0 \\
            0 & \tilde{\Lambda}_2
          \end{array} \right]
              [\tilde{\BFU}_1 \tilde{\BFU}_2]^{\top}.
\end{align}
Then, we can apply Theorem~\ref{the:Davis70} by setting $\BFX_1 = \tilde{\BFU}_1$, $\BFX_2 = \tilde{\BFU}_2$, $\BFL_1 = \tilde{\Lambda}_1$, $\BFL_2 = \tilde{\Lambda}_2$, $\BFM = \Lambda_1$ and $\BFZ = \BFU_1$. Computing the residual matrix $\BFR$ gives:
\begin{equation*}
  \BFR = \tilde{\BFA} \BFU_1 - \BFU_1 \Lambda_1 = \BFA \BFU_1 + \BFE \BFU_1 - \BFU_1 \Lambda_1 = \BFE \BFU_1.
\end{equation*}
The Frobenius norm of the residual is $\|\BFR\|_{F}^{2} = tr(\BFU_1^{\top} \BFE^{\top} \BFE \BFU_1) \leq \sum_i \lambda^{(\rho)}_i(\BFE^{\top} \BFE)$ where we have used the Rayleigh-Ritz theorem to bound the trace \citep{lutkepohl1997handbook}. With the spectral norm we have $\|\BFE \BFU_1\|_2 = \sigma_{\max}(\BFE \BFU_1) \leq \sigma_{\max} (\BFE)$ using a similar process. It is easy to show that $\delta$ is computed as $\tilde{\lambda}_{\rho+1} - \lambda_{\rho}$ provided $\tilde{\lambda}_{\rho+1} > \lambda_{\rho}$. 
\end{proof}

It follows that we want to choose a $\rho$ to make the ``eigengap" $\tilde{\lambda}_{\rho+1} - \lambda_{\rho}$ as large as possible whilst having a residual matrix with small maximum eigenvectors. The eigengap corresponds to the quality of the clusters \citep[see][for example]{Luxburg07}, while the perturbation represents the residual eigenvalues whilst moving from one Laplacian to the next. In the following part of this subsection we will elaborate on the perturbation by showing how it relates to edges in a graph.

It is worth noting that when we add vertices by going from a core to a lower core, the dimensions of the eigenspace corresponding to the new vertices will be zero in the higher core. Therefore, we restrict the analysis of the eigenspace to existing dimensions. In other words, we are interested in the following matrix: 

\begin{equation}
  \Delta \BFL^{(i)} = \hat{\BFL}^{(i)} - \BFL^{(i+1)},
\end{equation}
where $\hat{\BFL}^{(i)}$ is the submatrix of $\BFL^{(i)}$ corresponding to the vertices in the $(i+1)^{th}$ core. The individual elements of the normalized Laplacian are computed as:
\begin{equation}
  \BFL_{xy} = \left\{
  \begin{array}{ll}
    1 & \text{if } x = y \text{ and } \degree(v_x) \neq 0 \\
    -\frac{1}{\sqrt{\degree(v_x) \degree(v_y)}} & \text{if } x \neq y \text{ and } (v_x, v_y) \in E(G) \\
    0  & \text{otherwise}
  \end{array} \right.
\end{equation}
so we have $\BFL_{xy} \in [-1, 1]$ for all $x, y$. Note that off-diagonal elements are non-positive. This allows us to introduce our main result.

\begin{theorem}
Let $\BFL^{(i)} \in \mathbb{R}^{n_i \times n_i}$ be the Laplacian matrix corresponding to the $i^{th}$ core of a graph, and $\hat{\BFL}^{(i)}$ be the submatrix of $\BFL^{(i)}$ corresponding to the vertices in the $(i+1)^{th}$ core. Similarly, $E^{(i)}$ is the set of edges in the $i^{th}$ core, and $\hat{E}^{(i)}$ is the subset of $E^{(i)}$ with one vertex in the $(i+1)^{th}$ core. Then the following results hold with $\BFU_1$ and $\tilde{\BFU}_1$ as the matrices of the first $\rho$ eigenvectors of  $\hat{\BFL}^{(i)}$ and $\BFL^{(i+1)}$: 
 \begin{equation}
  \| sin\Theta(\mathcal{R}(\BFU_1), \mathcal{R}(\tilde{\BFU}_1))\|_F \leq \frac{\sqrt{|J_i|} |i-k|}{ik(\tilde{\lambda}_{\rho+1} - \lambda_{\rho})},
\end{equation}
\begin{equation}
  \| sin\Theta(\mathcal{R}(\BFU_1), \mathcal{R}(\tilde{\BFU}_1))\|_2 \leq \frac{\sqrt{n_{i+1}} |i-k|}{ik(\tilde{\lambda}_{\rho+1} - \lambda_{\rho})},
\end{equation}
where $J_i = \bigcup_{(v_x, v_y) \in \Delta E^{(i)}} N_{G_i}(v_x, v_y)$ is the number of changes in the Laplacian, $\Delta E^{(i)} = \hat{E}^{(i)} - E^{(i+1)}$ is the change in the edges, $N_G(v_x, v_y)$ is the set of edges adjacent to and including the edge $(v_x, v_y)$ in $G$ and $\tilde{\lambda}_{\rho+1} > \lambda_{\rho}$.
\end{theorem}
\begin{proof}
The change in Laplacian elements between cores can be bounded for $x \neq y$ as
\begin{eqnarray*}
 \Delta \BFL^{(i)}_{xy} &=& -\frac{1}{\sqrt{\degree_{G_i}(v_x) \degree_{G_i}(v_y)}} + \frac{1}{\sqrt{\degree_{G_{i+1}}(v_x) \degree_{G_{i+1}}(v_y)}} \\ 
 &\geq& \frac{i-k}{ik},
\end{eqnarray*}
where $G_i = \core_{i}(G)$. It follows that the Frobenius norm of the change in Laplacians is bounded as: 
\begin{eqnarray*}
 \|\Delta \BFL^{(i)}\|^2_F = \sum_{x,y} (\Delta \BFL^{(i)}_{xy})^2 \leq |J_i| \left(\frac{i-k}{ik}\right)^2.
\end{eqnarray*}
Note also that $\|\Delta \BFL^{(i)}\|^2_F = \sum_j \lambda^2_j(\Delta \BFL^{(i)})$. The spectral norm can be written in terms of maximising the Rayleigh quotient: 
\begin{eqnarray*}
 \|\Delta \BFL^{(i)}\|^2_2 &=& \max \frac{ \textbf{u}^T (\Delta \BFL^{(i)})^2\textbf{u}}{\textbf{u}^T\textbf{u}} \\ 
 &\leq& n_{i+1} \left(\frac{i-k}{ik}\right)^2,
\end{eqnarray*}
where $\textbf{u} \in \mathbb{R}^{n_{i+1}}$. Putting the pieces together with Lemma \ref{lem:pert} gives the required result.
\end{proof}

According to this theorem, we can readily show the scenarios in which moving from the $i^{th}$ core to the $(i+1)^{th}$ core does not significantly alter the eigenspaces. The $\frac{i-k}{ik}$ term implies that for a fixed degeneracy $k$ the core transitions close to $k$ result in small eigenspace perturbations. In the case of Frobenius norm, the eigenspaces grow apart at a rate proportional to the neighbourhood of edge changes within the $(i+1)^{th}$ core. On the other hand, for the spectral norm we see that the eigenspaces diverge according to the size ($n_{i+1}$) of the $(i+1)^{th}$ core. 

\subsection{Cluster Expansion}

The analysis so far has considered changes between cores however an important part of the \textsc{CoreCluster} algorithm is how clusters are grown in a greedy fashion according to Alg. \ref{alg:Select}. In this section we show a close relationship between the selection criterion and the the notion of graph cuts, which motivate many spectral clustering approaches. To see the connection, first we define the cut of a graph as follows: 
\begin{equation}
 cut(C_1, C_2) = \sum_{v_i \in C_1, v_j \in C_2} \BFW_{ij},
\end{equation}
and it corresponds to the number of edges between disjoint subsets $C_1, C_2 \subset V(G)$. Based on this definition the cut for a partition into $\rho$ clusters is defined as:
\begin{equation}
 cut(C_1, \ldots, C_{\rho}) = \sum_{i=1}^{\rho} cut(C_i, \bar{C}_i),
\end{equation}
where $\bar{C}$ is the set of vertices in $G$ that are not in $C$. Consider adding a new vertex $v$ to the graph, then the change in the cut term corresponding to candidate cluster $C_j$ is given by
\begin{eqnarray*}
 \Delta cut_j &=& cut(C_j', \bar{C'}_j) - cut(C_j, \bar{C}_j) \\
 &=& cut(C_j, \bar{C}_j) + |N_G(v) \cap V(\bar{C}_j)| - cut(C_j, \bar{C}_j)  \\
 &=& |N_G(v) \cap V(\bar{C}_j)|   \\
 &=& |N_G(v)| - |N_G(v) \cap V(C_j)|,   \\
\end{eqnarray*}
where $C_j' = C_j \cup \{v\}$. The connection to the selection criteria of Eq. \ref{eq:selection_criterion} can be seen by noting that the minimization of $\Delta cut_j$ corresponds to choosing a cluster $j$ for the vertex $v$ which maximises $|N_G(v) \cap V(C_j)|$. In this sense, the values of $\alpha|N_G(v)|$ and $\beta$ can be seen respectively as confidence thresholds on the number of neighbours in the chosen cluster and the degree of the vertex at hand.

\section{Complexity} \label{sec:timecomplexity}

The proposed \textsc{CoreCluster}  procedure executes the graph clustering algorithm ({\bf Cluster}$(G)$)  on the subgraphs induced by the subsets $S_{k},\ldots,S_{0}$. 
Each $S_{i}$ corresponds to a subset of vertices of $V_{i}$ that cannot be assigned  to already existing clusters according to the selection procedure (see Step 8 of Alg. \ref{alg:CoreClusterFramework}).
Thus, the selection step makes it possible for some of the vertices in $V_{i}$  to be incorporated in an already created cluster, reducing  the burden of the computation of ${\bf Cluster}(G[S_{i}])$.

In this way, it becomes apparent that the speed up of the algorithm is attributed to the fact that the \textsc{CoreCluster}$(G)$ runs in $k+1$ disjoint subgraphs of $G$ instead of $G$ itself. 
As the $i$-th  selection phase requires $O(|V(G_{i})|^3)$ steps, we conclude that the running time of \textsc{CoreCluster}$(G)$ is bounded by 
\begin{eqnarray}
  \sum_{i=k,\ldots,0} O(|S_{i}|^{3})\leq\!\!\!\!\sum_{i=k,\ldots,0}
  O(|V_{i}|^{3})\leq O(k\cdot n_{\rm max}^{3})\label{bound},
\end{eqnarray}
where $n_{\rm max}\!=\!\max\{|V_{k}|,\!\ldots,\!|V_{0}|\}$.
In the above bound, the first  equality holds only in the extremal case where no selection occurs during the selection phases.
Clearly, the general bound in Eq. \ref{bound} is the best possible when $|V_{k}|,\!\ldots,\!|V_{0}|$ tend to be equally distributed (which would accelerate the running time by a factor of $k^{2}$).
According to the first inequality of Eq.~\ref{bound} the running time of the algorithm is proportional to $(k+1)\cdot n_{\rm max}^{3}$, where $n_{\rm max}=$ $\max\{|S_{k}|,$ $\ldots,|S_{0}|\}$.
Let $\rho_{G} =\max\left\{\frac{|V(G)|}{|S_{i}|} \mid i=0,\ldots,k\right\}$ and $\mu_{G}=\max\left\{\frac{|V(G)|}{|V_{i}|}\mid i=0,\ldots,k\right\}$.
Therefore, we observe that $\rho_{G}\geq \mu_{G}$.
Notice that the discrepancy between $\rho_G$ and $\mu_G$ is a measure of the acceleration of the algorithm because of the selection phases. 

Concluding, the acceleration of \textsc{CoreCluster} is upper bounded by 

\begin{eqnarray*}
  \sum_{i=k,\ldots,0} O(|S_{i}|^{3})= O\bigg(\frac{k}{\rho_{G}^{3}}\cdot n^{3}\bigg)
  \label{boundfinal}.
\end{eqnarray*}

\noindent This estimation is purely theoretical and its purpose is to expose the general complexity contribution of our algorithmic machinery.
In practice, the acceleration can be {\em much better} and this also depends on the heuristics that are applied at the selection phase.

\section{Experimental Evaluation} \label{sec:experiments}

The \textsc{CoreCluster} framework is intended to be used with algorithms of high computational complexity as its' primal purpose is to ``lower'' the cost of such algorithms and apply them only on the ``important parts'' of a graph. 
This can be demonstrated simply by utilizing an expensive algorithm (see Section~\ref{sec:spectal_alg}) as we do in our experiments. The computational cost becomes obviously smaller as the sum of the parts (in processing time) is smaller than the total for algorithms with high complexity (e.g., $O(n^3)$).

The degeneracy behaviour of graphs might differ along with their other properties (e.g., density, etc.).
Thus, in this section, we introduce additional experiments to study the effectiveness of \textsc{CoreCluster} framework under different scenarios.
The purpose of these experiments is to identify in a clear manner the properties of the graphs and the utilized algorithms that best optimize its' performance.

\subsection{Datasets description}\label{sec:datasets}

While real networks are the objective, actual datasets lack ground truth which leaves only evaluation metrics of the quality of clustering as an option and not
direct comparison. 
On the other hand, artificial networks  offer ground truth and a large variety of properties that can be parameterized to produce different ``types'' of networks. 
The evaluation of our framework is conducted on both real and artificial networks in order to have  complete and decisive results.

\subsubsection{Artificial Networks}\label{sec:artfc}
We exploit the graph generator by Fortunato \cite{rohtua}  to produce graphs with a clustering structure which is available to the tester (ground truth).
This graph generator provides a wide range of input parameters. 
We used the parameters presented on Table~\ref{table:first} and tuned them accordingly, by considering various combinations in order to get a wide range of graphs with different features. Thus, the testing of our approach is credible as it is evaluated in essentially hundreds of graphs with different properties and quality of the clustering structure. 
The parameters used are: $N$ is 
the size of the graph, $max_d$ is the maximum node degree, $min_d$ is 
the minimum node degree and  $\mu$ is the mixing parameter representing the
overlapping between clusters, i.e., each node shares a fraction $1-\mu$ of its
links with the other nodes of its community and a fraction $\mu$ with the other
nodes of the network. Graphs produced by the generator contain inherent
clusters and the cluster assignment is offered by the generator, enabling thus
usage of these data sets for evaluating graph clustering algorithms. In order to ensure a thorough and robust evaluation, for each of the combinations among the parameters, we generated 10 distinct graphs. 

Table~\ref{table:first} depicts the various parameters' values for the three main different settings in our experiments. 
It becomes apparent that the most important parameter is the $min_d$,  as it is the one differentiating the overall density of the graph. 

\begin{table}[h]
		\begin{center}
			\begin{tabular}{cccc}
				\toprule
				&D1&D2&D3\\ 
				\midrule
				$max_d$ & & & \\  
				(node max degree)& 10\%, 30\%, &10\%, 30\%,  50\%& 200 edges\\
	
				\midrule
				$min_d$ &	$\sim$5 && \\
				(node min degree)&(the absolutely minimum)&	 7	&20\\ 
				&  & &\\ 
				\midrule
				$\mu$ &	1\% -- 43\%& 	3\% -- 43\% &	3\% -- 43\%\\
				(mixing parameter)&(in 7 equal steps)&(in 6 equal steps)& (in 6 equal steps)\\
				
				\midrule
				$N$  &&&	\\
				(graph size in nodes) &100$-$3600& 	100$-$3600& 	100$-$3600\\
				
				\bottomrule
				
			\end{tabular}
		\end{center}
		\caption{Parameters' values for the artificial graphs.}\vspace{-4mm}
		\label{table:first}
\end{table}

\subsubsection{Real Networks}
We also perform evaluations to a subset of the \textit{Facebook 100} dataset provided by \citet{facebook100}. 
This is a collection of friendship networks of \textit{Facebook} from 2005, for 100 US Universities (i.e., 100 individual networks). The evaluations were not
performed to the full extend of this dataset as hardware limitation did not
allow us to evaluate, with spectral clustering, networks with more than $13K$
nodes (the \textsc{CoreCluster} framework could handle much larger networks).
About half of the networks from this dataset were used for the final evaluation.

\subsection{Evaluation Metrics}\label{sec:metrics}

To evaluate the clustering results we use different metrics for the artificial and the real (\textit{Facebook 100}) data.
\begin{itemize}
	\item {\bf{\em Artificial networks:}} Given the ground truth, we measure the quality of the clustering results in terms of the widely used Normalized Mutual
	Information \citep[NMI,][]{DBLP:books/daglib/0021593}. NMI measures how ``clear'' each found cluster is in relevance to the known clusters (taking into account the size of the cluster as well).
	
	Let $ \Omega = \{\omega_1, \omega_2, \ldots , \omega_k\}$ be the set of detected clusters and $\mathbb{C}=  \{c_1, c_2, \ldots , c_j\}$ the set of known clusters (ground-truth). Then, NMI is given as follows:
	\begin{equation}
	\textrm{NMI}(\Omega,\mathbb{C})= \frac{I(\Omega,\mathbb{C})}
	{[H(\Omega)+H(\mathbb{C})]/2},
	\end{equation}
	where $I$ is the mutual information:
	\begin{equation}
	I(\Omega;\mathbb{C})=\sum_{k}\sum_{j}\frac{|\omega_k \cap c_j|}
	{N}\log\left(\frac{N|\omega_k \cap c_j|}{|\omega_k||c_j|}\right) ,
	\end{equation}
	and $H$ is the Entropy:
	\begin{equation}
	H(\Omega)= - \sum_{k}\frac{|\omega_k|}{N}\log\left(\frac{|\omega_k|}{N}\right).
	\end{equation}
	
	\item {\bf{\em Facebook:}}  Since the networks of this dataset lack ground truth, we choose to evaluate the results with the 
	evaluation criterion of conductance. Given a graph $G$ and a cut $(S,\overline{S})$, conductance is defined as:
	\begin{equation*}
	\phi(S)=\frac{\sum_{i\in{S},j\notin{S}}{A_{ij}}}{\min{(a(S),a(\overline{S}))}},
	\end{equation*}
	where $A_{ij}$ are the entries in the adjacency matrix $\mathbf{A}$ of  G and
	$a(S)=\sum_{i\in{S}}{\sum_{j\in{G}}{A_{ij}}}$. Informally, conductance  measures
	(for a cluster) the ratio of internal to external connectivity. It has been used
	widely to examine clustering quality (e.g., \cite{leskovec-www10}) and has a
	simple and intuitive definition.
\end{itemize}
 Both NMI and Conductance measures takes values in the range of $(0,1)$. It is also important to be noticed that:
 \begin{enumerate}
 	\item For NMI, \textbf{higher} values are better.
	\item For Conductance, \textbf{lower} values are indicating better clustering quality.
\end{enumerate} 

\subsection{Algorithms for Additional Evaluation}\label{sec:Algorithms}
  
Apart from the spectral clustering algorithm proposed by \citet{NgJW01onsp}, we expand the study of our framework to a wide variety of algorithms. The purpose of this extended study is to evaluate the clustering performance of \textsc{CoreCluster}  various properties these algorithms might display. At the same time, we examine whether the properties of the graphs  affect the framework in a consistent manner (i.e., the performance is good for the same ``type'' of graphs across different algorithms). The latter is justified through this extended analysis and establishes clear conditions for the use \textsc{CoreCluster}.

 We considered algorithms indiscriminately of their computational complexity, focusing more on how they approach the clustering problem (i.e., what they try to optimize). In the following, we list in sort the clustering  algorithms that have been considered in our analysis.
\begin{itemize}
	\item {\bf{\em{InfoMap}}} \citep{rosvall2009map}: An information theoretic approach that tackles the problem of graph clustering as a compression problem. For random walks over a graph the probability of flow is converted to a description. By minimizing the description length of the nodes in the flow's path, InfoMap optimizes the network's partitions. 
	\item {\bf{\em{Leading Eigenvector}}} \citep{newman2006finding}: In order to optimize modularity, this method uses the eigenvector corresponding to the largest eigenvalue of the modularity matrix. While this is also a spectral method, the calculation of the leading eigenvector is much more efficient as it can be done with the power iteration method.
	\item {\bf{\em{Fast Greedy}}} \citep{PhysRevE.70.066111}: This is another modularity optimization algorithm that works in a greedy manner. It follows an hierarchical agglomerative process, starting from each node being a unique community and joining communities that produce the highest modularity. 
	\item {\bf{\em{MCL}}} \citep{van2000cluster}: The Markov Cluster ({\em{MCL}}) algorithm follows a simple iteration of two steps over the transition probability matrix $T$ of  graph : i) raise $T$ to an integer power and ii) raise the values in the cells of the latter resulting matrix in a real valued power $p$. The two steps ``represent'':  i) the probability that a random walker takes a specific path in $p$ steps and ii)  enhancing  the probability that two nodes are connected in  with ``greater strength'' for larger values from step a. 
	\item {\bf{\em{Metis}}} \citep{metis}: Metis relies on a multilevel approach where the graph is coarsened (vertices are unified and edge weights are adjusted accordingly). Then follows the partitioning of the coarsened graph and the projection of the clusters back to the original one. It is a well established algorithm with low computational cost that has been shown to be very effective. 
	\item {\bf{\em{MultiLevel}}} \citep{blondel2008fast}: The term {\em MultiLevel} is used to refer to the greedy optimization algorithm introduced by \citet{blondel2008fast}. This approach is similar to {\em Fast Greedy} algorithm with the difference that, at each iteration, joined vertices are replaced with single nodes (similarly to metis).
	\item {\bf{\em{SpinGlass}} } \citep{reichardt2006statistical}: A cross-discipline approach, that interprets the community structure of a graph as the spin configuration with the minimum energy of the spin glass (spin glass: disordered magnet, spin: magnetic orientation, the spin states represent the community indices). In essence, this algorithm deals with the adhesion between vertices which is the equivalent of edge betweenness and provides a different approach to the work of \citet{PhysRevE.70.066111}
	\item {\bf{\em{Walktrap}}} \citep{pons2005computing}: The main intuition behind this algorithm is that short random walks will tend to stay in the same communities. Thus the transition probability between two vertices will be high if they are in the same community (for {\em short} random walks and adjusting for the degree of a vertex). By using this property as distance, clustering can be applied with an agglomerative approach (for efficiency).
\end{itemize}

\subsection{Performance}

\begin{figure}[h!]
  \begin{subfigure}{\textwidth}
    \centering
    \includegraphics[scale=0.24]{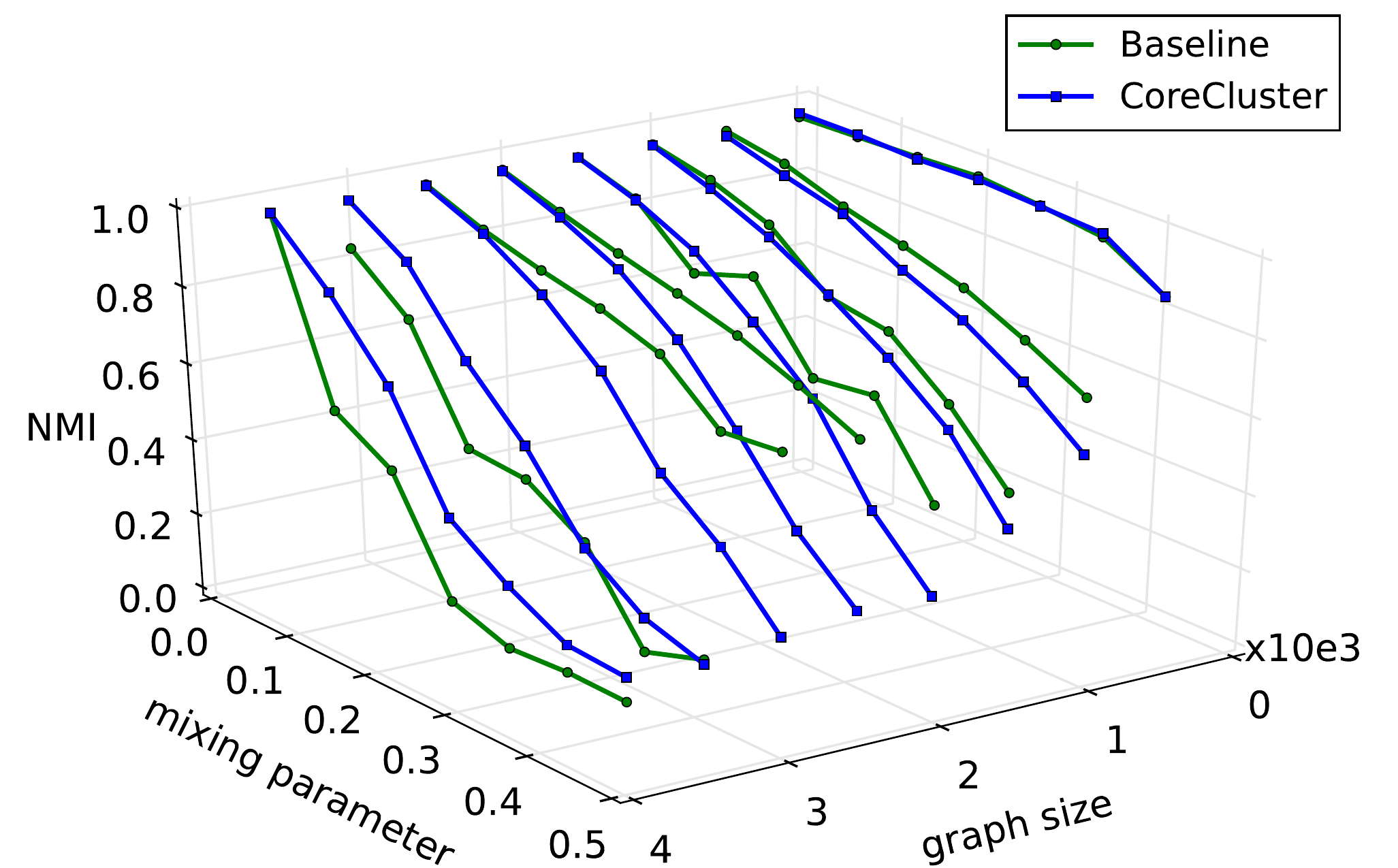}
    \includegraphics[scale=0.24]{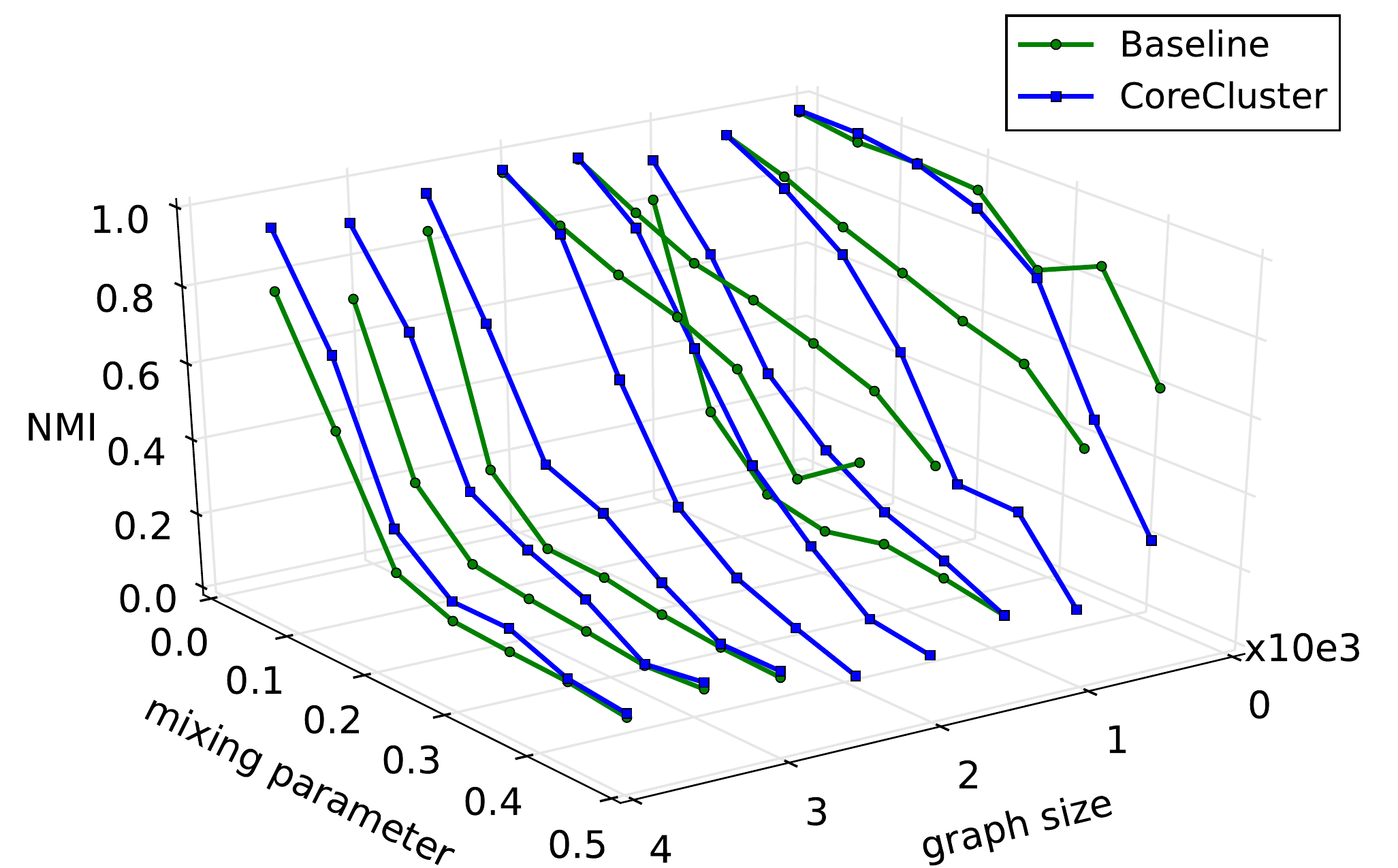}
    \includegraphics[scale=0.24]{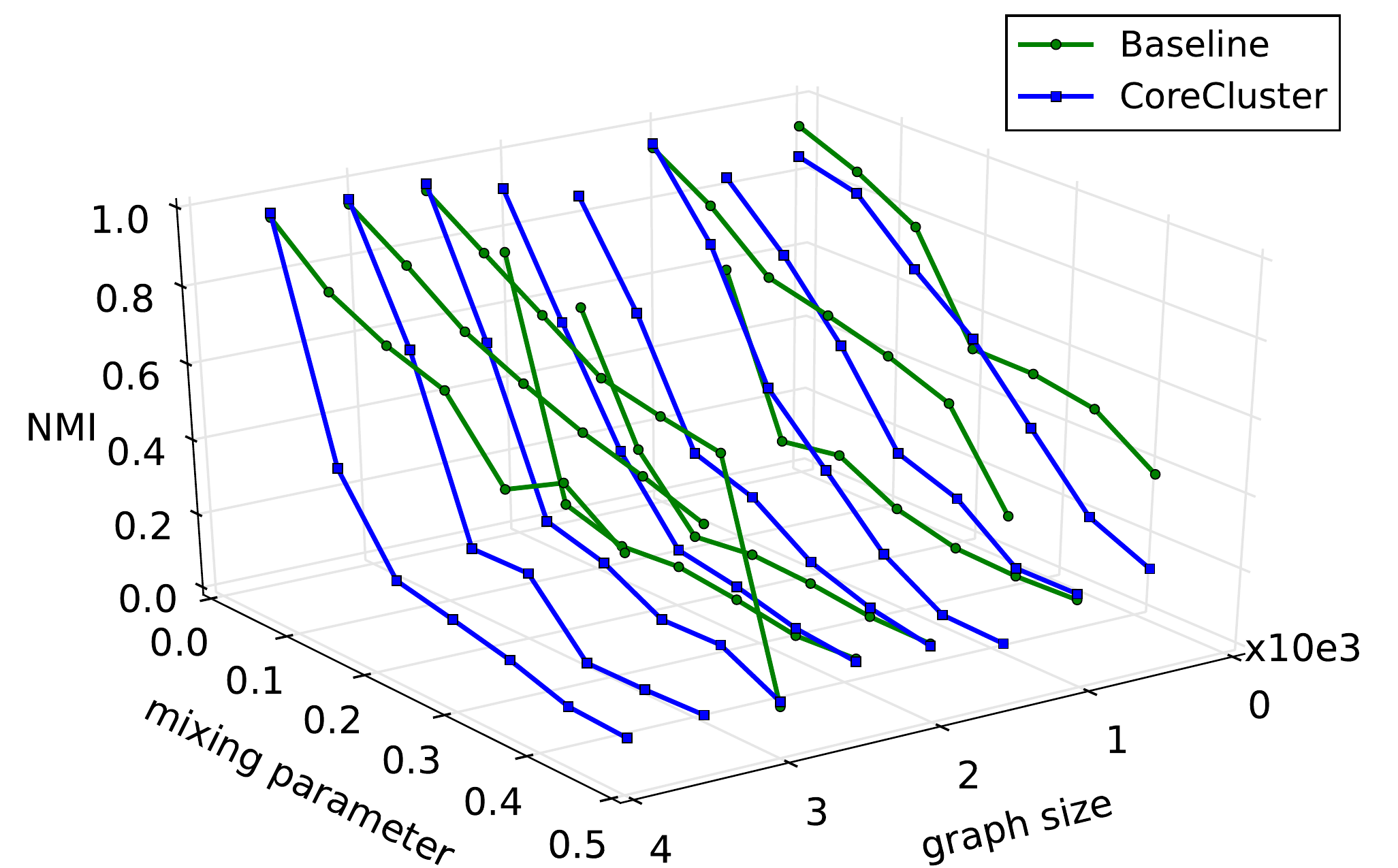}
    \caption{Spectral Clustering}
  \end{subfigure}

  \begin{subfigure}{\textwidth}
    \centering
    \includegraphics[scale=0.24]{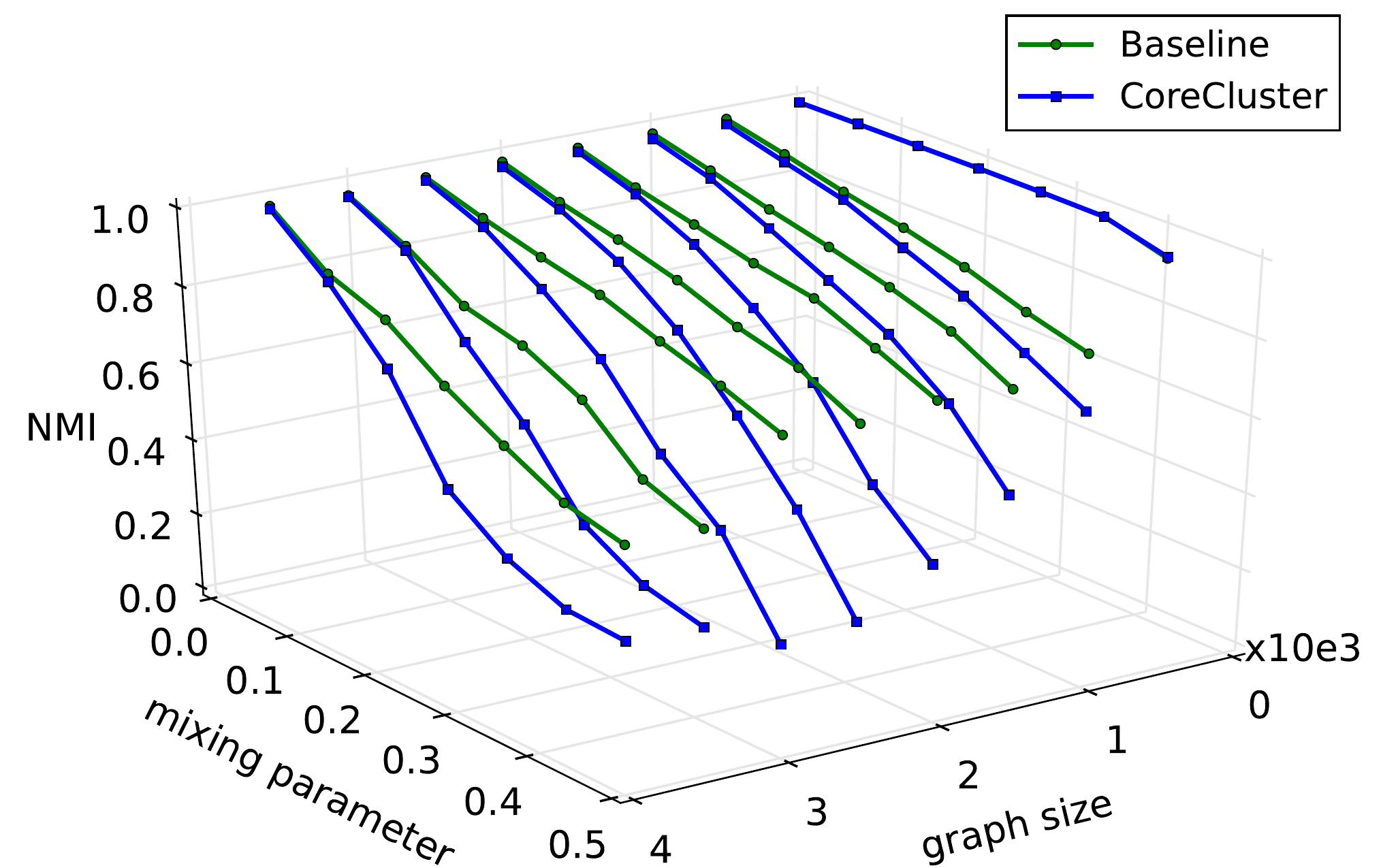}
    \includegraphics[scale=0.24]{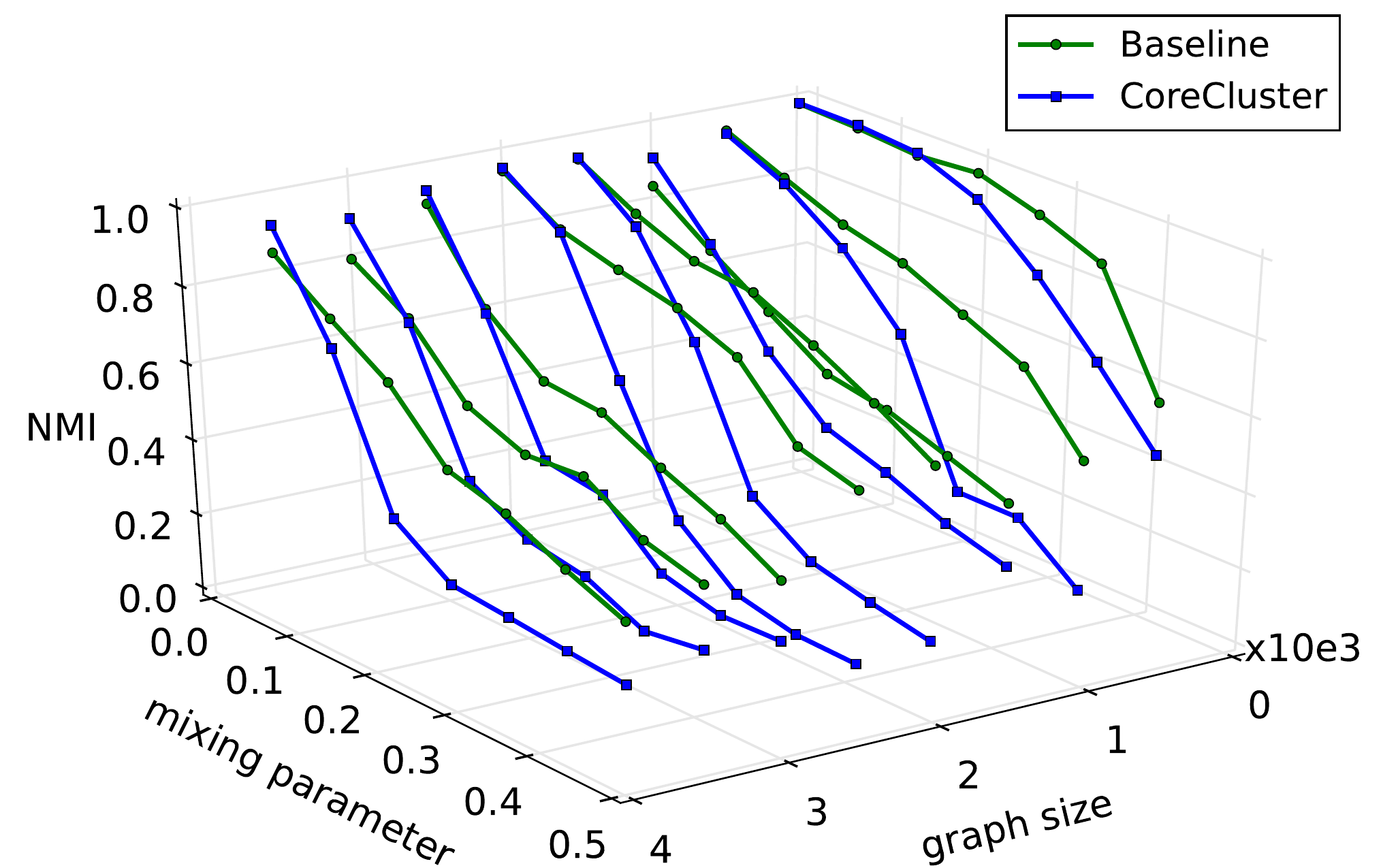}
    \includegraphics[scale=0.24]{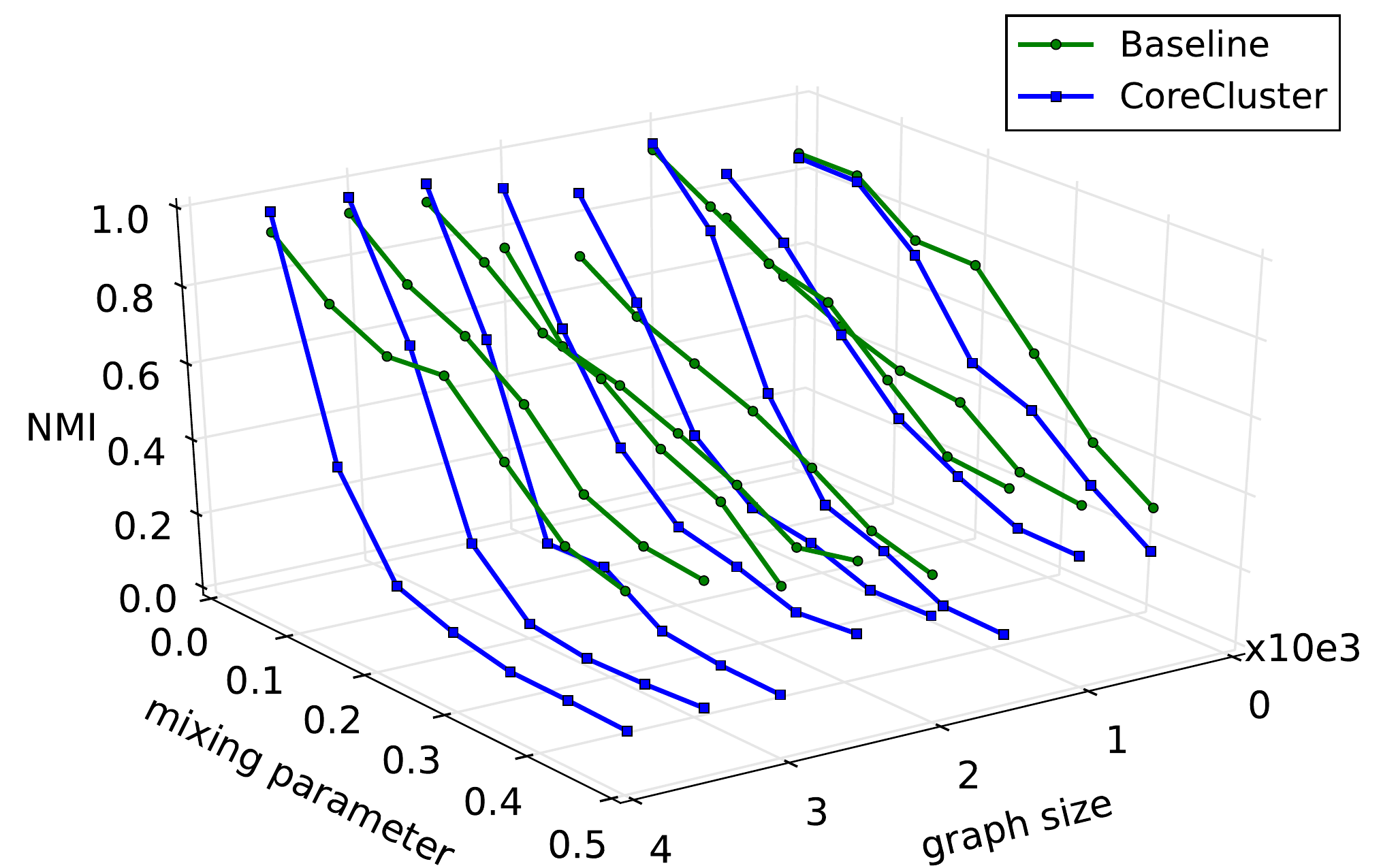}
    \caption{InfoMap}
  \end{subfigure}
  
  \begin{subfigure}{\textwidth}
    \centering
    \includegraphics[scale=0.24]{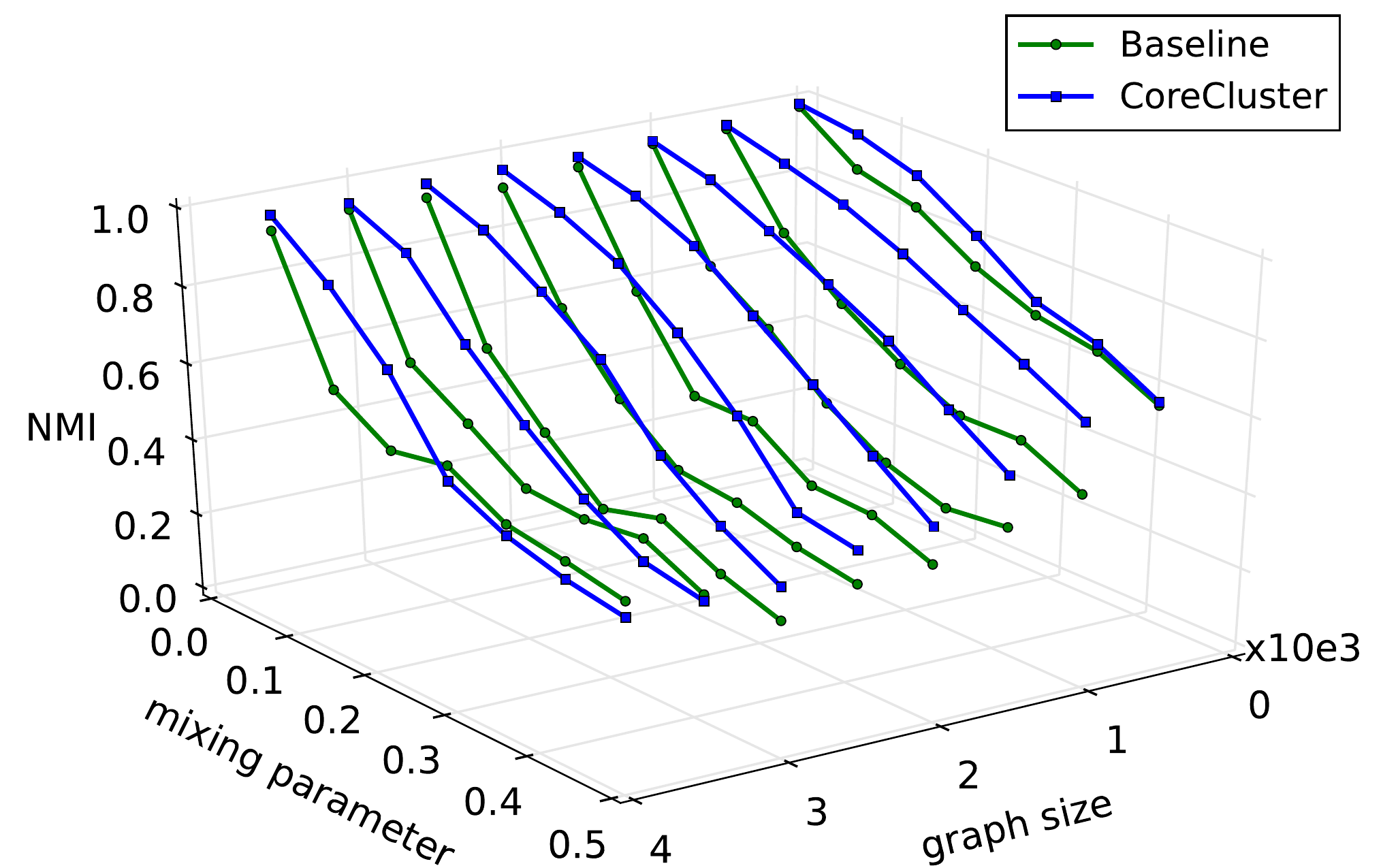}
    \includegraphics[scale=0.24]{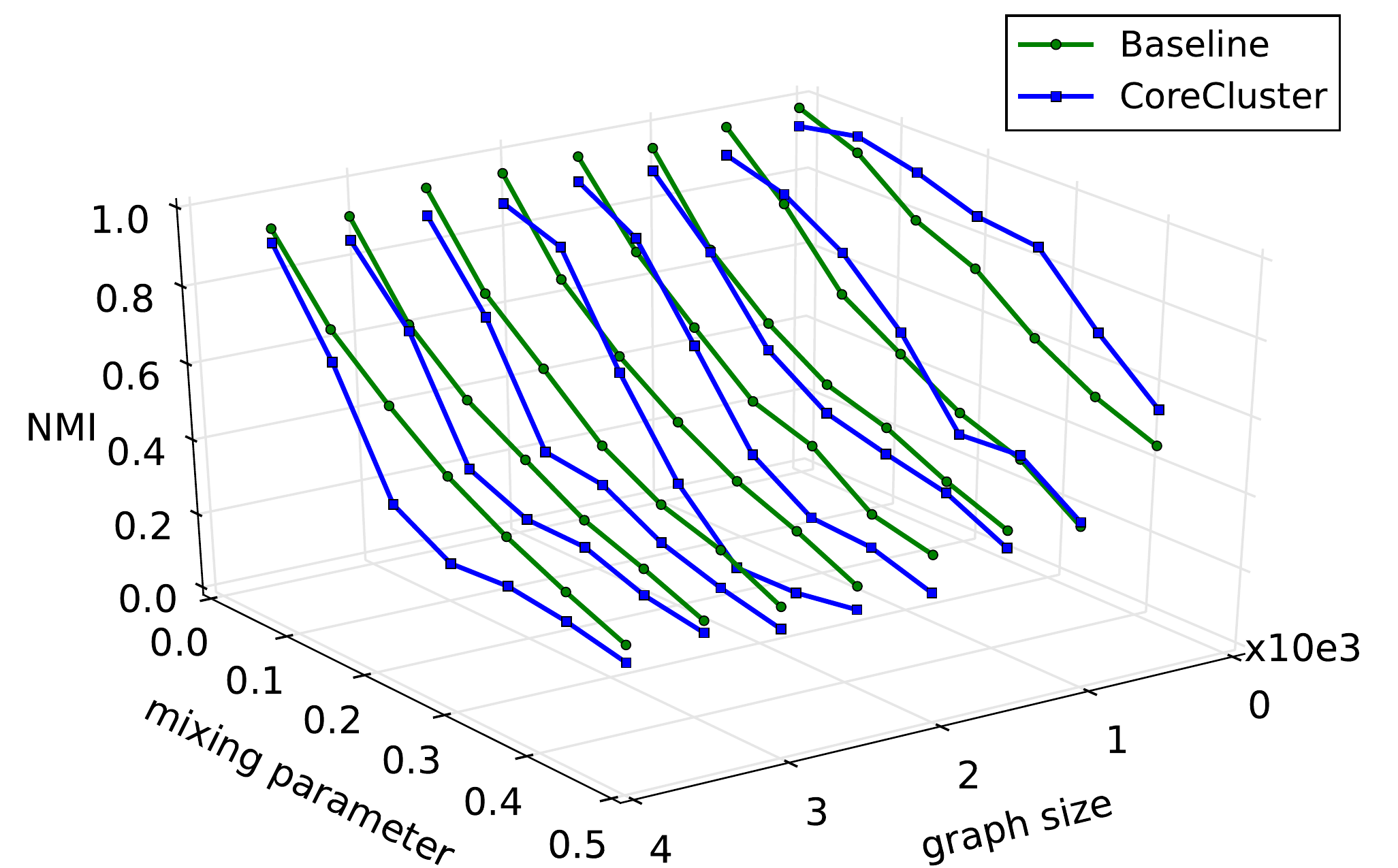}
    \includegraphics[scale=0.24]{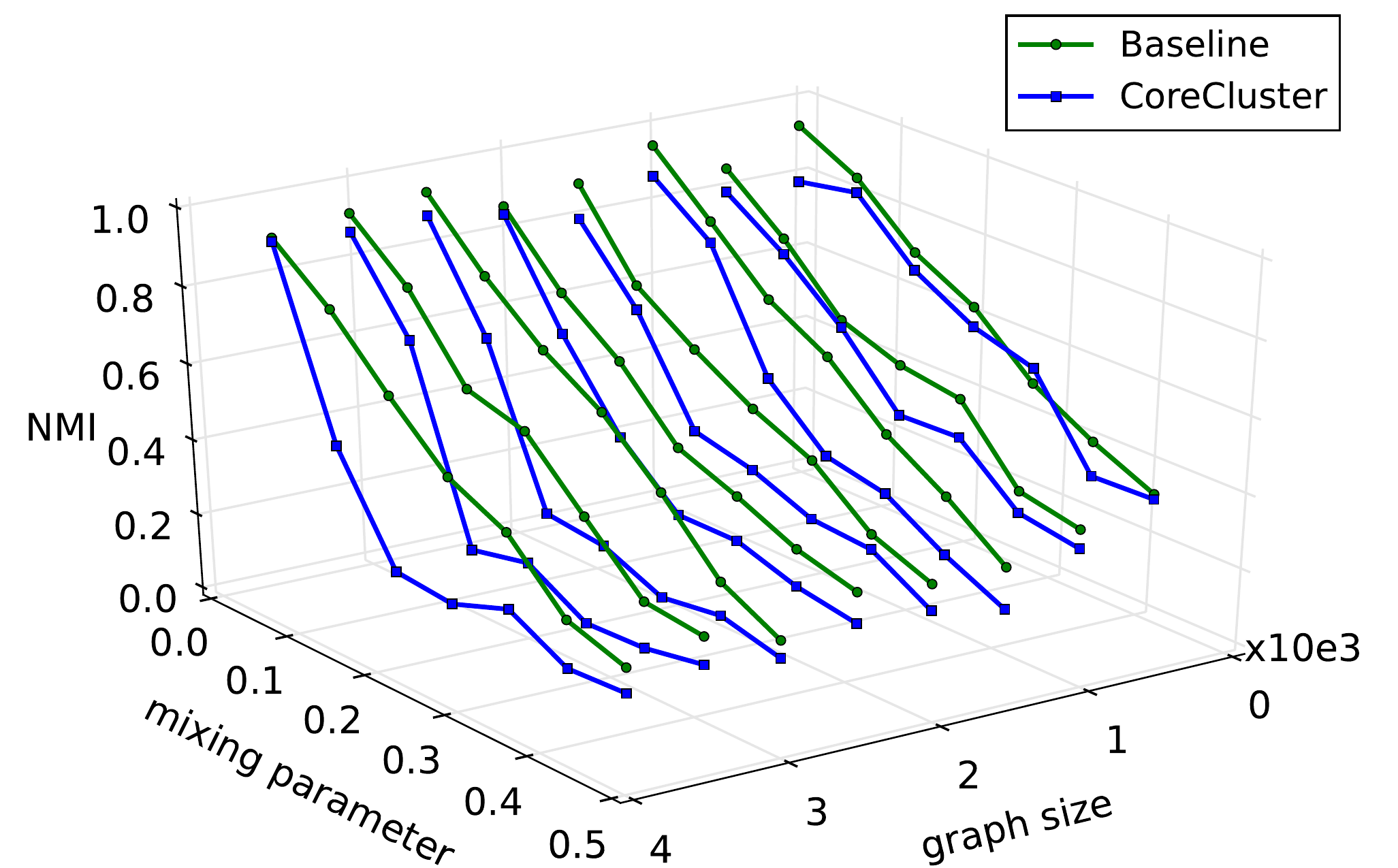}
    \caption{Leading Eigenvector}
  \end{subfigure}

  \begin{subfigure}{\textwidth}
    \centering
    \includegraphics[scale=0.24]{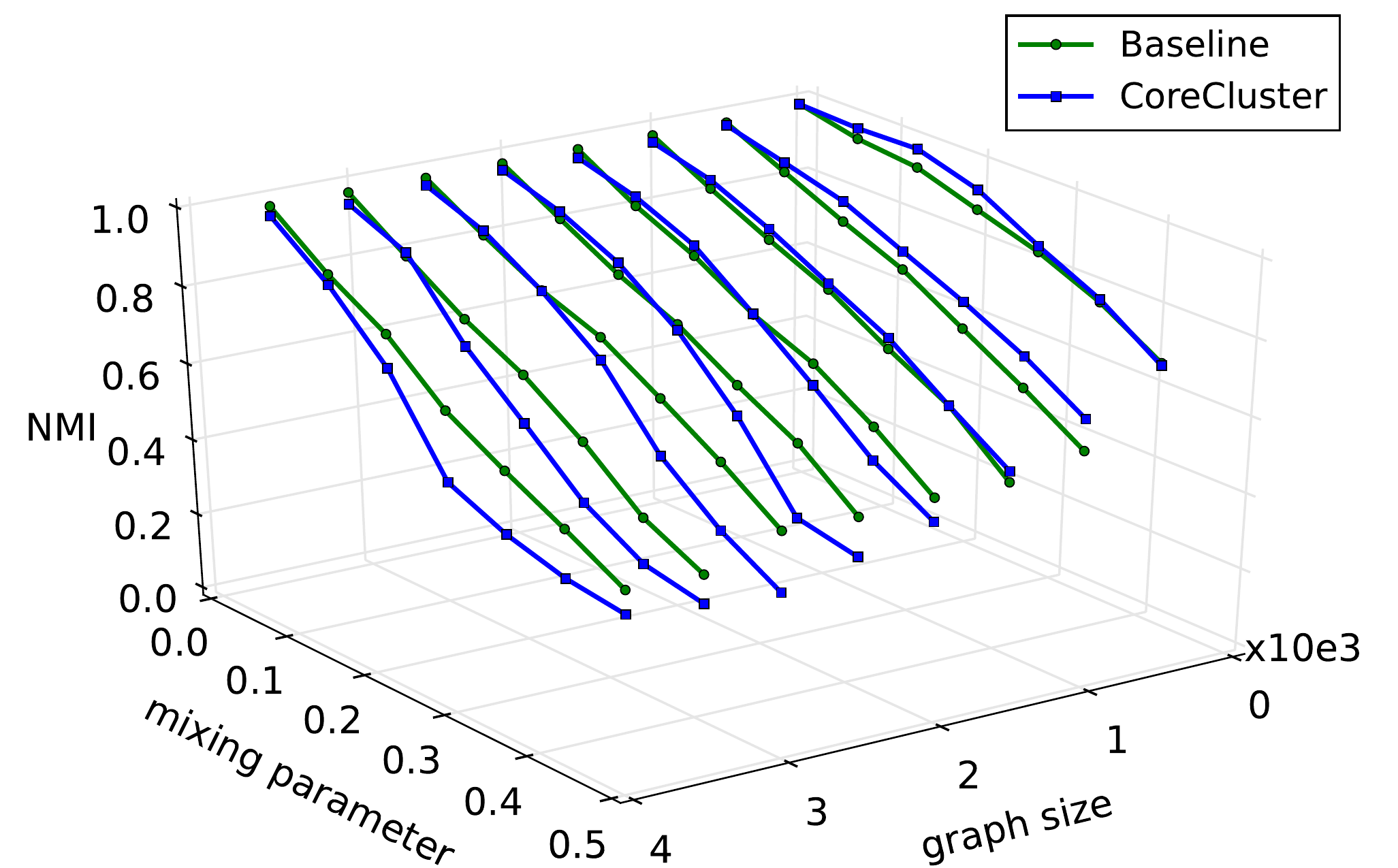}
    \includegraphics[scale=0.24]{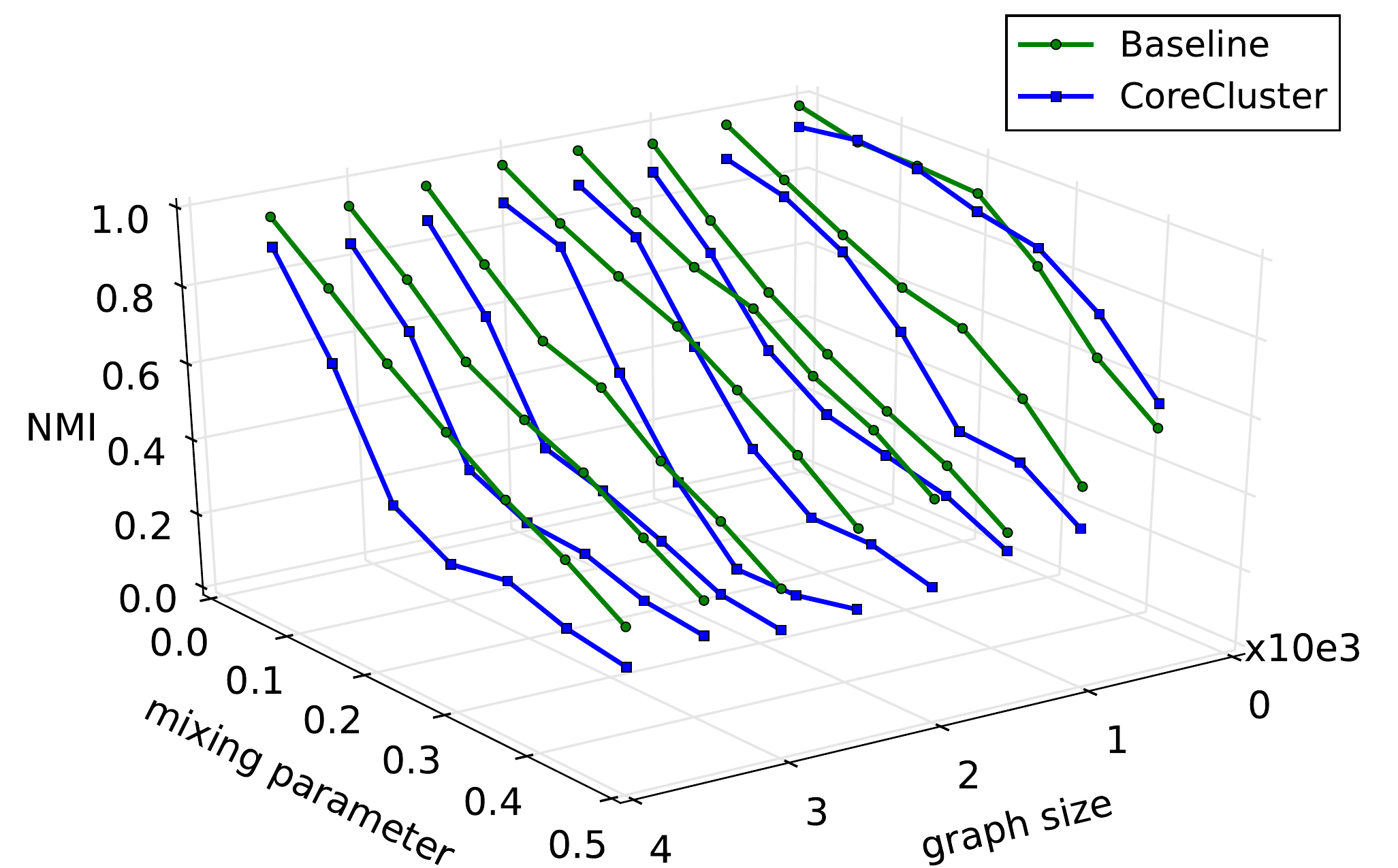}
    \includegraphics[scale=0.24]{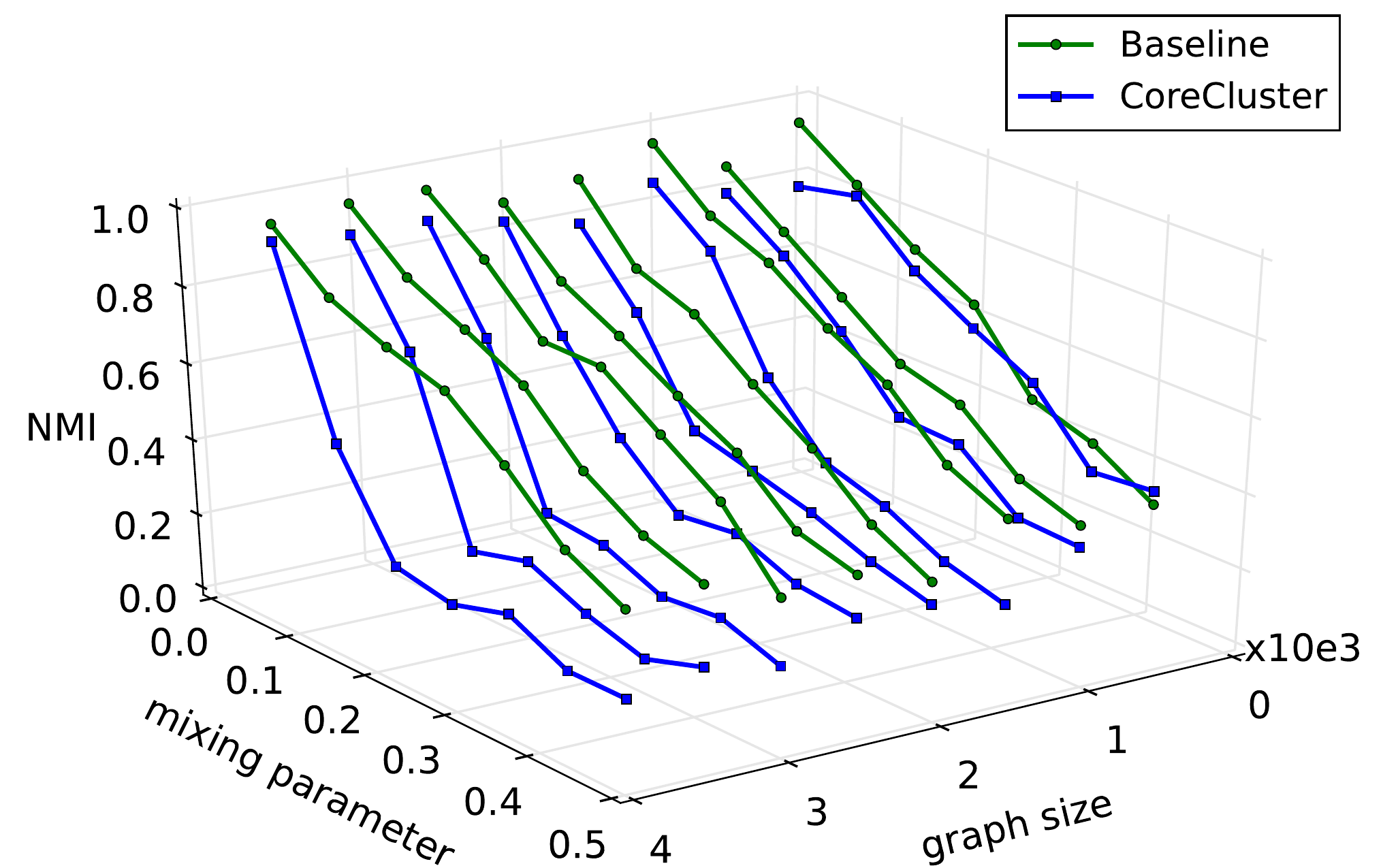}
    \caption{Fast Greedy}
  \end{subfigure}

  \begin{subfigure}{\textwidth}
    \centering
    \includegraphics[scale=0.24]{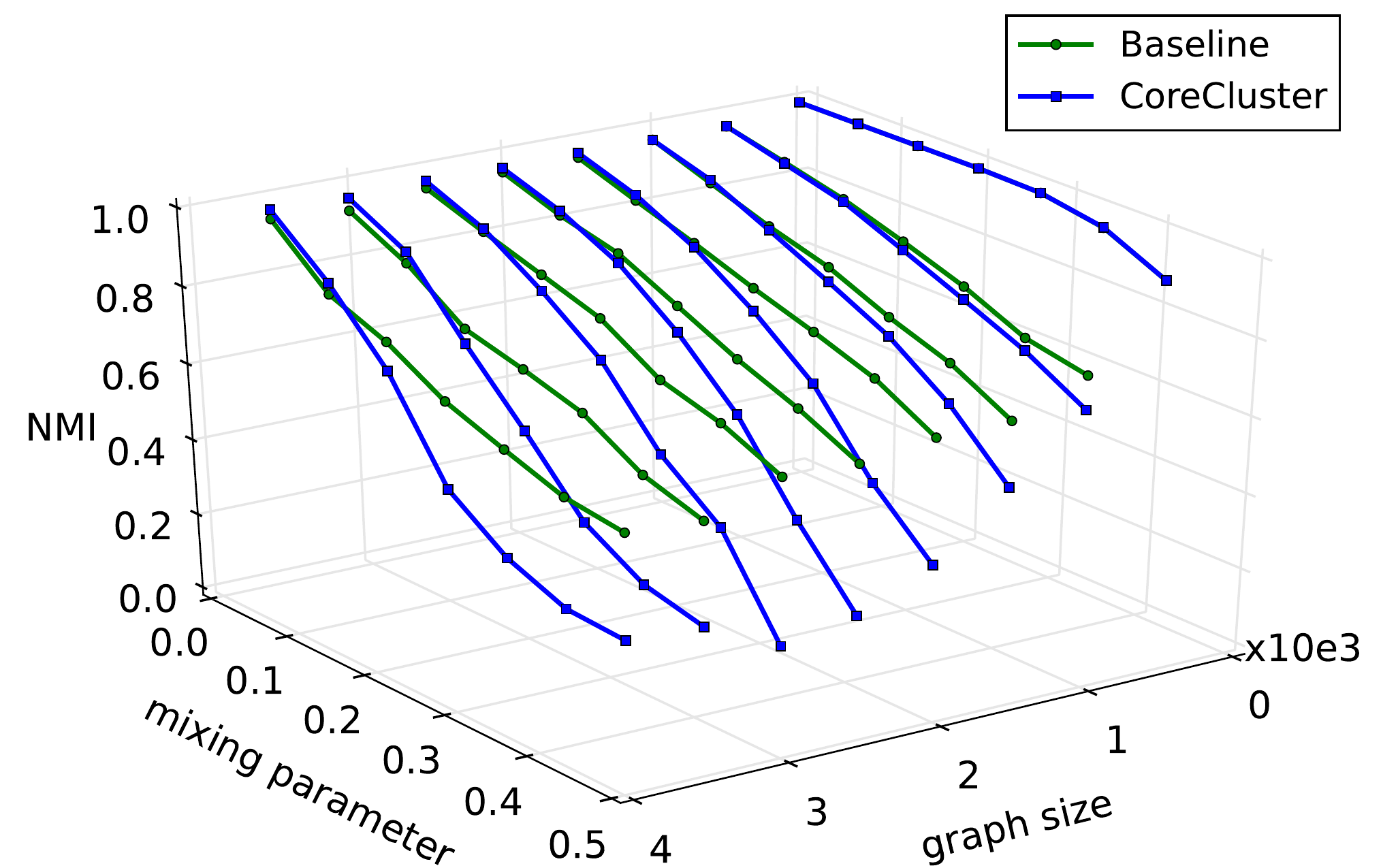}
    \includegraphics[scale=0.24]{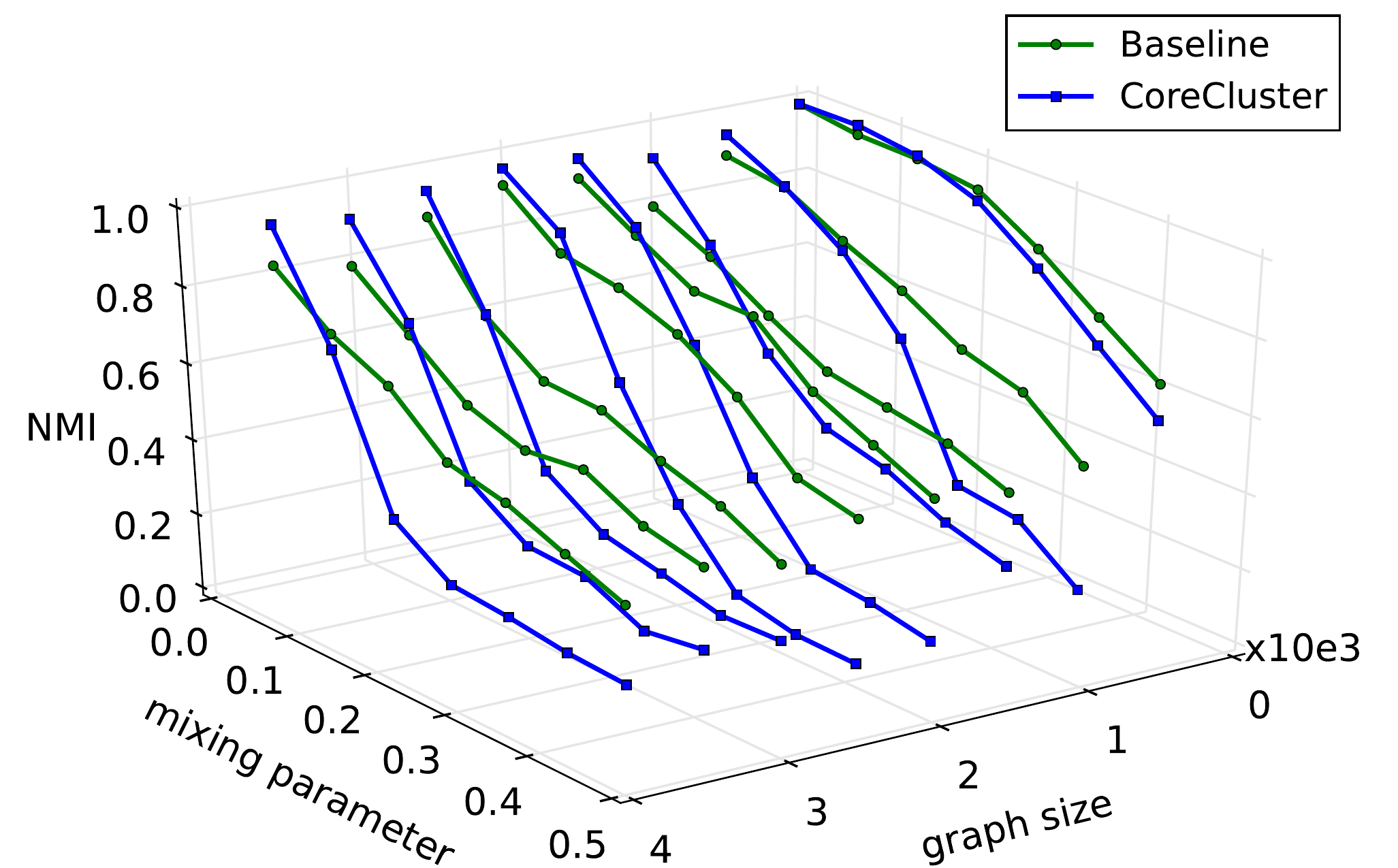}
    \includegraphics[scale=0.24]{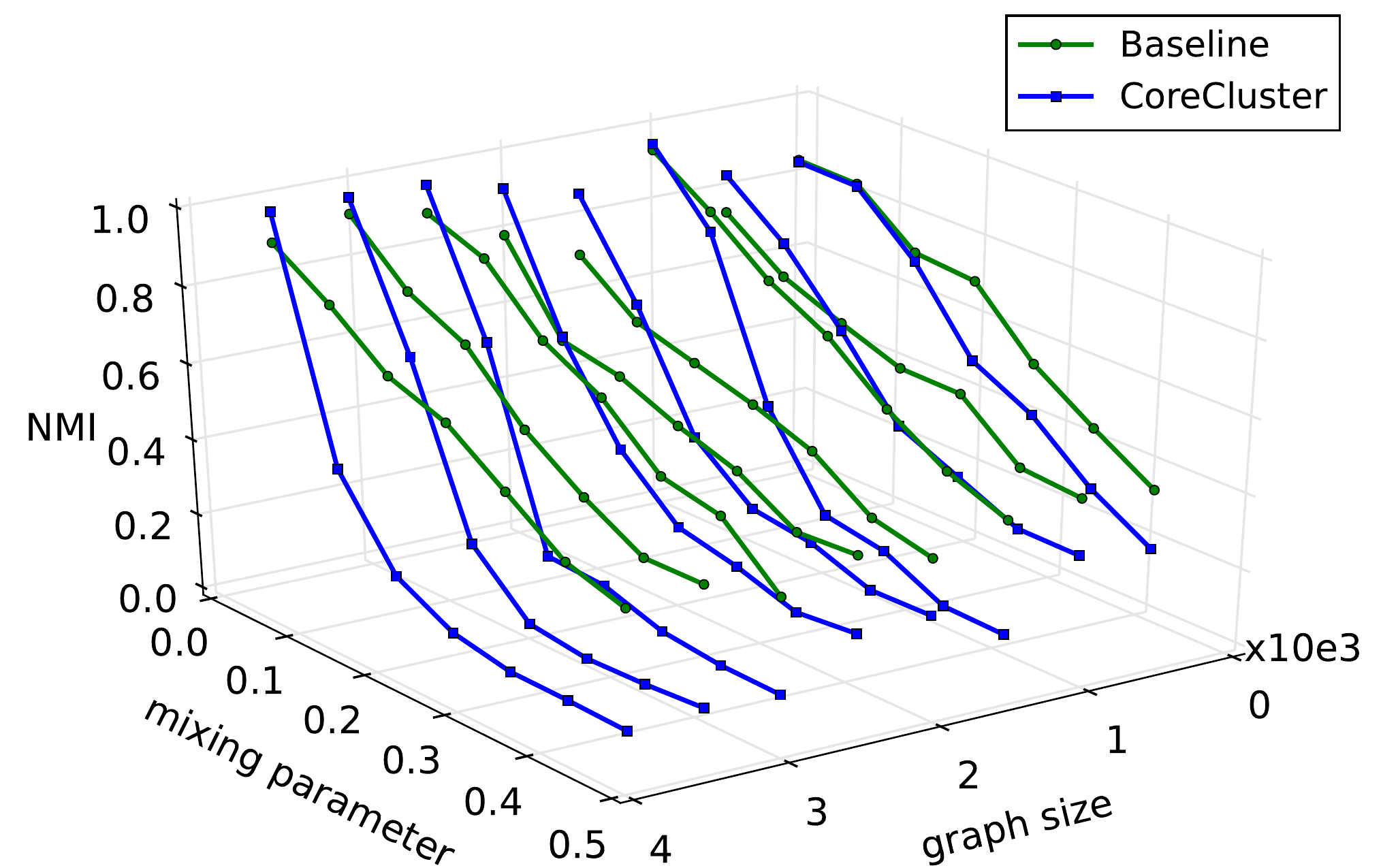}
    \caption{MCL}
  \end{subfigure}
  \caption{Clustering quality comparisons in terms of NMI for the three artificial graphs (D1, D2, and D3) (Part I).}
  \label{fig:NMIcomp}
\end{figure}

\begin{figure}[h!]
  \ContinuedFloat 
  \begin{subfigure}[b]{\textwidth}
    \centering
    \includegraphics[scale=0.24]{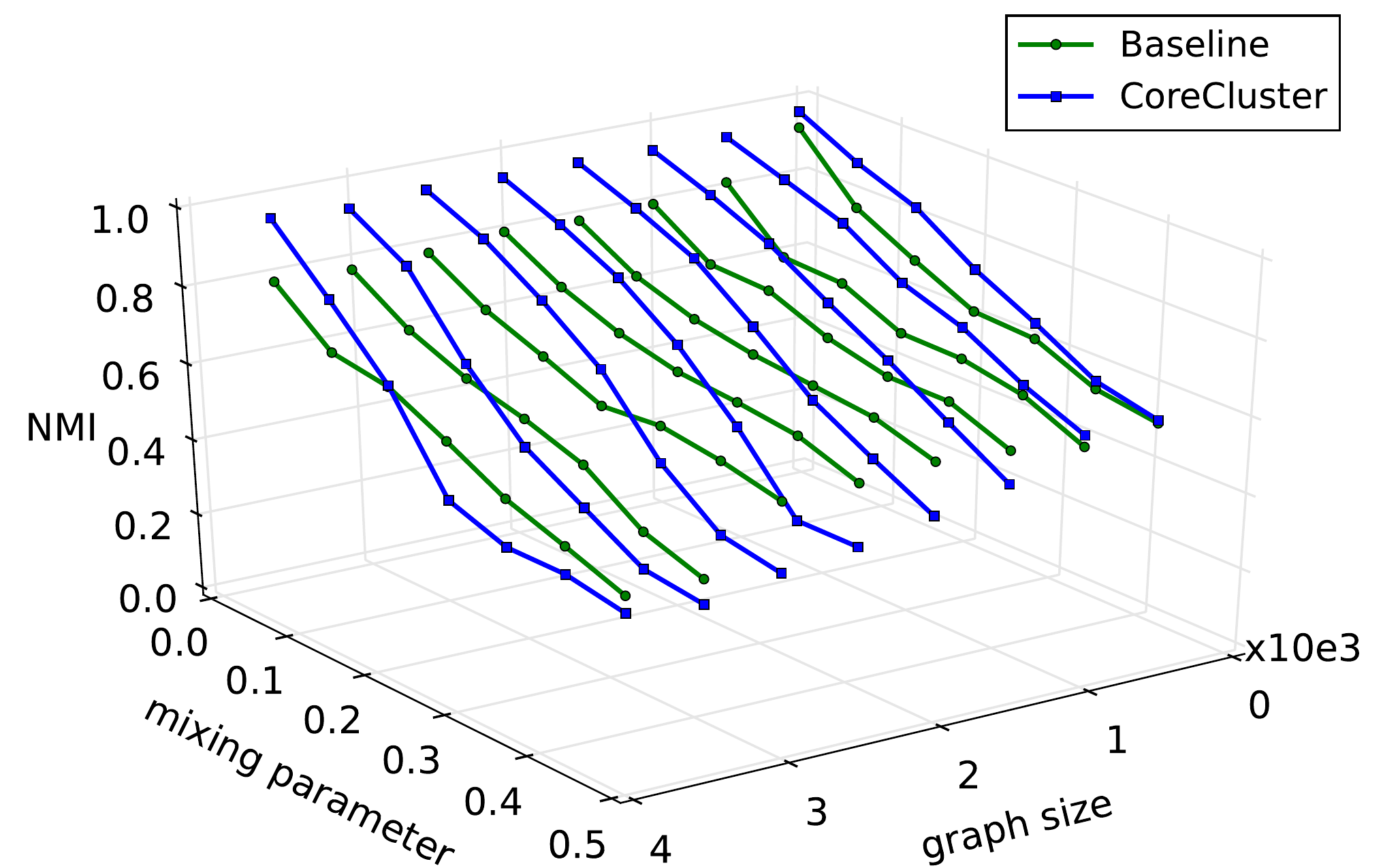}
    \includegraphics[scale=0.24]{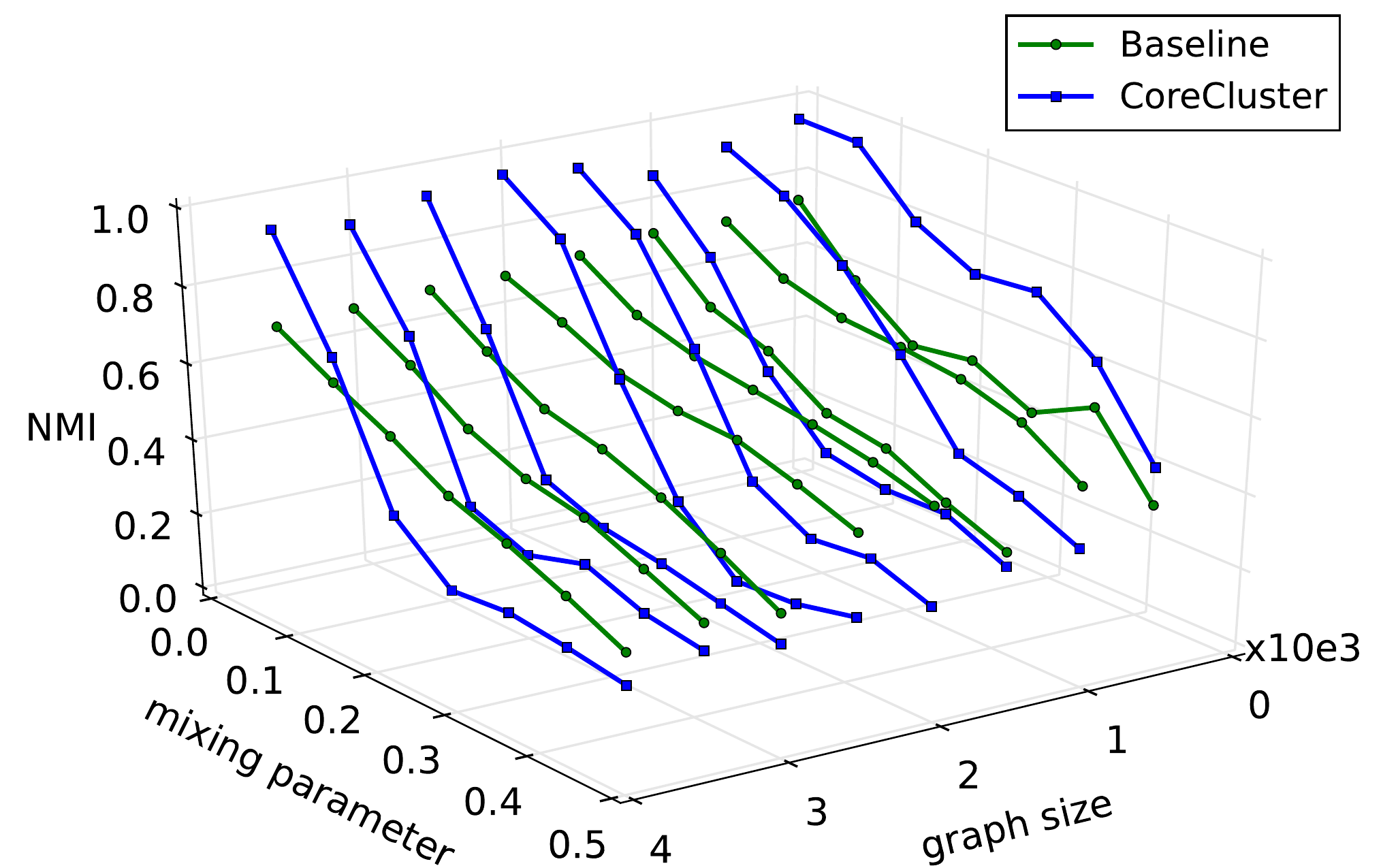}
    \includegraphics[scale=0.24]{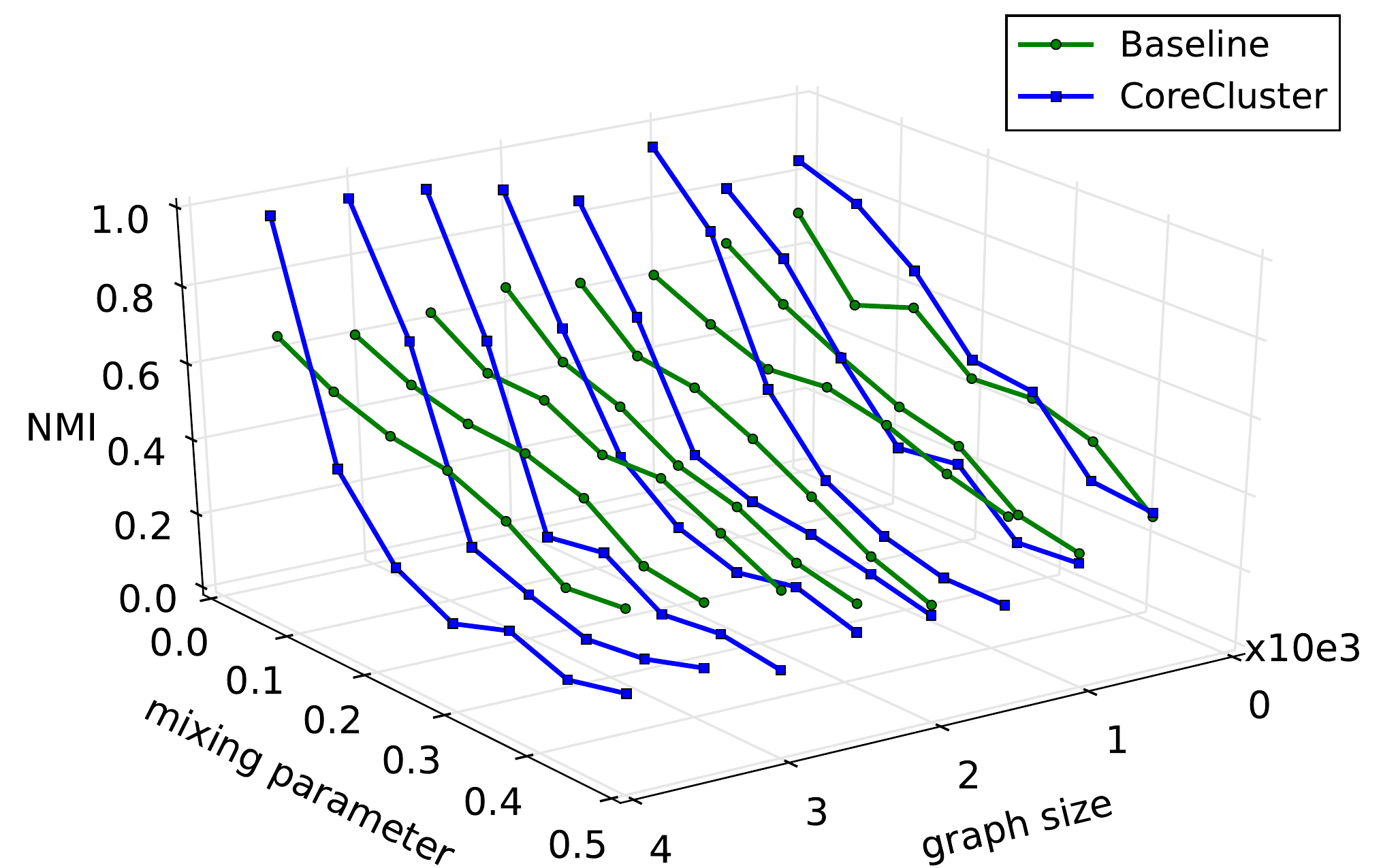}
    \caption{Metis}
  \end{subfigure}

  \begin{subfigure}{\textwidth}
    \centering
    \includegraphics[scale=0.24]{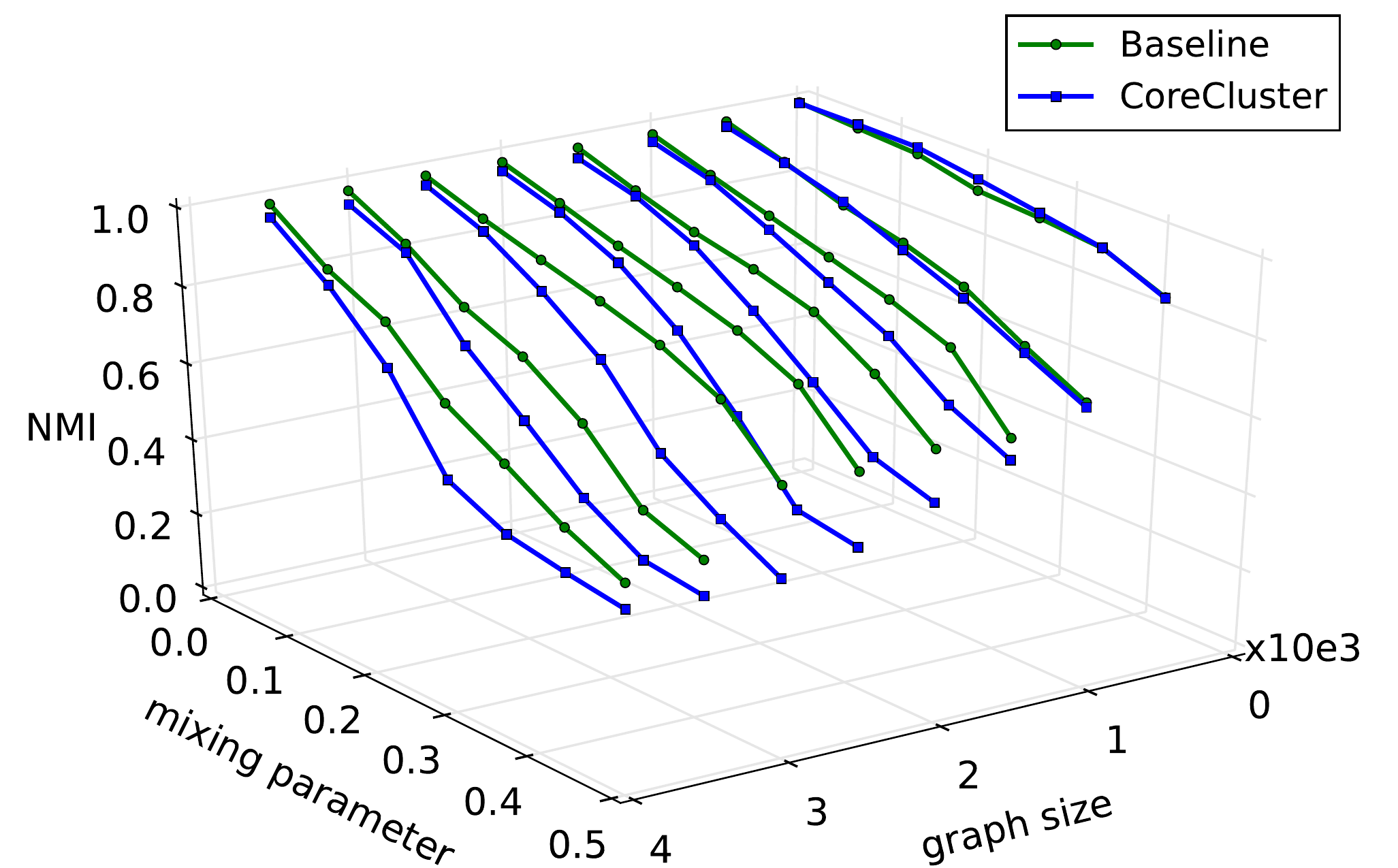}
    \includegraphics[scale=0.24]{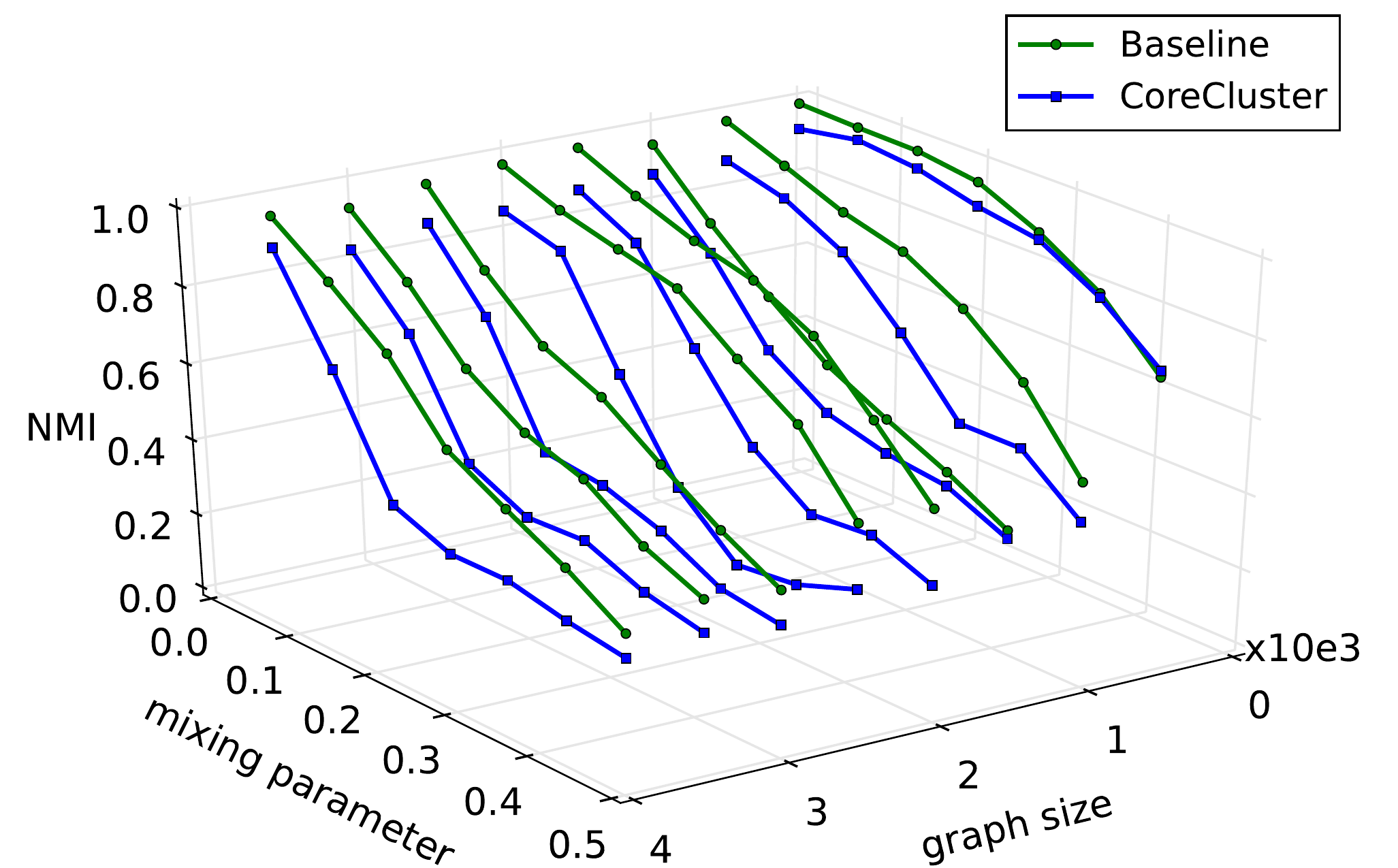}
    \includegraphics[scale=0.24]{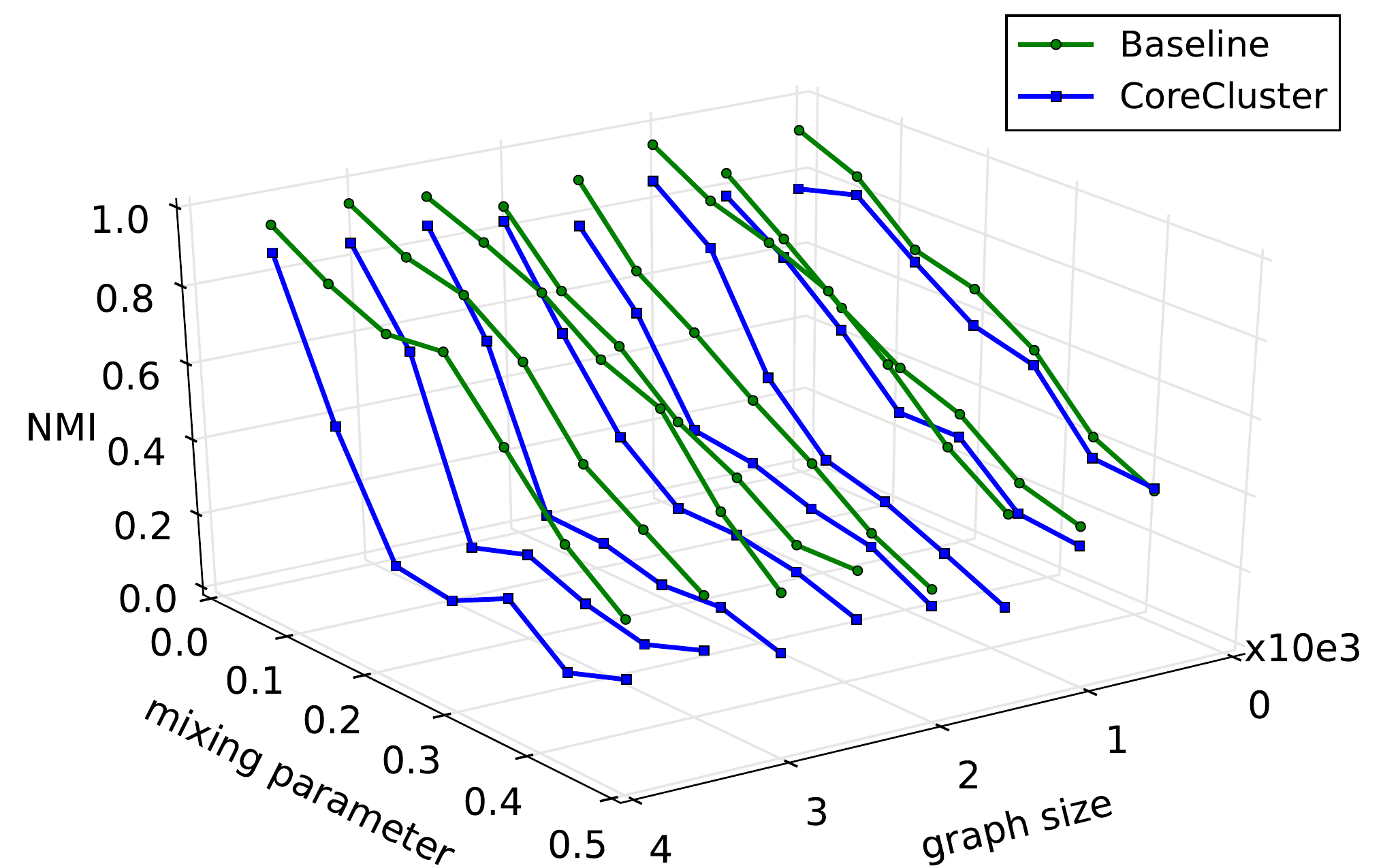}
    \caption{MultiLevel}
  \end{subfigure}

  \begin{subfigure}{\textwidth}
    \centering
    \includegraphics[scale=0.24]{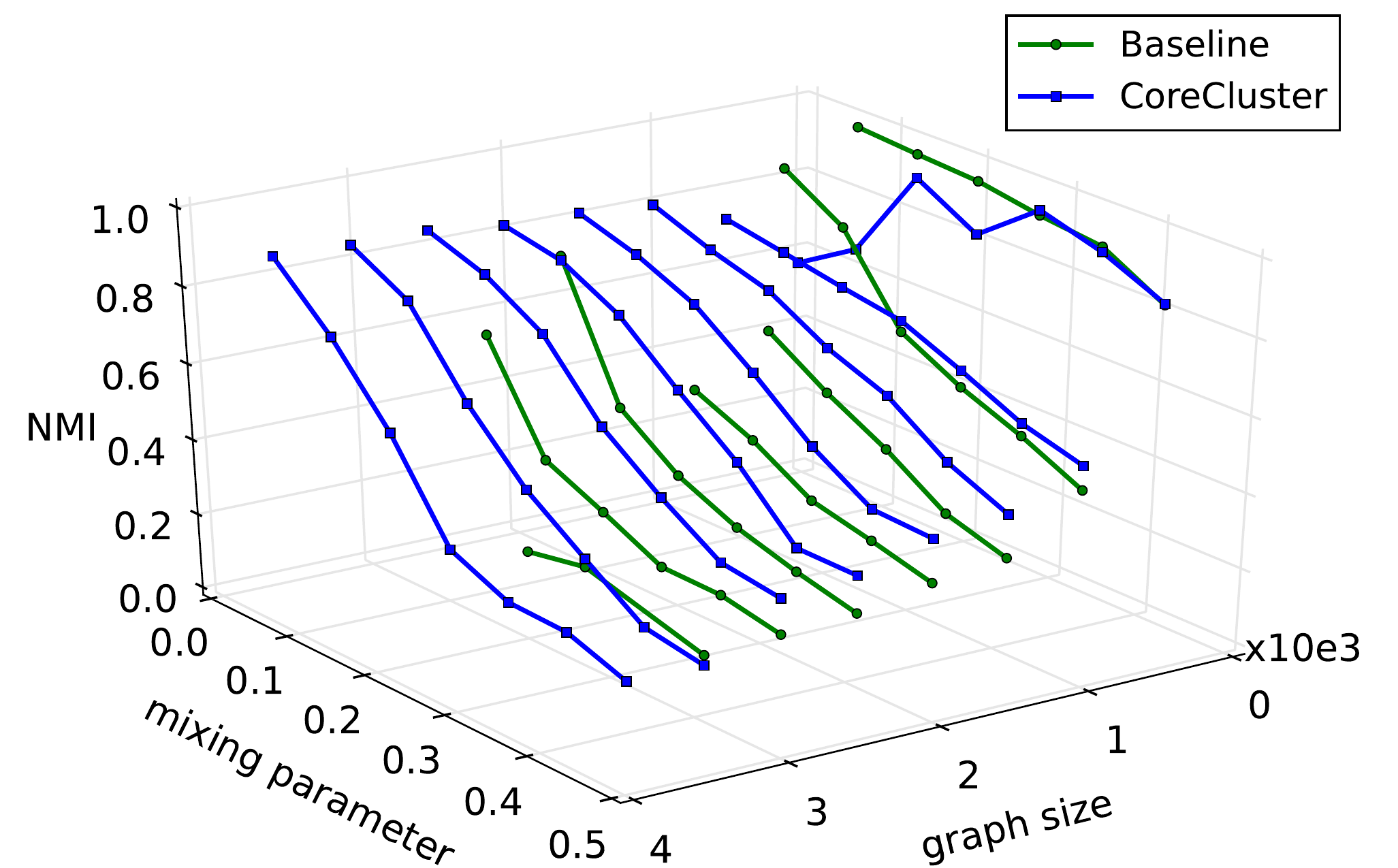}
    \includegraphics[scale=0.24]{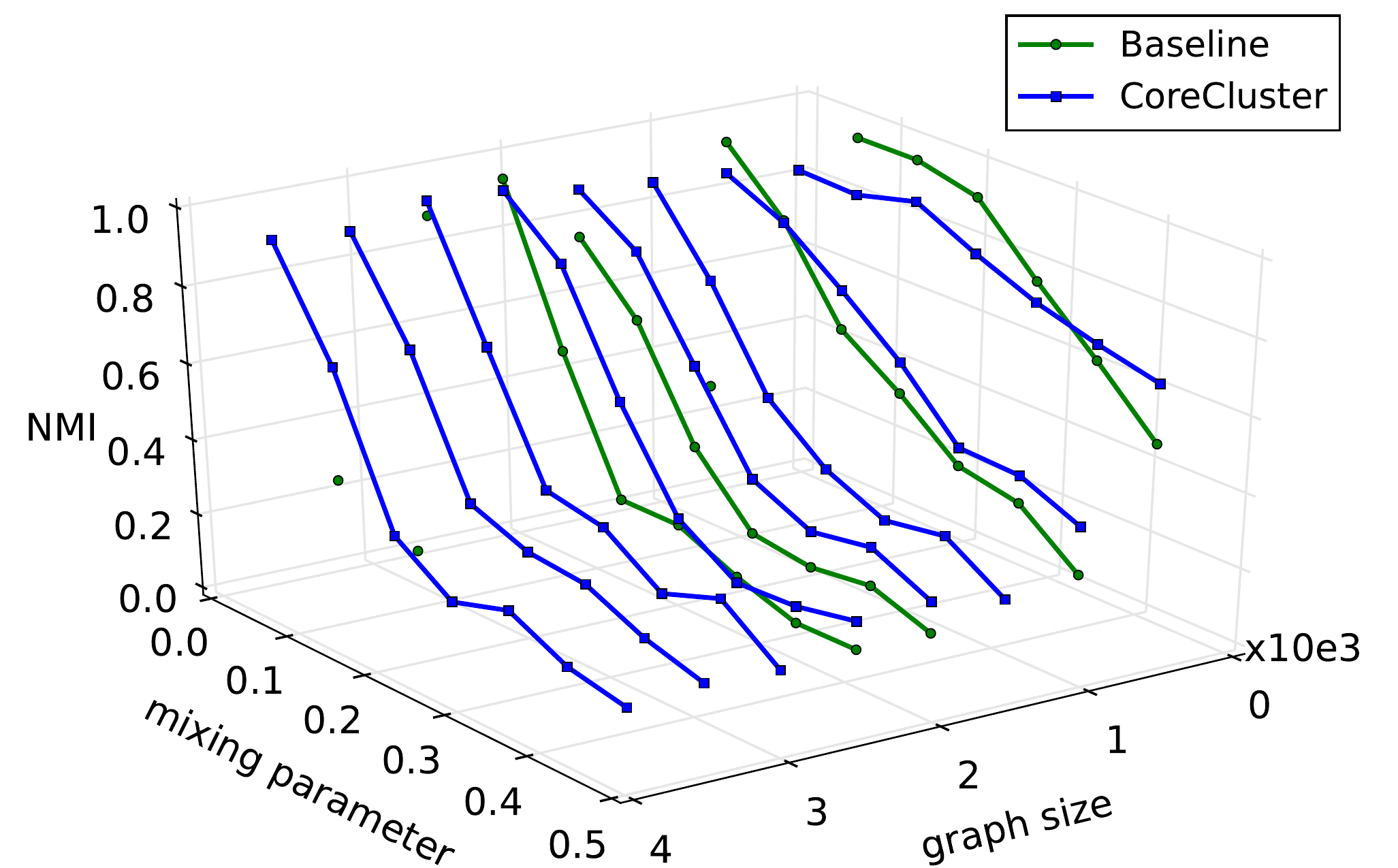}
    \includegraphics[scale=0.24]{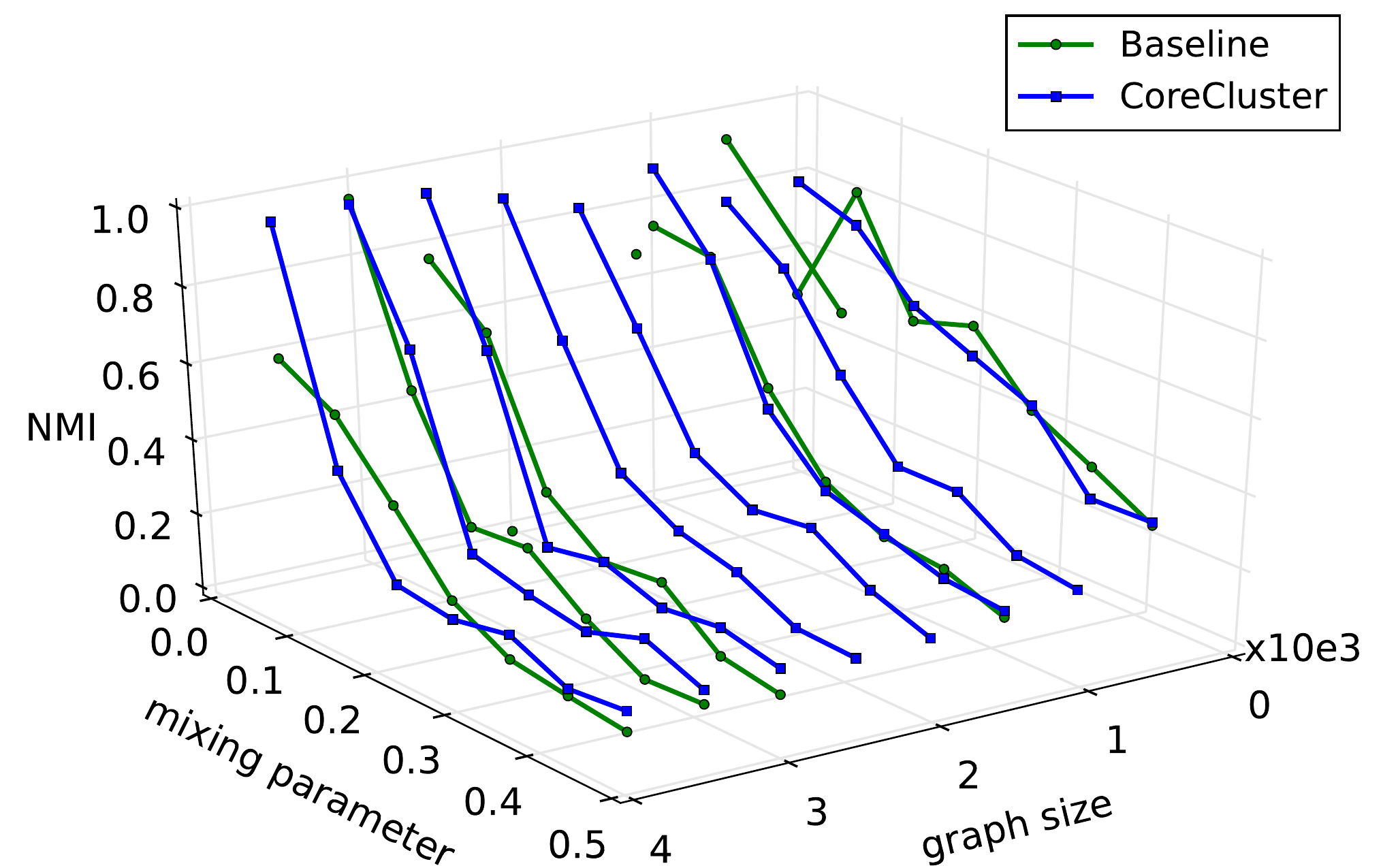}
    \caption{SpinGlass}
  \end{subfigure}
  
  \begin{subfigure}{\textwidth}
    \centering
    \includegraphics[scale=0.24]{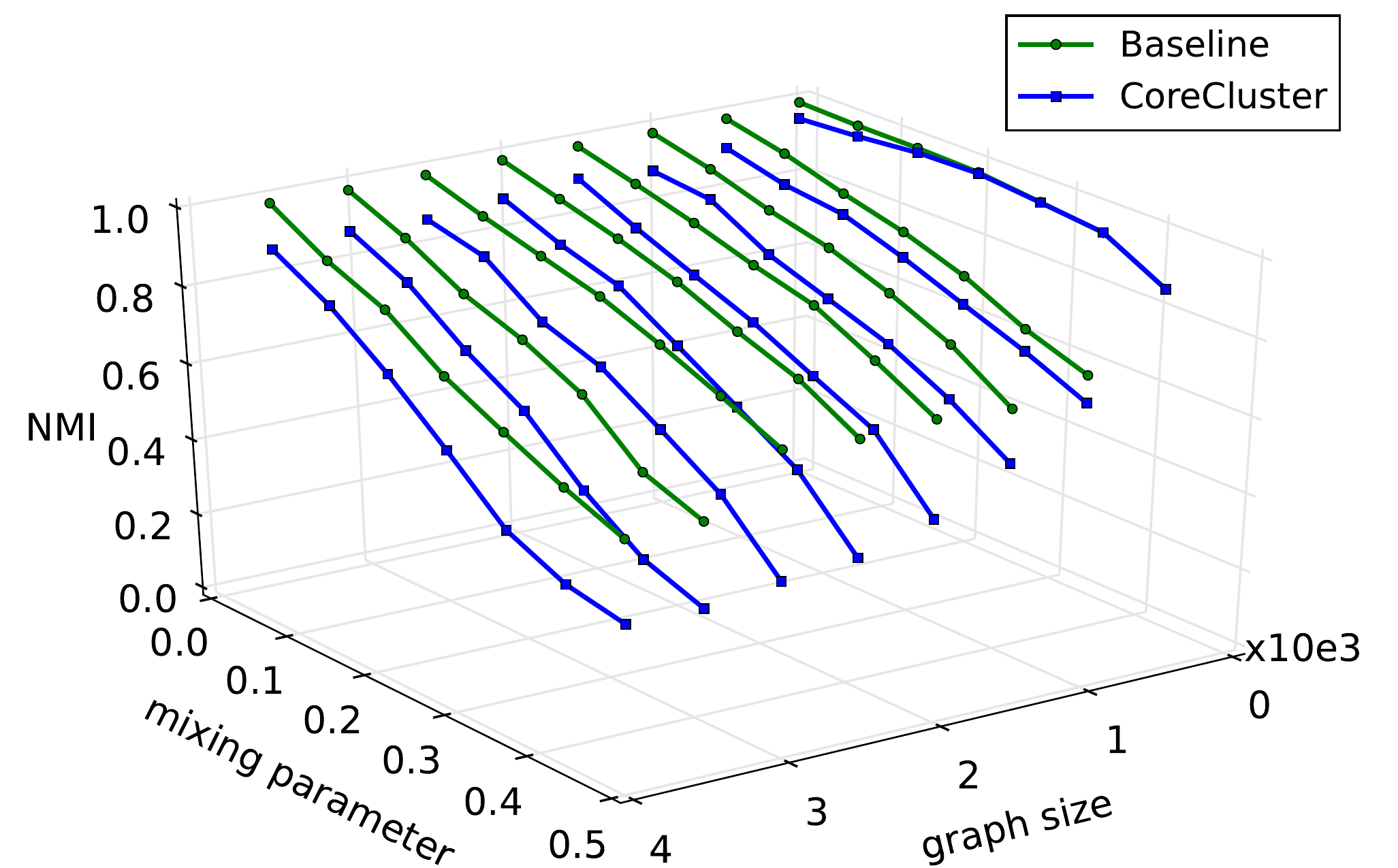}
    \includegraphics[scale=0.24]{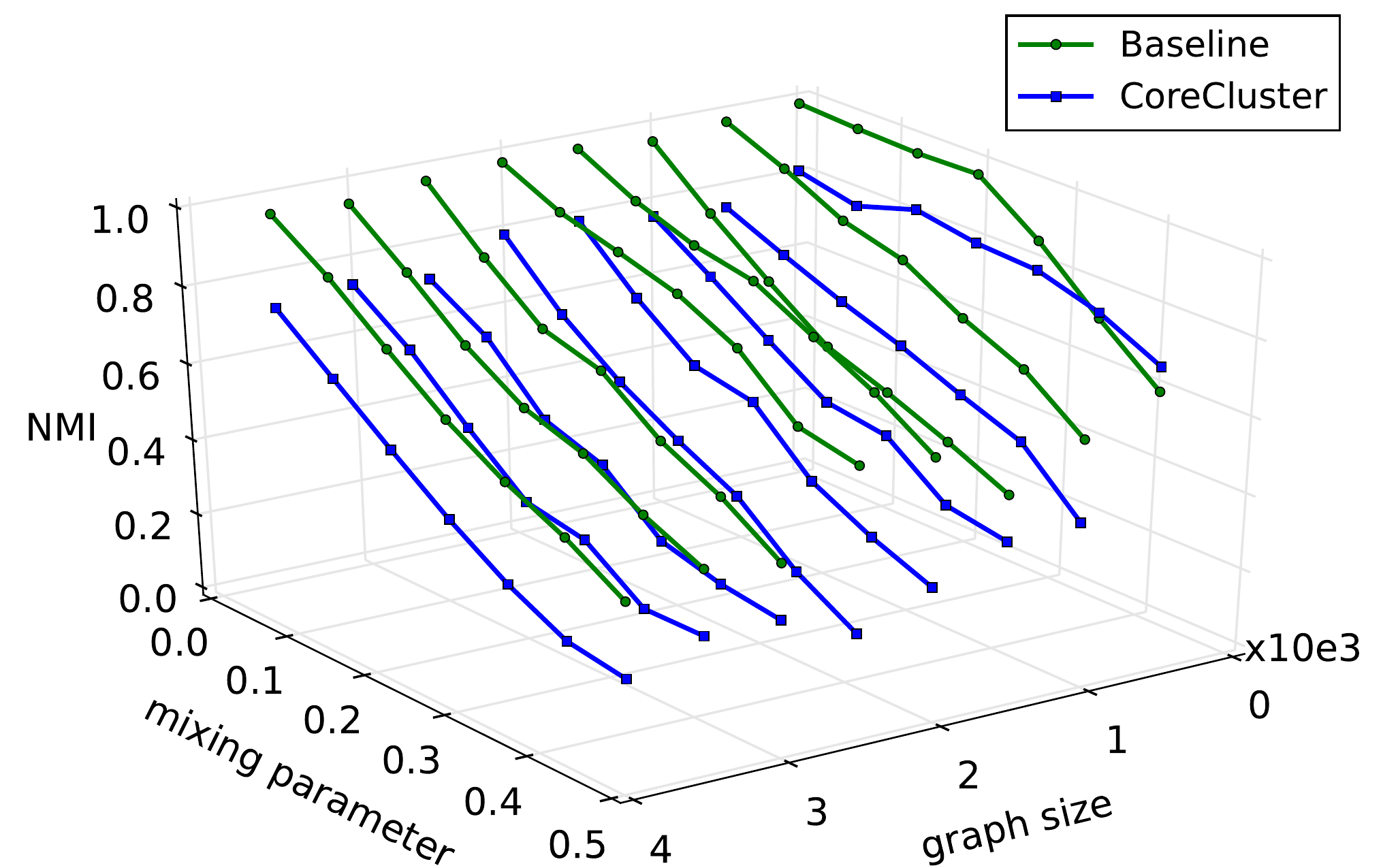}
    \includegraphics[scale=0.24]{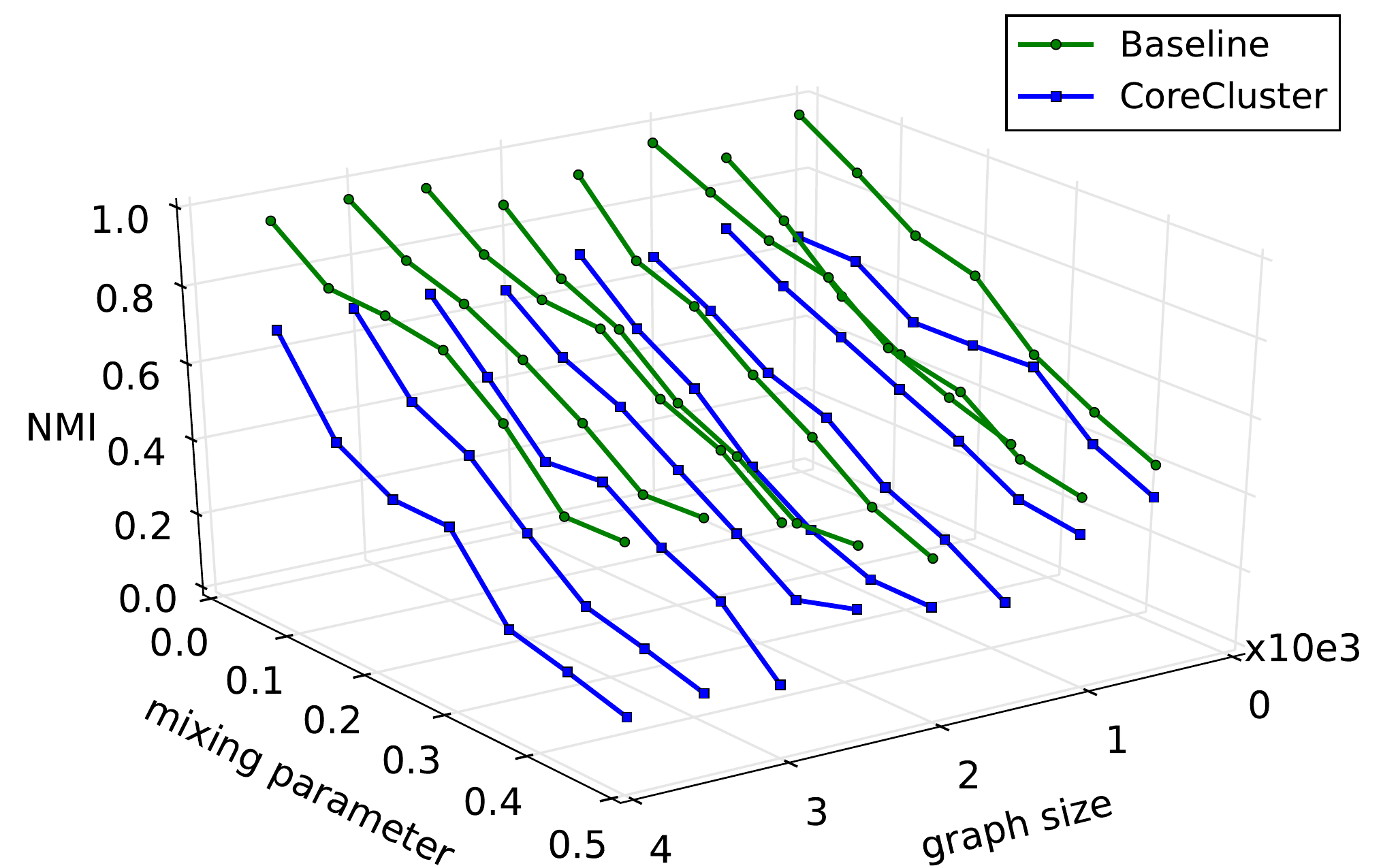}
    \caption{Walktrap}
  \end{subfigure}
  \caption{Clustering quality comparisons in terms of NMI for the three artificial graphs (D1, D2, and D3) (Part II).}

\end{figure}
\subsubsection{Artificial Data, NMI comparison}\label{sec:NMI}

As mentioned on the previous sections, we wish to evaluate the performance of our framework in multiple aspects. We start here with the artificial data, comparing the performance in relevance to the {\em NMI} metric. This comparison is separate for each of the three groups of the artificial data. Moreover, we compare across the parameters of the data generator (Table~\ref{table:first}) as those affect crucial aspects of the graph's structure (density, cluster clarity, etc.). These parameters also affect the degeneracy properties of the graph (number of nodes in the maximum $k$-core, number of core partitions etc.).

We organize the comparisons per algorithm. For each of the aforementioned clustering algorithms, we display the evaluation of the ``original version''  (which we will refer to as {\em Baseline} in all comparisons) in comparison to the evaluation for the corresponding \textsc{CoreCluster} ``combination'' (\textsc{CoreCluster} utilizing the {\em Baseline} in place of the \textbf{ Cluster} algorithm as described in Section~\ref{sec:method}). 

Thus for each algorithm we have three figures (one for each of the aforementioned artificial data sets - Section~\ref{sec:artfc}) displaying the NMI clustering evaluation metric in relevance to the cluster size as well as the mixing parameter (that affects the ``clarity'' of the clusters). 

We start this comparison at  Figure~\ref{fig:NMIcomp} with the Spectral Clustering algorithm. In this case, we observe that the \textsc{CoreCluster} solution  performs quite close to the Baseline and in fact outperforms it in many occasions. We further make an initial note on the effect of graph density to the quality of the results for our framework.
Thus, we can remark that the largest gap between the Baseline and \textsc{CoreCluster} across all algorithms is observed for the D3 dataset, which has the biggest density.

This is an expected result as the global density of the graph affects the maximum $k$-core number and the partitions generated by the $k$-core decomposition algorithm. Graphs that are very dense will have a high maximum core number but there will be no partition of significance. In simpler terms, when all the nodes are highly connected, they are all likely to belong in high-core partitions and subsequently the biggest portion of the graph could belong to the maximum core. The case of the D3 dataset displays this point as it was intentionally generated with high density. On the contrary, D1 was generated as a case closer to the density of real graphs and D2 was generated as an intermediate case. We see consistently along all experiments that the general graph density affects more or less the same our framework. The cases in D3 where \textsc{CoreCluster} performs in a similar manner with the Baseline are of those with low mixing parameter. The reason for this is twofold:
\begin{enumerate}
	\item The clusters are better separated from each other.
	\item While the minimum and the maximum node degree are the main parameters that affect directly the density of the generated graph, the mixing parameter contributes as well. Thus, for low mixing parameters, we can get lower densities as well in all of the datasets.  
\end{enumerate}

Our framework is designed based on analysis of real networks and their Degeneracy \citep{giatsidis2011d,giatsidis2011evaluating} and for this reason datasets with ``abnormal'' behaviour in the Degeneracy aspect are expected to give bad evaluations. Intuitively, in order to provide acceleration to the ``input'' algorithm ({\bf Cluster}), the size of $\core_i$ should be (significantly) smaller than the size of $\core_{i-1}$. Otherwise, the incremental nature of our framework would not make sense.

Moving to the rest of the algorithms, we see similarly that \textsc{CoreCluster} performs quite close to the Baseline for most cases. In addition to the cases of high density, we would like to study in detail other parameters that also affect the divergence between the two algorithms(the Baseline and \textsc{CoreCluster}). For this reason in the following sections, we examine in depth properties of both the graph and the Baseline algorithms in order to clearly outline the evaluation of the \textsc{CoreCluster} framework.  

It is also worth noting that the {\em SpinGlass} algorithm is missing evaluations for the larger graphs in the Baseline part. Its execution was conducted with the implementation provided by igraph\footnote{\url{http://igraph.org/c/}} and it was riddled with abruptly terminations on these datasets. Only a small portion of the data was processed successfully with just the Baseline and for this reason we present only the parts that were covered thoroughly (i.e., more than one graph per combination passed the processing with SpinGlass).  

\subsubsection{Analysis: Graph and Algorithm Properties - Artificial Data}
In this part of our work, we present a thorough analysis of the evaluations in regards to the following two aspects:
\begin{itemize}
	\item \textbf{Performance analysis in relevance to the coverage of the maximum-core:} How much of the complete graph exists in the max-core versus the NMI from \textsc{CoreCluster}.
	\item   \textbf{``Final Performance'' vs ``Performance at the Maximum-Core'': } The relation of the performance of the Baseline on the max-core versus that of  \textsc{CoreCluster} at the entire graph.
\end{itemize}
Similarly to the performance evaluations presented in Section~\ref{sec:NMI}, the aforementioned evaluations are conducted per Dataset (D1,D2,D3).

\paragraph{How to use the information bellow}
\mbox{}\\
Most tools and algorithms in the Machine Learning domain require a lot of experimentation and fine tuning before deployment. Since \textsc{CoreCluster} is a meta algorithmic framework, we are limiting the additional complexity in this process by clearly marking properties that are easy and efficient to detect before the deployment of our framework.
Instead of general remarks on the performance, the next two sections will provide a clear roadmap to anyone who wishes to apply \textsc{CoreCluster} on any data.

\paragraph{Performance analysis in relevance to the coverage of the maximum-core}
\begin{figure}
	\begin{subfigure}{1.0\textwidth}
		\centering
		\includegraphics[scale=0.5]{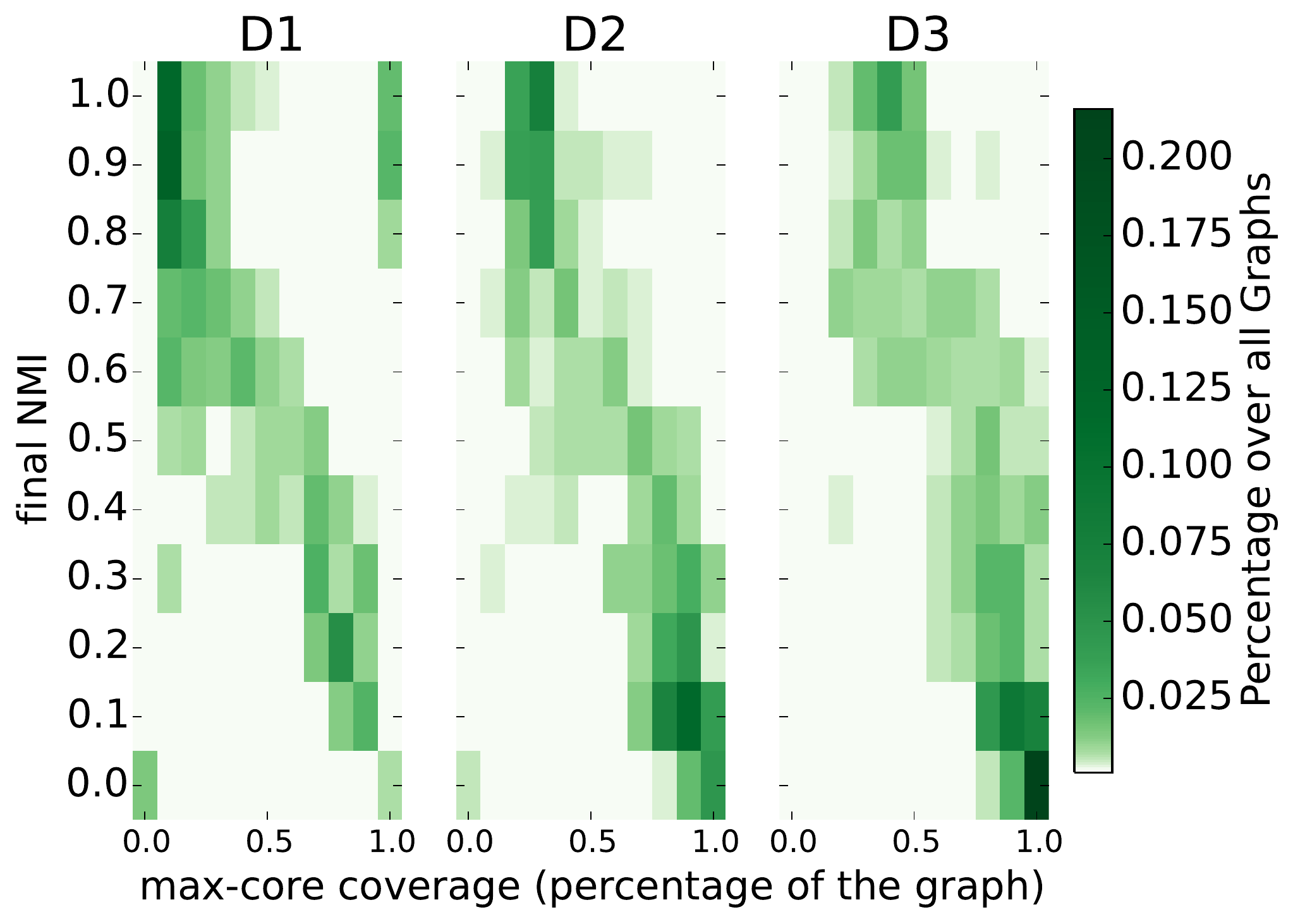}
		\caption{Spectral Clustering}
	\end{subfigure} 
	
	\begin{subfigure}{0.5\textwidth}
		\centering
		\includegraphics[scale=0.35]{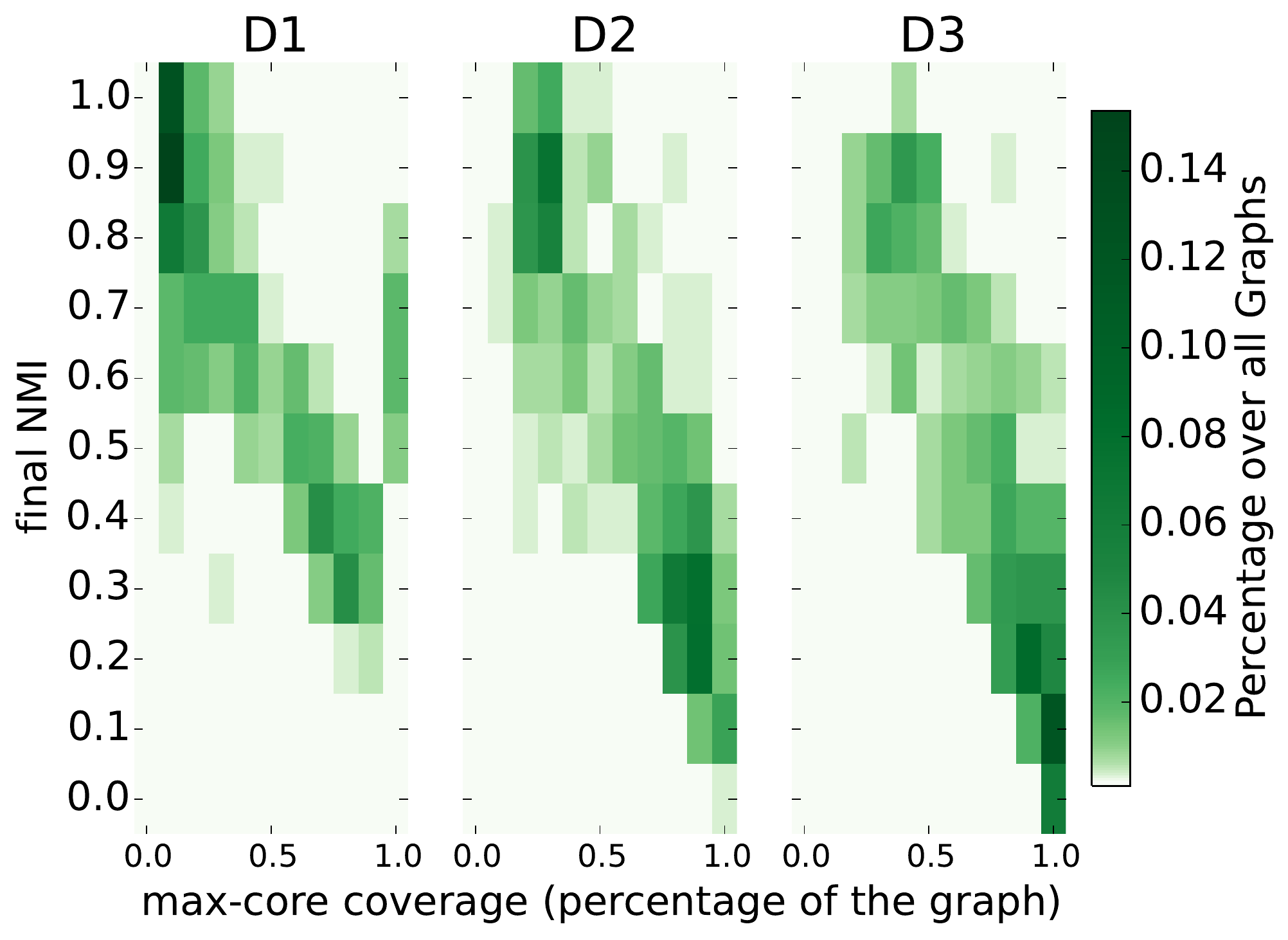}
		\caption{Leading Eigenvector}
	\end{subfigure}
	\begin{subfigure}{0.5\textwidth}
		\centering
		\includegraphics[scale=0.35]{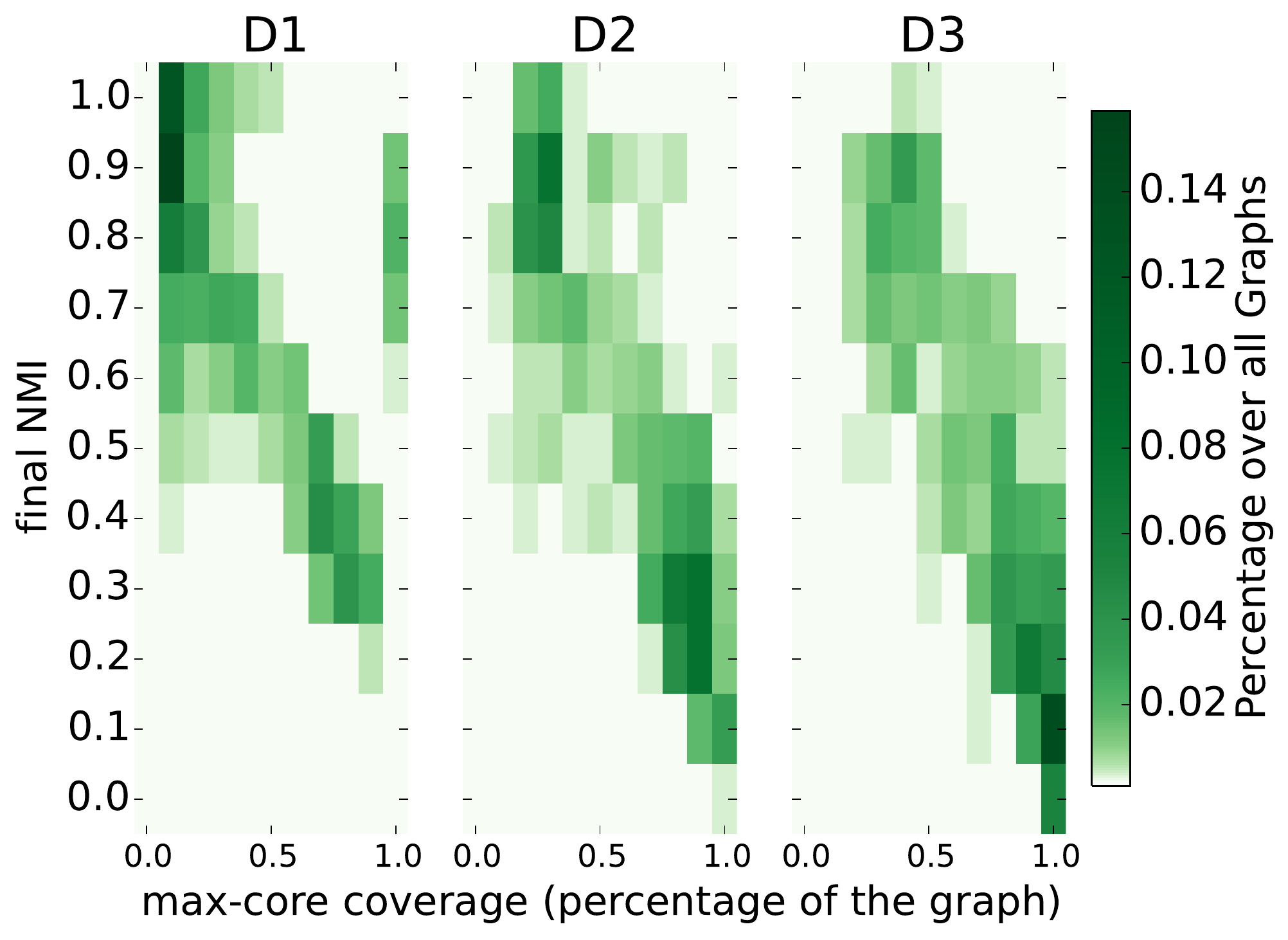}
		\caption{Fast Greedy}
	\end{subfigure}
	
	\begin{subfigure}{0.5\textwidth}
		\centering
		\includegraphics[scale=0.35]{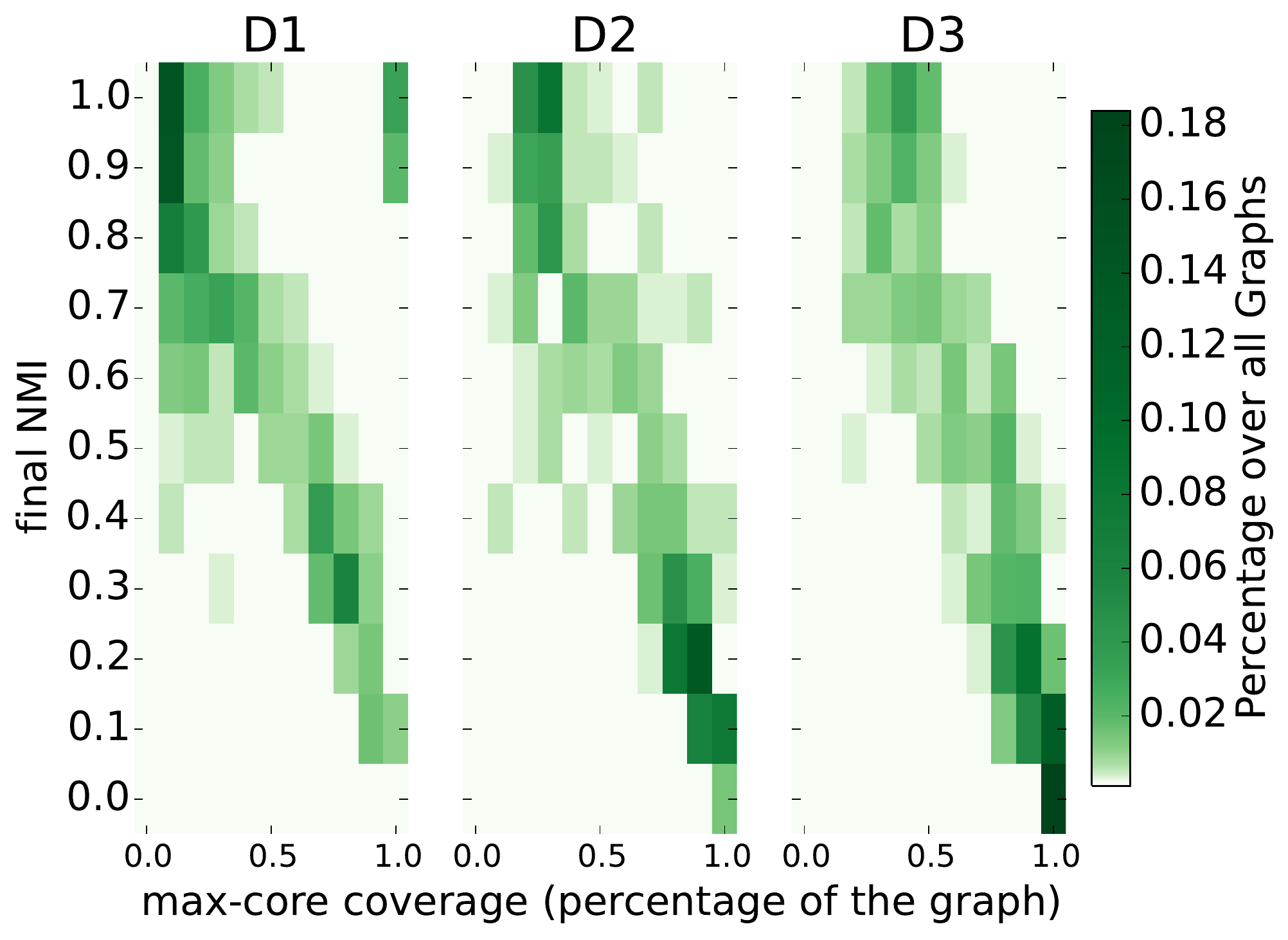}
		\caption{MCL}
	\end{subfigure}
	\begin{subfigure}{0.5\textwidth}
		\centering
		\includegraphics[scale=0.35]{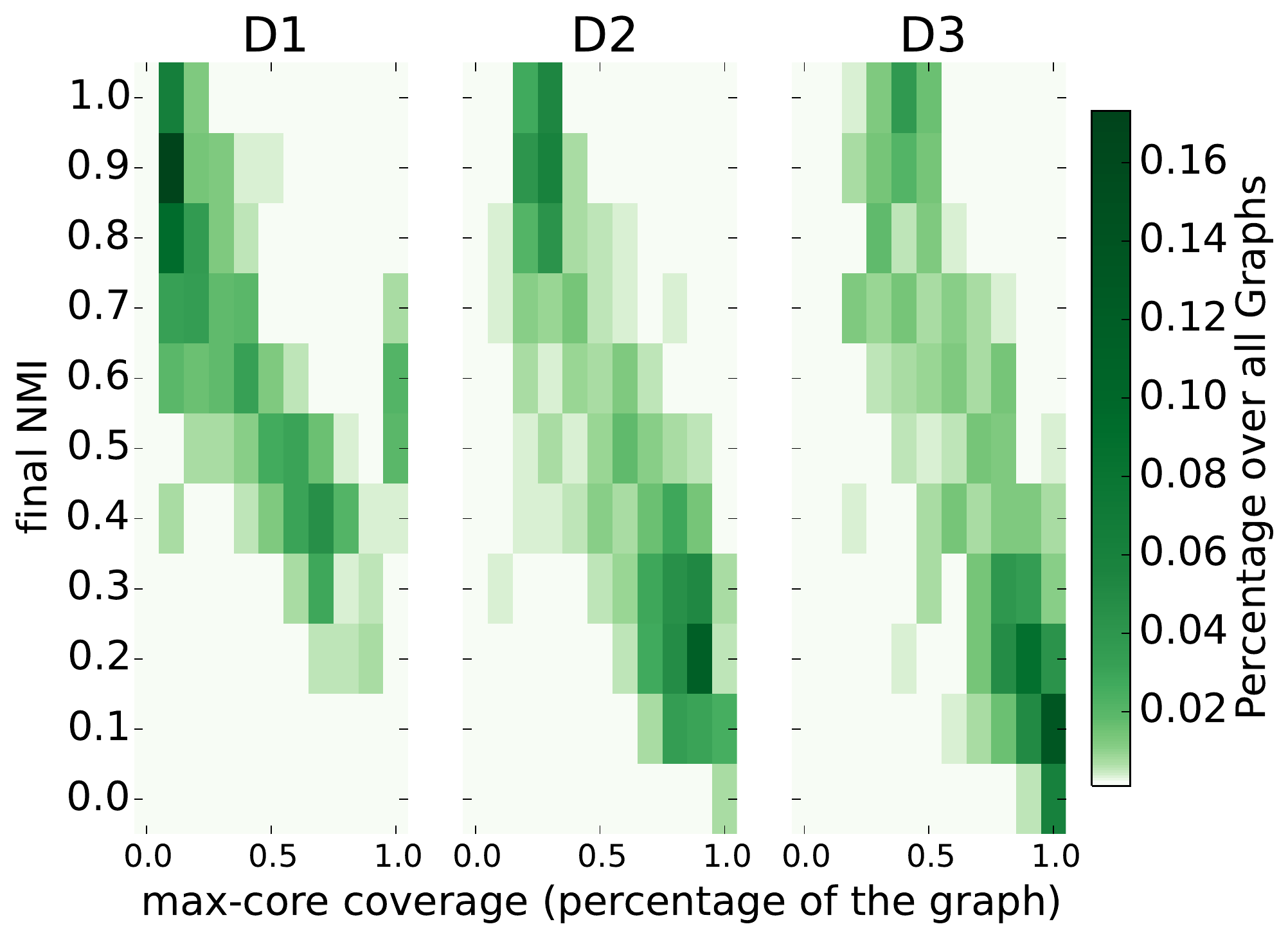}
		\caption{Metis}
	\end{subfigure}
	\caption{ Performance analysis in relevance to the coverage of the maximum-core for the three artificial graphs (D1, D2, and D3) (Part I)}
	\label{fig:coverage}
\end{figure}
\begin{figure}[t]
	\ContinuedFloat 
	\begin{subfigure}{0.5\textwidth}
		\centering
		\includegraphics[scale=0.35]{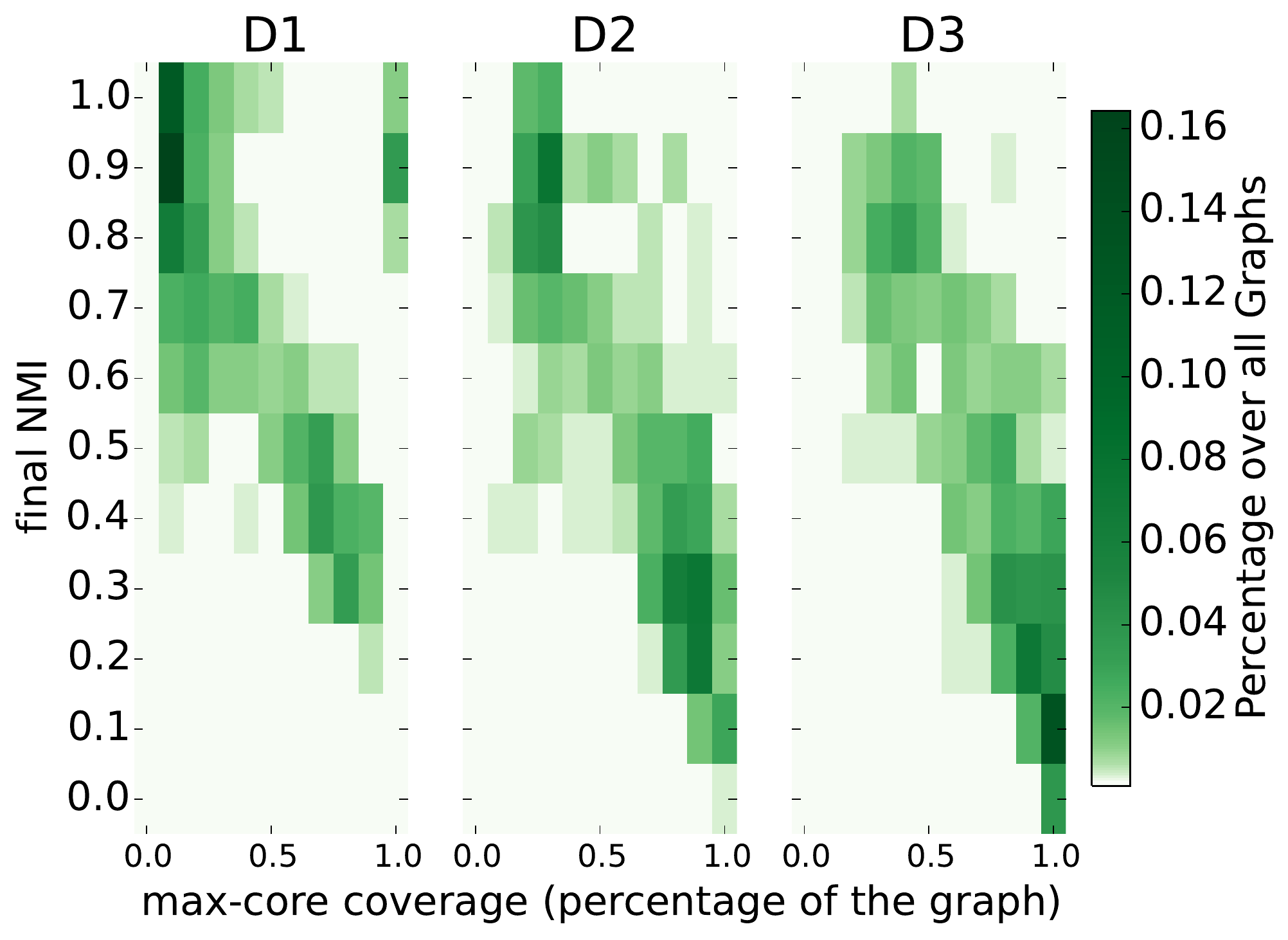}
		\caption{MultiLevel}
	\end{subfigure}
	\begin{subfigure}{0.5\textwidth}
		\centering
		\includegraphics[scale=0.35]{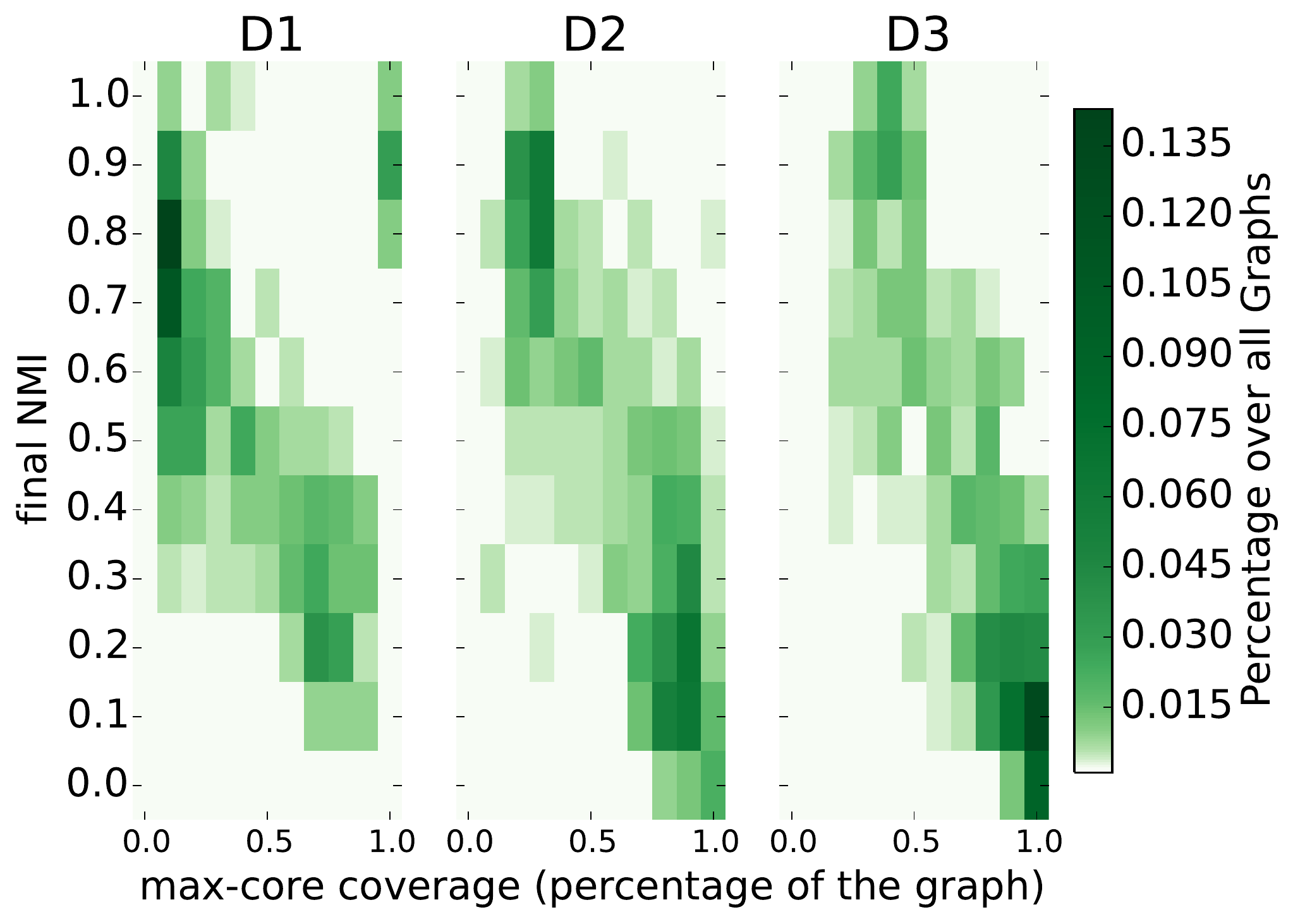}
		\caption{SpinGlass}
	\end{subfigure}
	
	\begin{subfigure}{0.5\textwidth}
		\centering
		\includegraphics[scale=0.35]{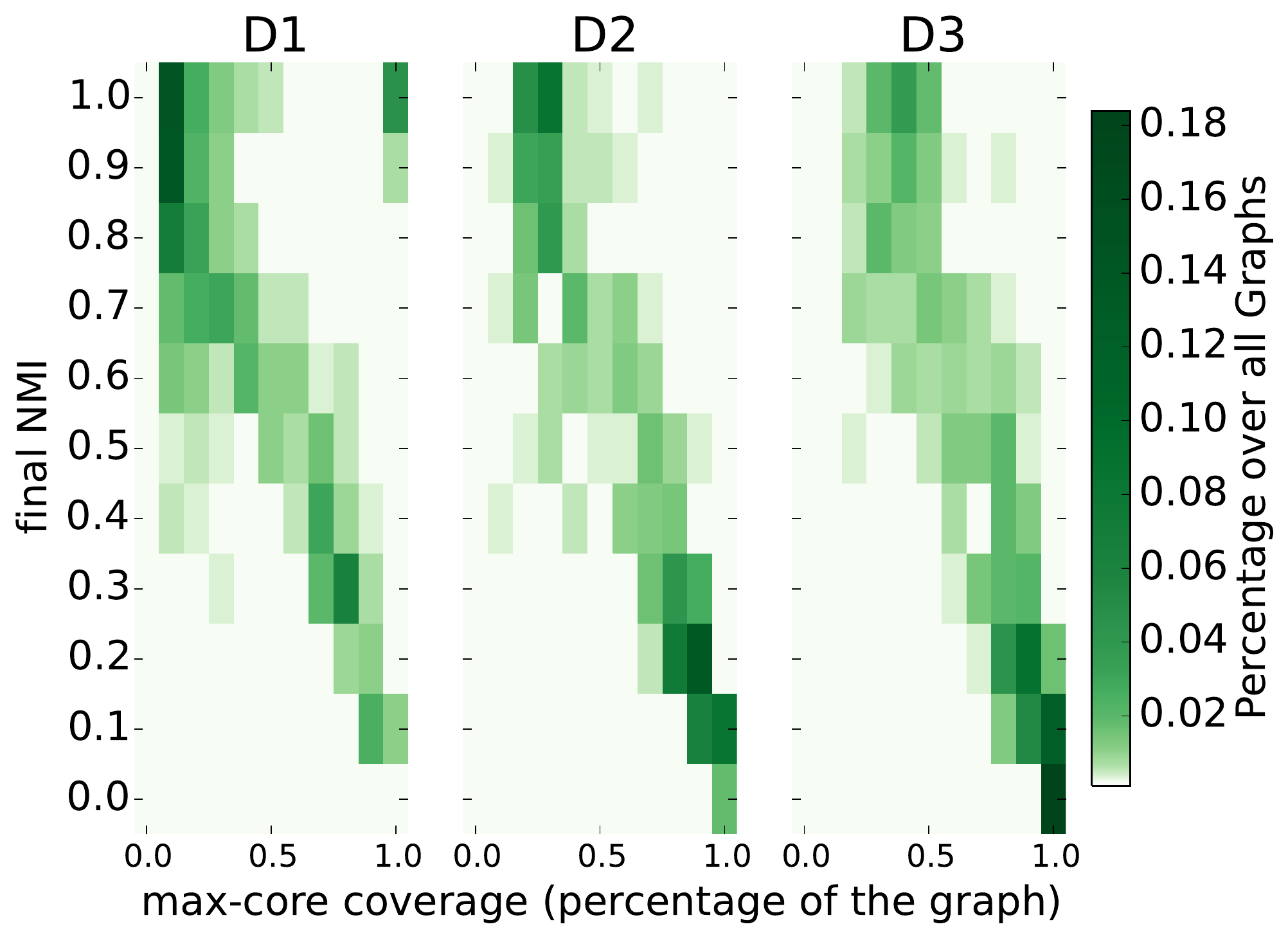}
		\caption{InfoMap}
	\end{subfigure}
	\begin{subfigure}{0.5\textwidth}
		\centering
		\includegraphics[scale=0.35]{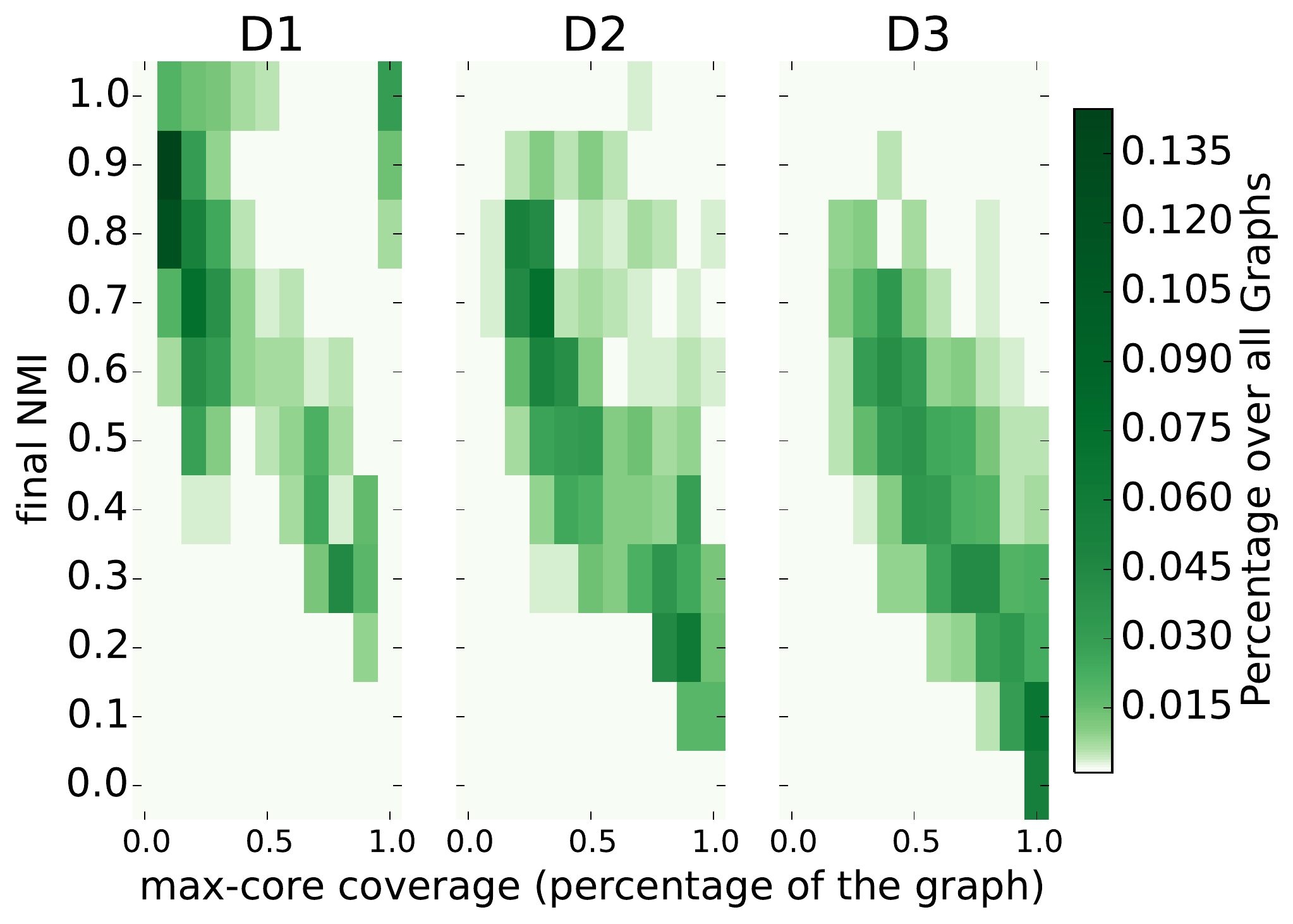}
		\caption{Walktrap}
	\end{subfigure}
	\caption{Performance analysis in relevance to the coverage of the maximum-core for the three artificial graphs (D1, D2, and D3) (Part I)}
	
\end{figure}

The degeneracy properties of a graph can vary greatly in two main aspects. The first one is the $k$-core number of the graph, i.e., the maximum of interconnectivity for which we can get a non empty part of the graph. The other one is the number of nodes for all the possible cores from $0$ to $k$. As graphs with communities are complex structures with dense and sparse areas, it would be disorienting to evaluate all possible properties for a graph in relevance to Degeneracy and the performance of  \textsc{CoreCluster}. For example, while density affects the core numbers, a graph with a high $k$-core value could be sparse as well.
For this reason, we focus on a key aspect that can  portray the graph in a major manner. This aspect is the percentage of the graph that is maintained in the maximal $k$-core. We call this dimension ``max-core coverage'' and we examine across it how well our framework performs (in terms of NMI). 

In Figure~\ref{fig:coverage}, we see the comparison between the {\itshape max-core coverage} and the NMI of our framework in the entire graph (referred here as {\itshape final NMI}). From this comparison, there is clearly a negative correlation between the two. This correlation is consistent across all algorithms which points out that it is a property of the graph that affects the performance of \textsc{CoreCluster}. In simple terms, our framework performs better when the {\itshape max-core} contains a small percentage of the total graph (and performs worse in the opposite case).  

This makes sense in an intuitive manner as well as the reasons for having a high percentage of the graph in the {\itshape max-core} coincide with those of having bad or no clustering structures. Regardless of the general density of the graph, the main reason for getting a high ``max-core coverage'', is that the distribution of edges among the  vertices is close to uniform.  

\paragraph{``Final Performance'' vs ``Performance at the Maximum-Core''}
\begin{figure}
	\begin{subfigure}{1.0\textwidth}
		\centering
		\includegraphics[scale=0.5]{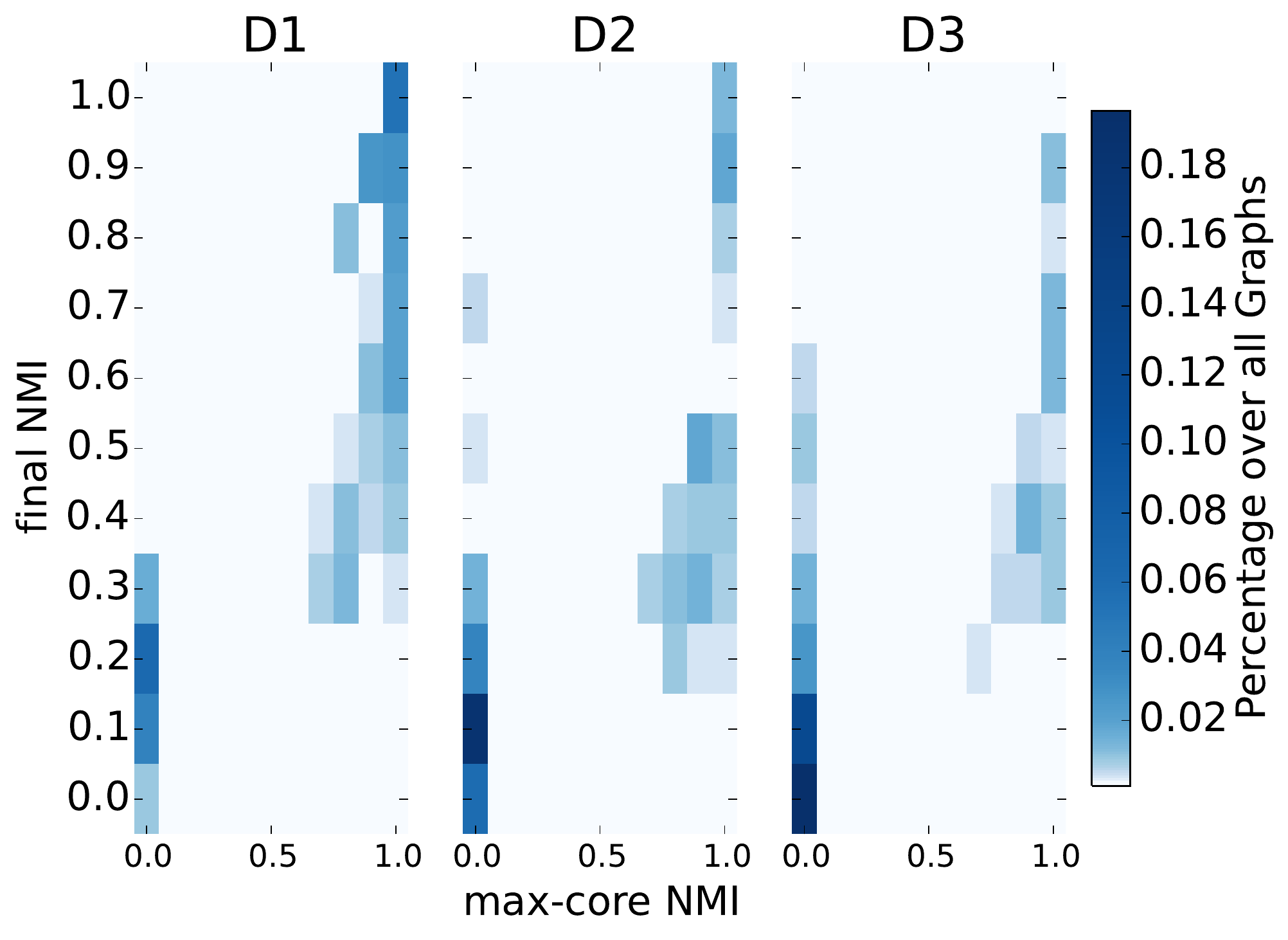}
		\caption{Spectral Clustering}
	\end{subfigure} 
	
	\begin{subfigure}{0.5\textwidth}
		\centering
		\includegraphics[scale=0.35]{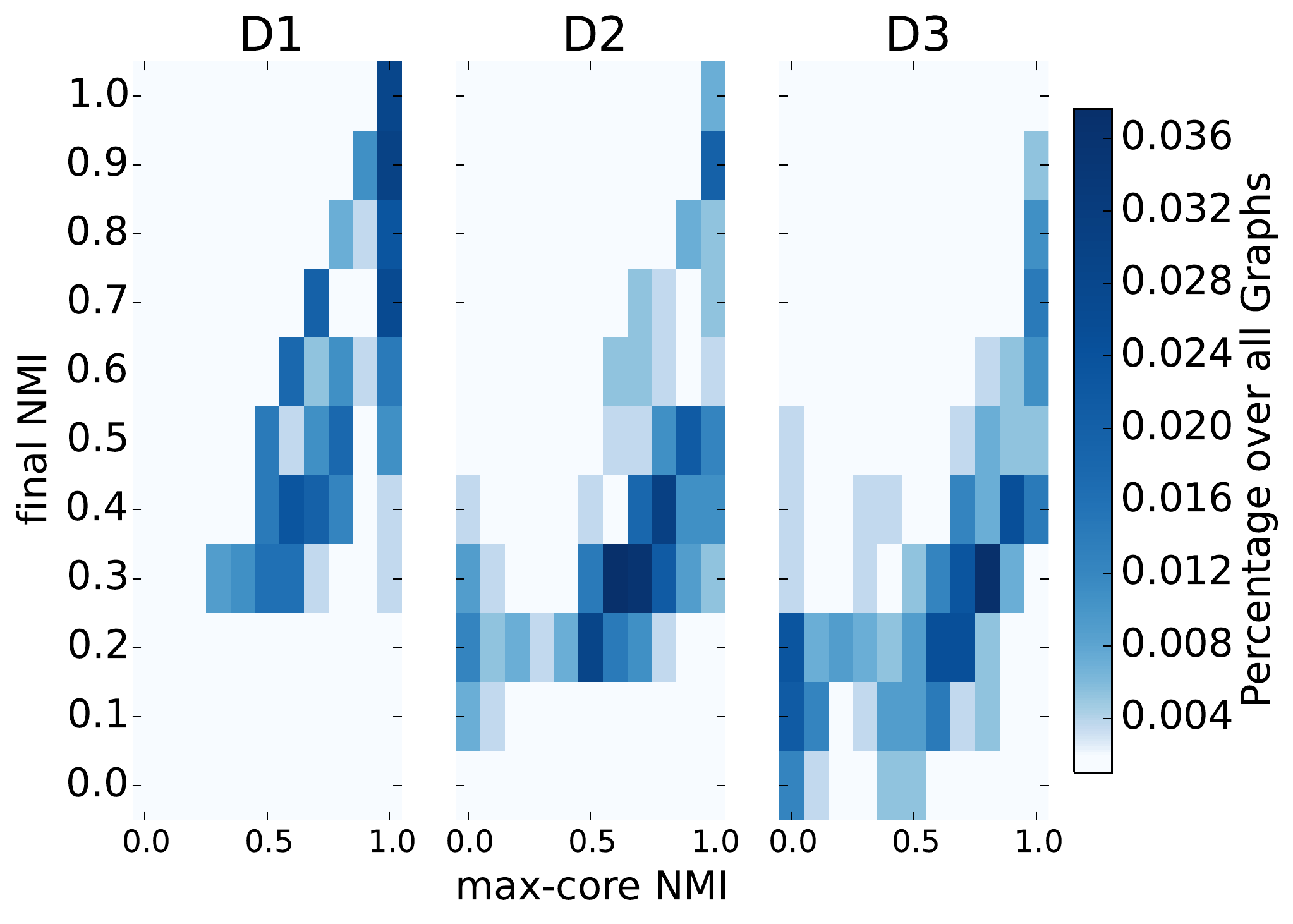}
		\caption{Leading Eigenvector}
	\end{subfigure}
	\begin{subfigure}{0.5\textwidth}
		\centering
		\includegraphics[scale=0.35]{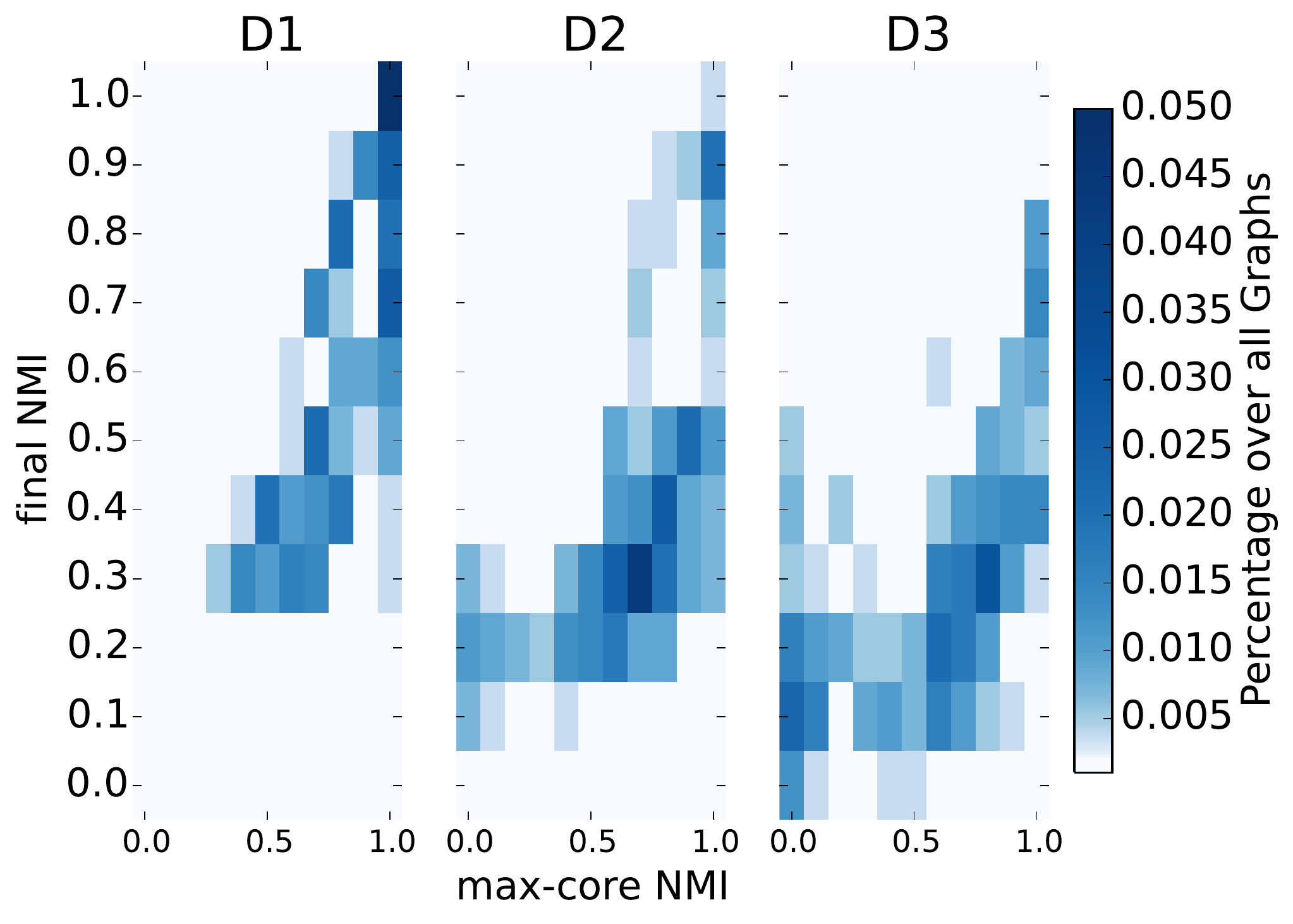}
		\caption{Fast Greedy}
	\end{subfigure}
	
	\begin{subfigure}{0.5\textwidth}
		\centering
		\includegraphics[scale=0.35]{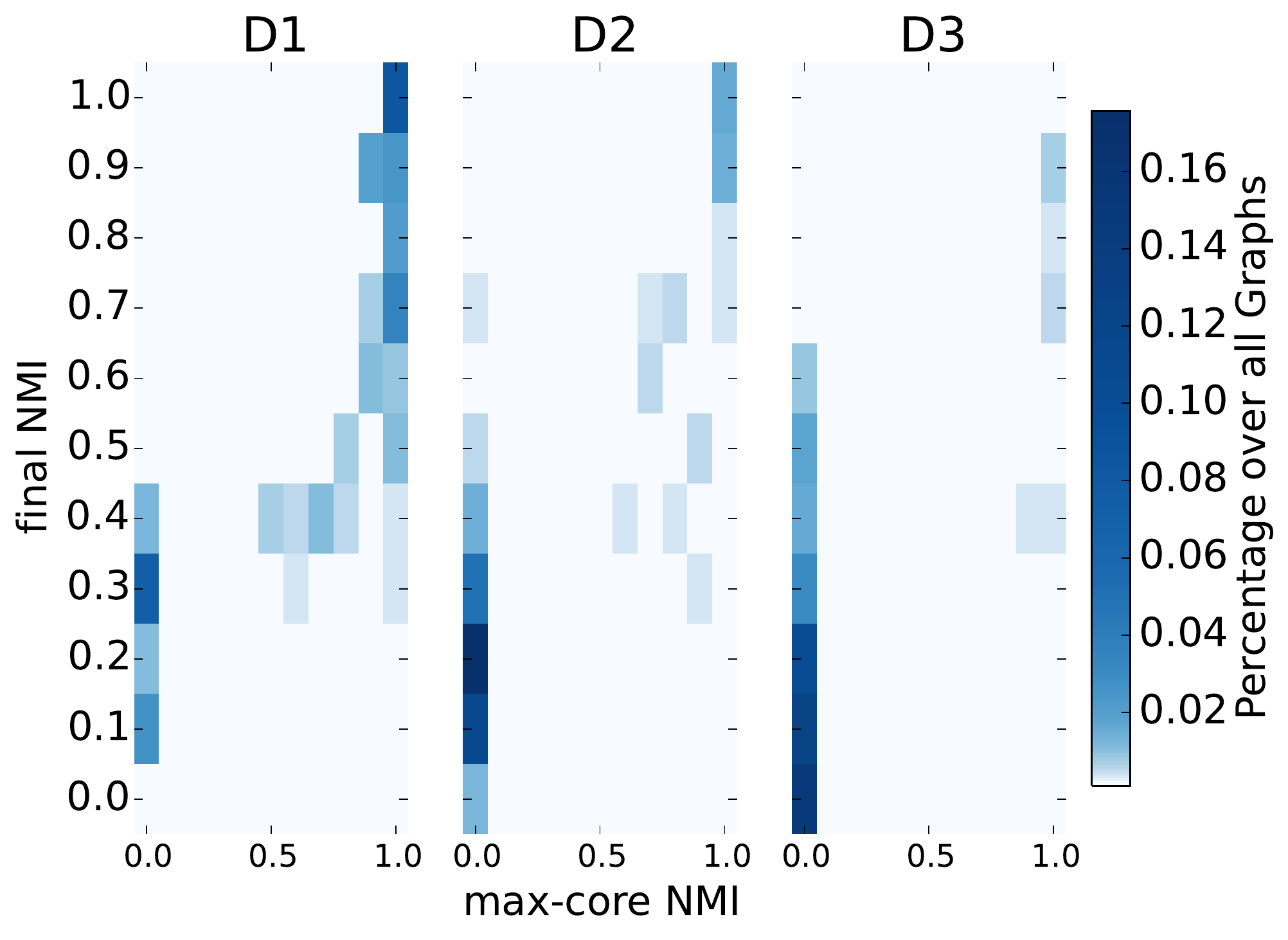}
		\caption{MCL}
	\end{subfigure}
	\begin{subfigure}{0.5\textwidth}
		\centering
		\includegraphics[scale=0.35]{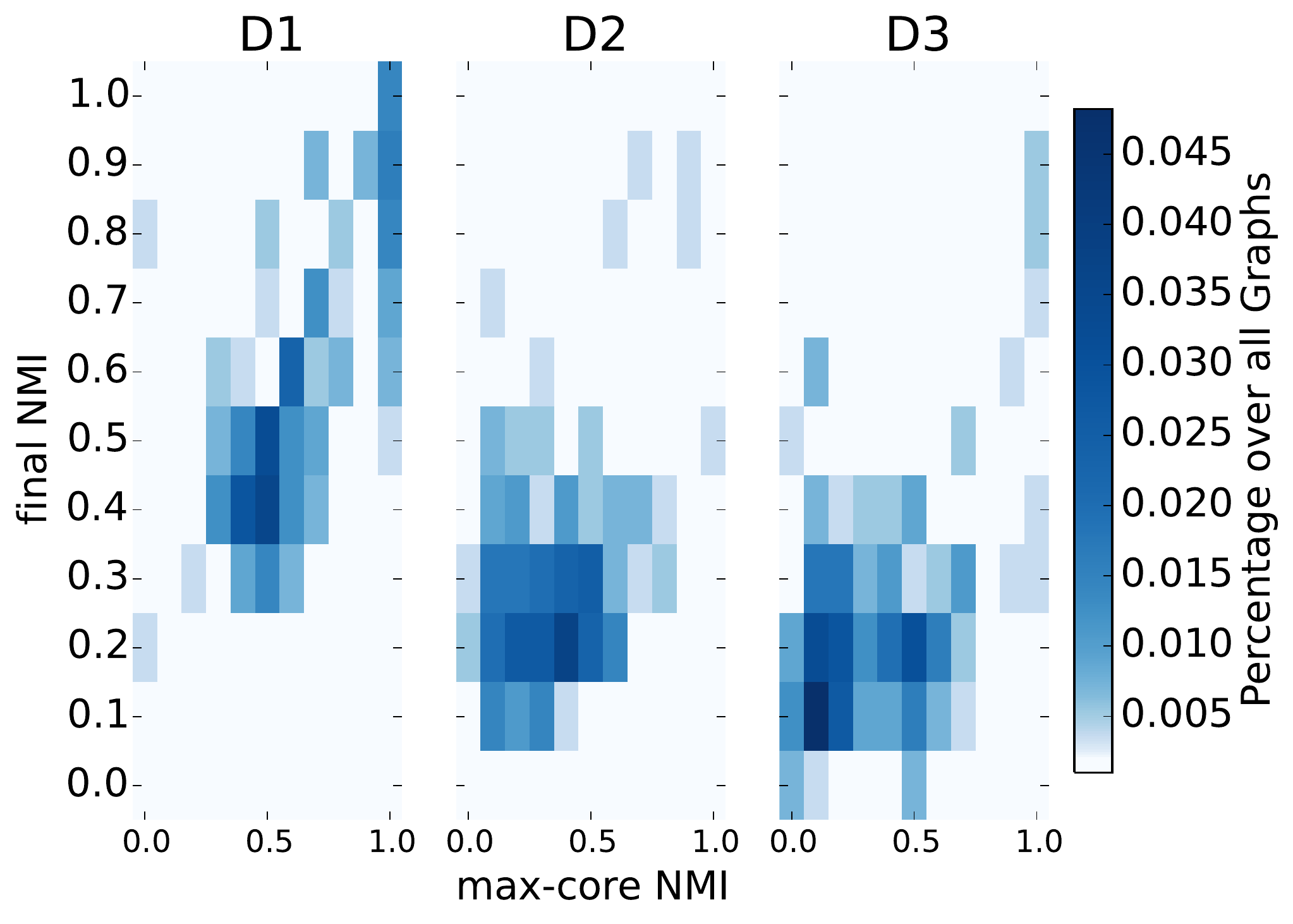}
		\caption{Metis}
	\end{subfigure}
	\caption{``Final Performance'' vs ``Performance at the Maximum-Core'' for the three artificial graphs (D1, D2, and D3)  (Part I).}
	\label{fig:maxvsfinalnmi}
\end{figure}
\begin{figure}[t]
	\ContinuedFloat 
	\begin{subfigure}{0.5\textwidth}
		\centering
		\includegraphics[scale=0.35]{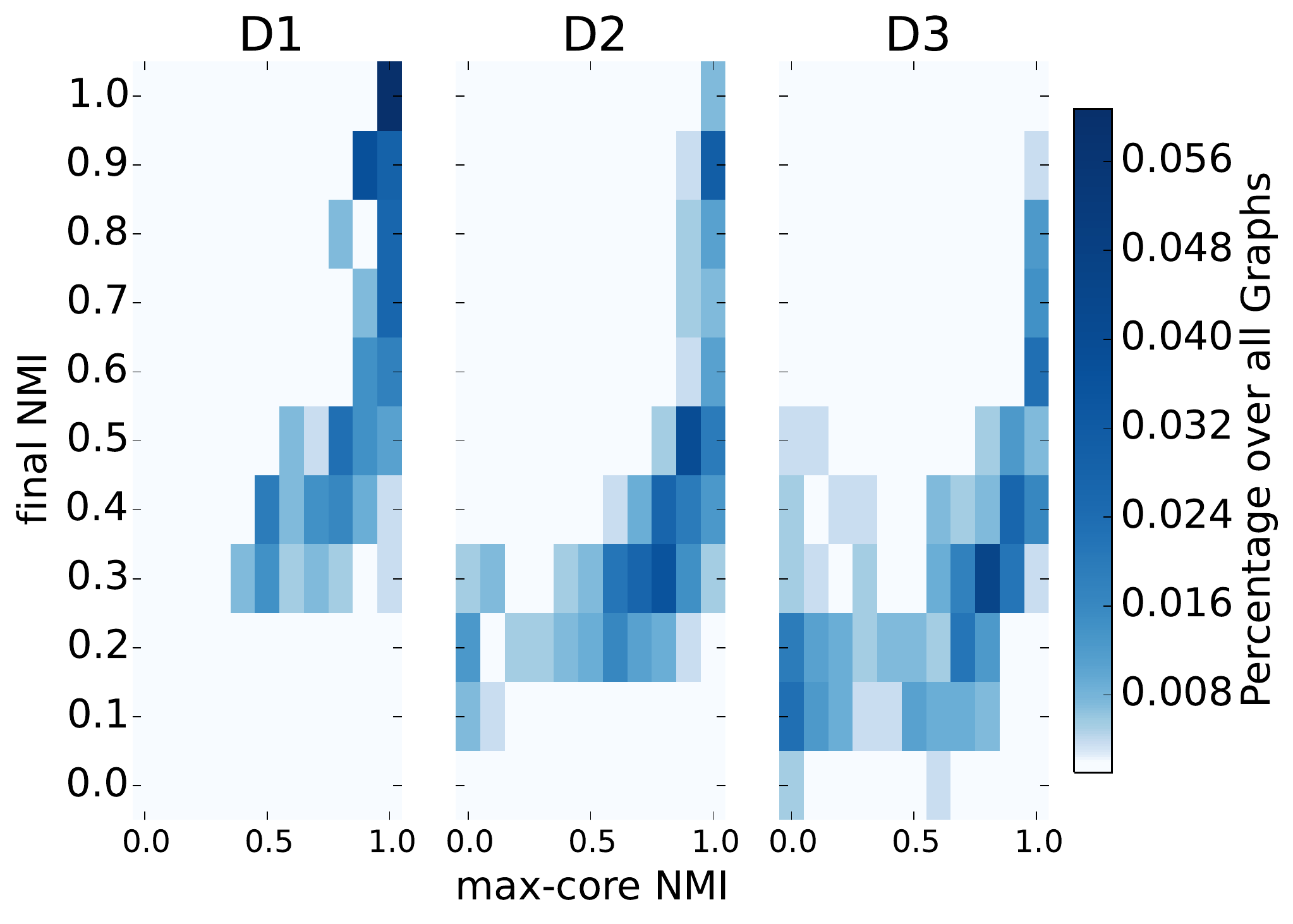}
		\caption{MultiLevel}
	\end{subfigure}
	\begin{subfigure}{0.5\textwidth}
		\centering
		\includegraphics[scale=0.35]{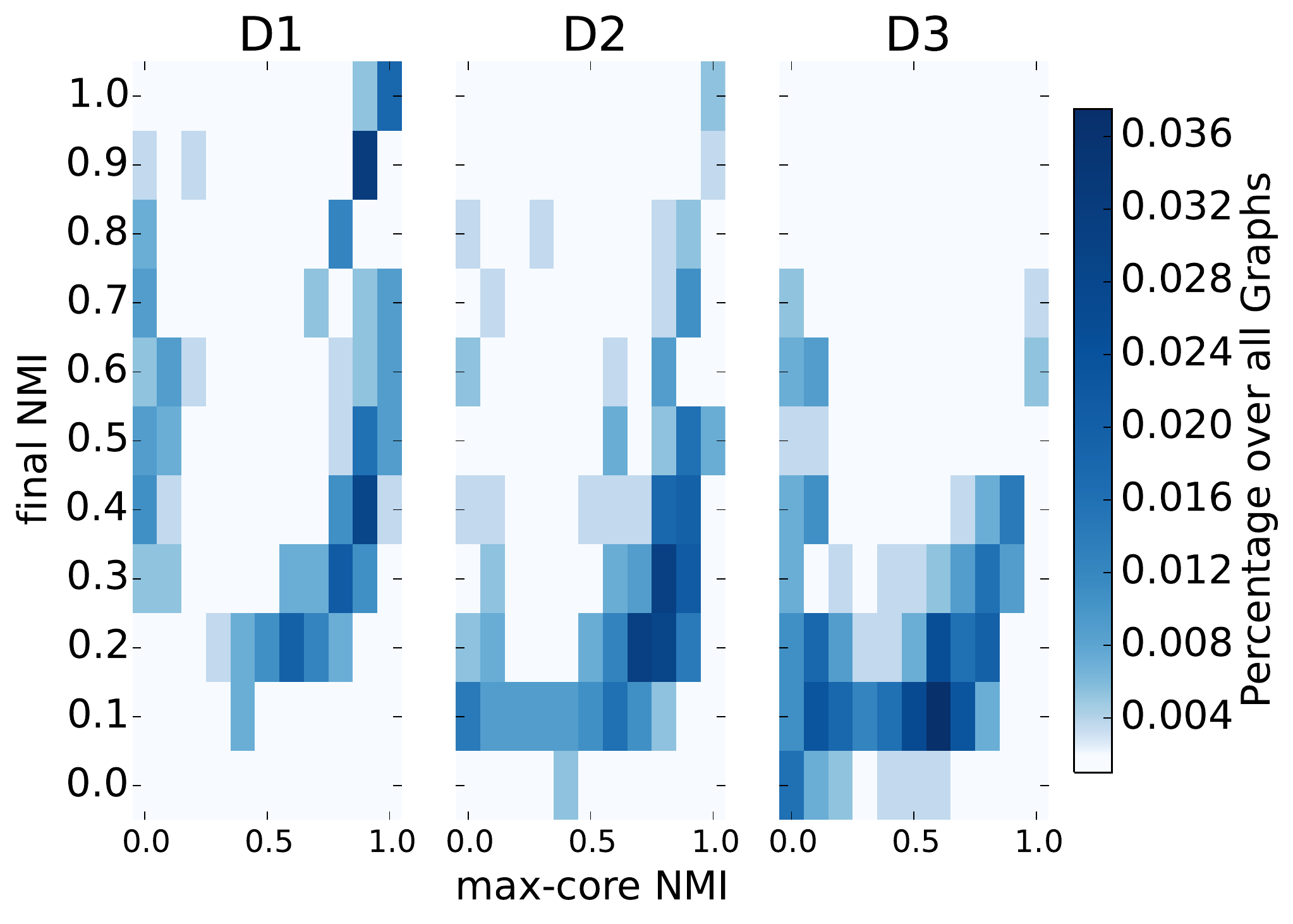}
		\caption{SpinGlass}
	\end{subfigure}
	
	\begin{subfigure}{0.5\textwidth}
		\centering
		\includegraphics[scale=0.35]{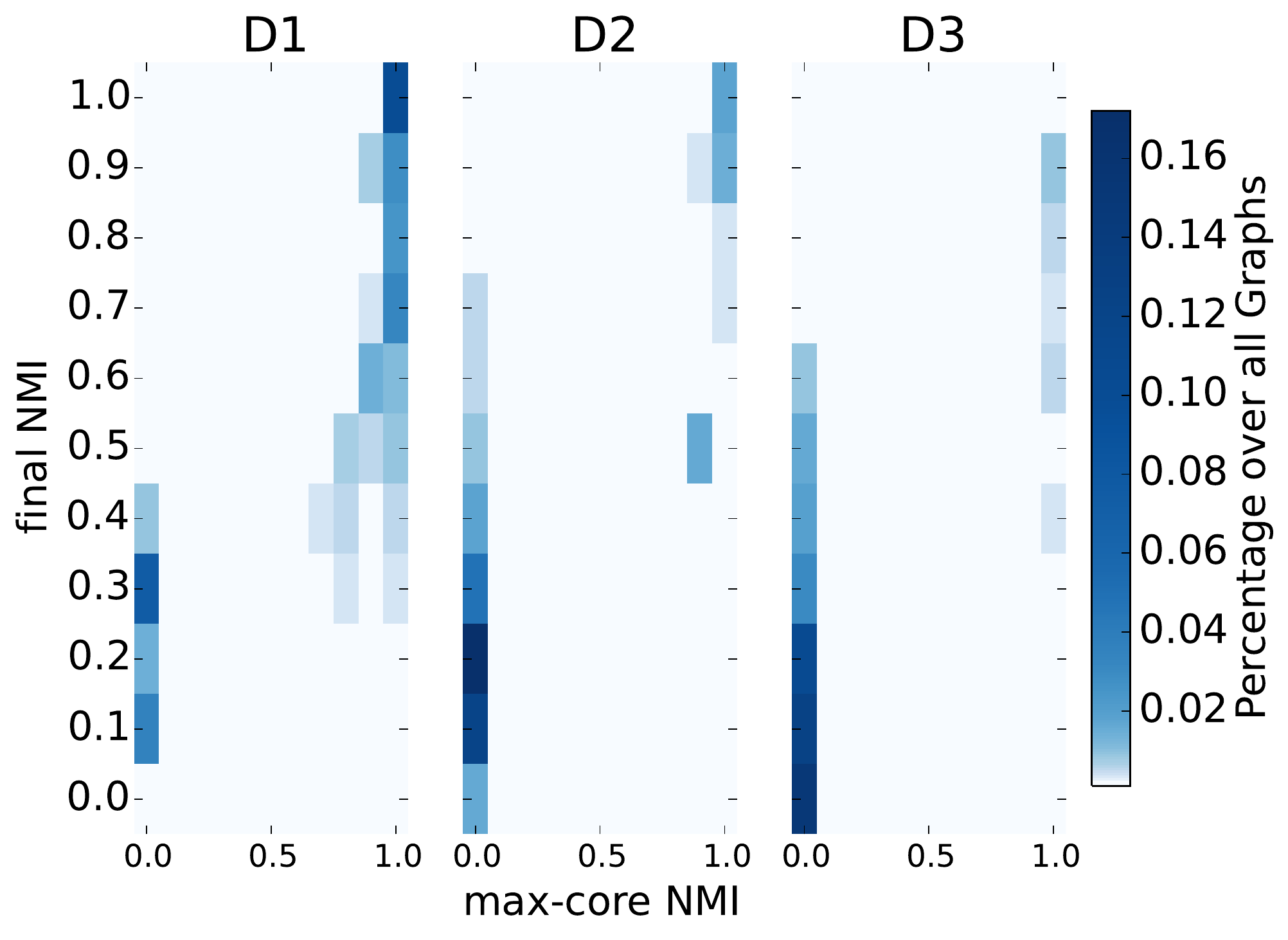}
		\caption{InfoMap}
	\end{subfigure}
	\begin{subfigure}{0.5\textwidth}
		\centering
		\includegraphics[scale=0.35]{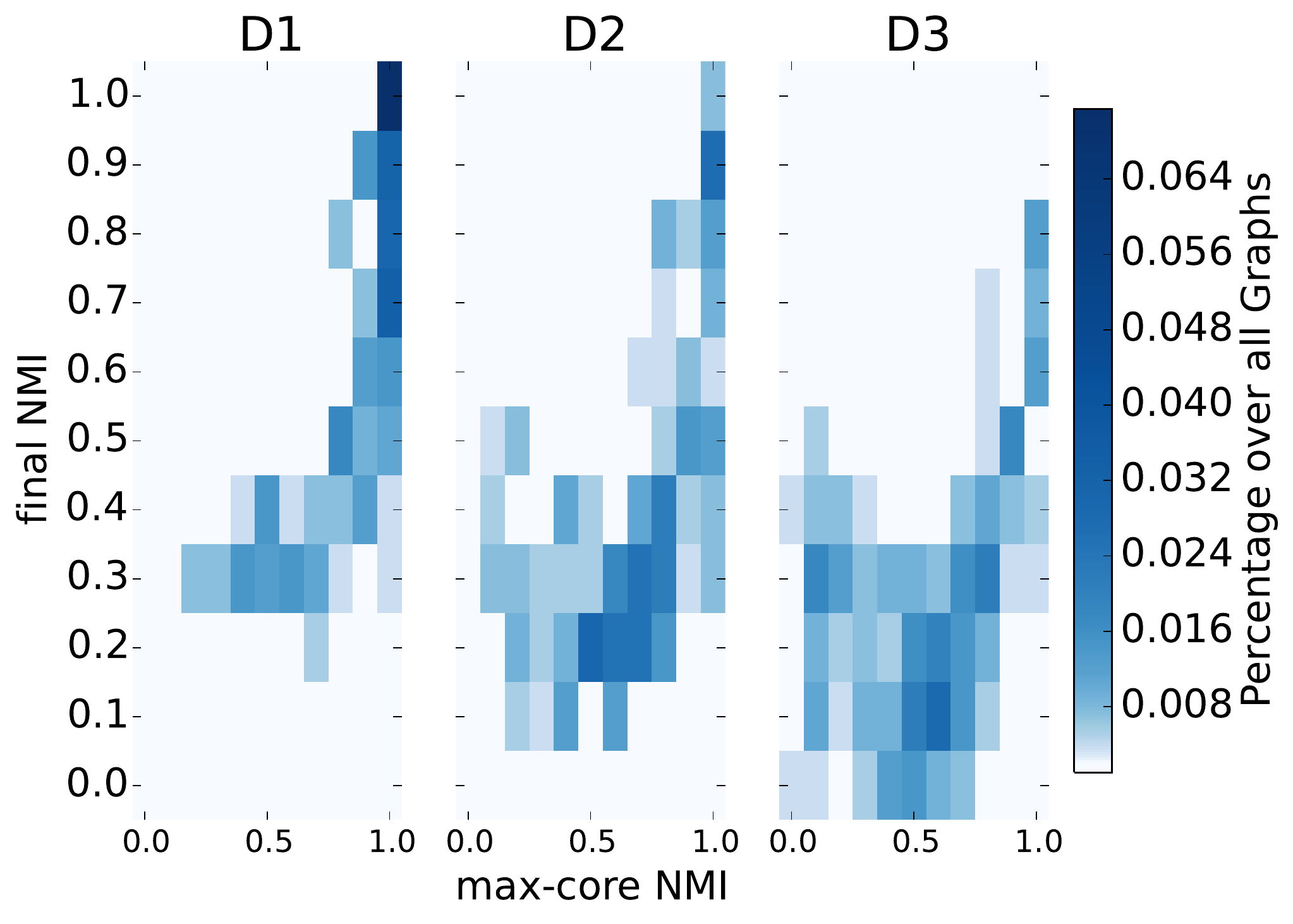}
		\caption{Walktrap}
	\end{subfigure}
	\caption{``Final Performance'' vs ``Performance at the Maximum-Core'' for the three artificial graphs (D1, D2, and D3)  (Part II).}
\end{figure}

Our proposed framework works in an incremental manner starting a cluster from the most cohesive part and adding “on top” of that. Following this hierarchical manner it is susceptible to errors propagating from the starting basis to the final clustering result. Thus, it is important to evaluate an algorithm at the most cohesive part in order to anticipate the behaviour at the entirety of the graph. 

In this section, we would like to study properties of the algorithm and in particular the behaviour of our framework in regards with the starting core (which is the application of the Baseline on the max-core). 
Specifically we would like to see how the ``starting'' performance at the maximum core affects the final performance in the entire graph. The performance at the maximum core is evaluated in the same manner as it would be for the entire graph but only for the sub-graph of the maximum core. We assume that:
\begin{itemize}
	\item the maximum core is the input graph, 
	\item the cluster membership is maintained,
\end{itemize}
and we simply apply any of the given clustering algorithms.
The maximum core contains a very dense part of the graph \citep{andersen-waw-2009}. Moreover, it contains the most cohesive parts of the clusters in respect to interconnectivity (i.e., a sub-graph where the number of neighbours k is -at least- the maximal one for all).

In our analysis of the results on the artificially generated data, we noticed that almost 50\% of the data contain only one cluster. Many graph clustering algorithms are implemented or originally designed with the assumption that the graph contains multiple clusters. In the current state of our framework, we apply the clustering algorithm in the maximum core without conducting any analysis on the number of the clusters. While the assumption that there will always be more than one cluster makes sense in general cases, in this specific scenario we see two distinctive behaviors.

Therefore, in order to study the effect of the performance at the maximum core, we partition our cases to data that have more that one cluster at the maximum core and data that have only one. For the first case, we report the relation of the NMI at the maximum core versus the NMI on the entire graph. In the second case, since the evaluation at the maximum core is meaningless, we report only the final evaluation on the entire graph.

Starting with Figure~\ref{fig:maxvsfinalnmi} (multiple clusters at the maximum core), we see a consistent behaviour:
\begin{enumerate}
	\item There is an obvious correlation between the performance at the maximum core and that at the entire graph.
	\item Combinations of datasets and clustering algorithms that \textsc{CoreCluster} was very close to the Baseline, are cases where the latter was performing relatively well on the max-core. 
	\item The performance at D1 dataset is consistently better. This validates that D1 contains structures which most algorithms would identify as clusters. 
\end{enumerate}
One example, of the second observation, is the case of the algorithm ``Leading EigenVector'' in D1 where the performance of our framework is identical to that of the Baseline. 
At the same time, Figure~\ref{fig:maxvsfinalnmi} indicates that this algorithm was really good at detecting the clusters at the maximum core. 
On the other hand, ``InfoMap'' is an example of an algorithm that had ``failures'' at the maximum core which would explain the greater divergence. 

\begin{figure}[t]
	\begin{subfigure}{1.0\textwidth}	
		\includegraphics[scale=0.42]{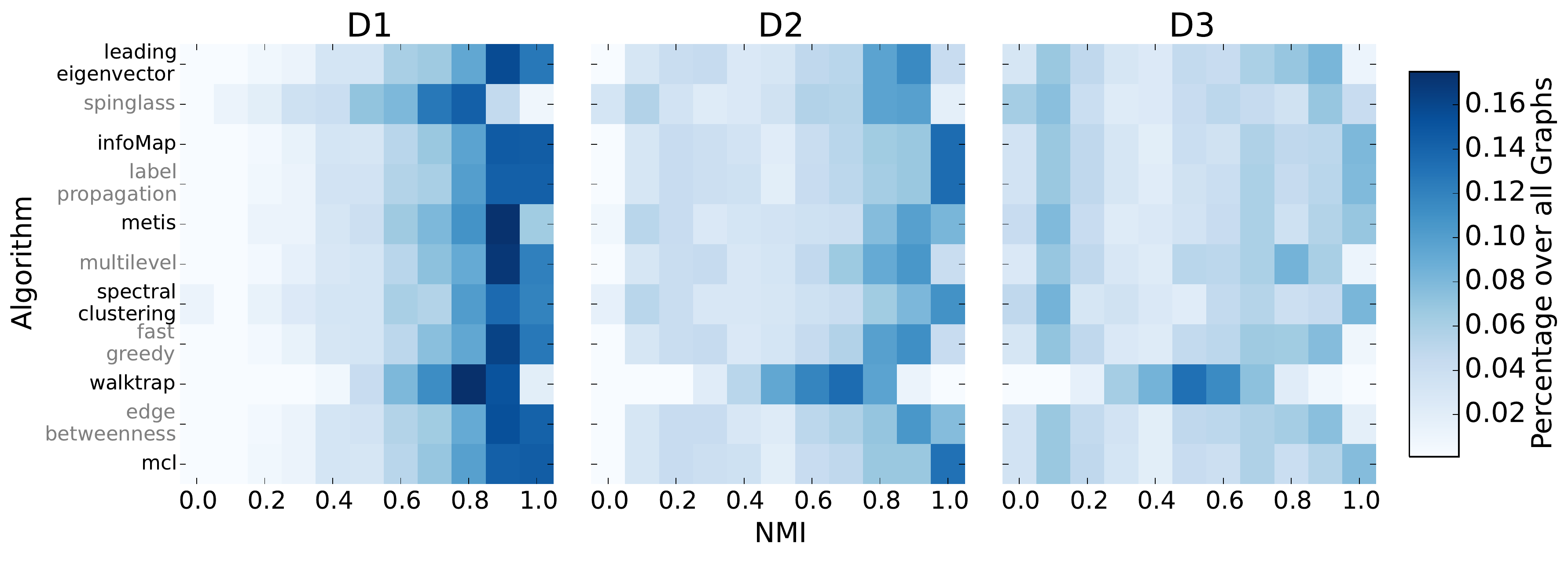}	
	\end{subfigure}
	\caption{Final NMI: \textbf{one cluster} at maximum core.}
	\label{fig:onefinalnmi}
\end{figure}

Additionally, we examine Figure~\ref{fig:onefinalnmi} to  provide further insight. With ``Walktrap'' as a counter example, we see that this algorithm behaves quite similarly to  ``Leading EigenVector'' in the case where we have multiple clusters in the maximum core. But the NMI evaluation is not that similar. We attribute this to the other 50\% of the data where we have only one cluster (at the maximum core). In this case, ``Walktrap'' performs distinctively worse. Upon close examination this algorithm was less capable at detecting that there is only one cluster in the maximum core (in comparison to other algorithms). We can clearly see the result of this in Figure~\ref{fig:onefinalnmi}.

\subsubsection{Real Data, Conductance comparison}\label{sec:NMI}
\begin{figure}
	\begin{subfigure}{1.0\textwidth}
		\centering
		\includegraphics[scale=0.5]{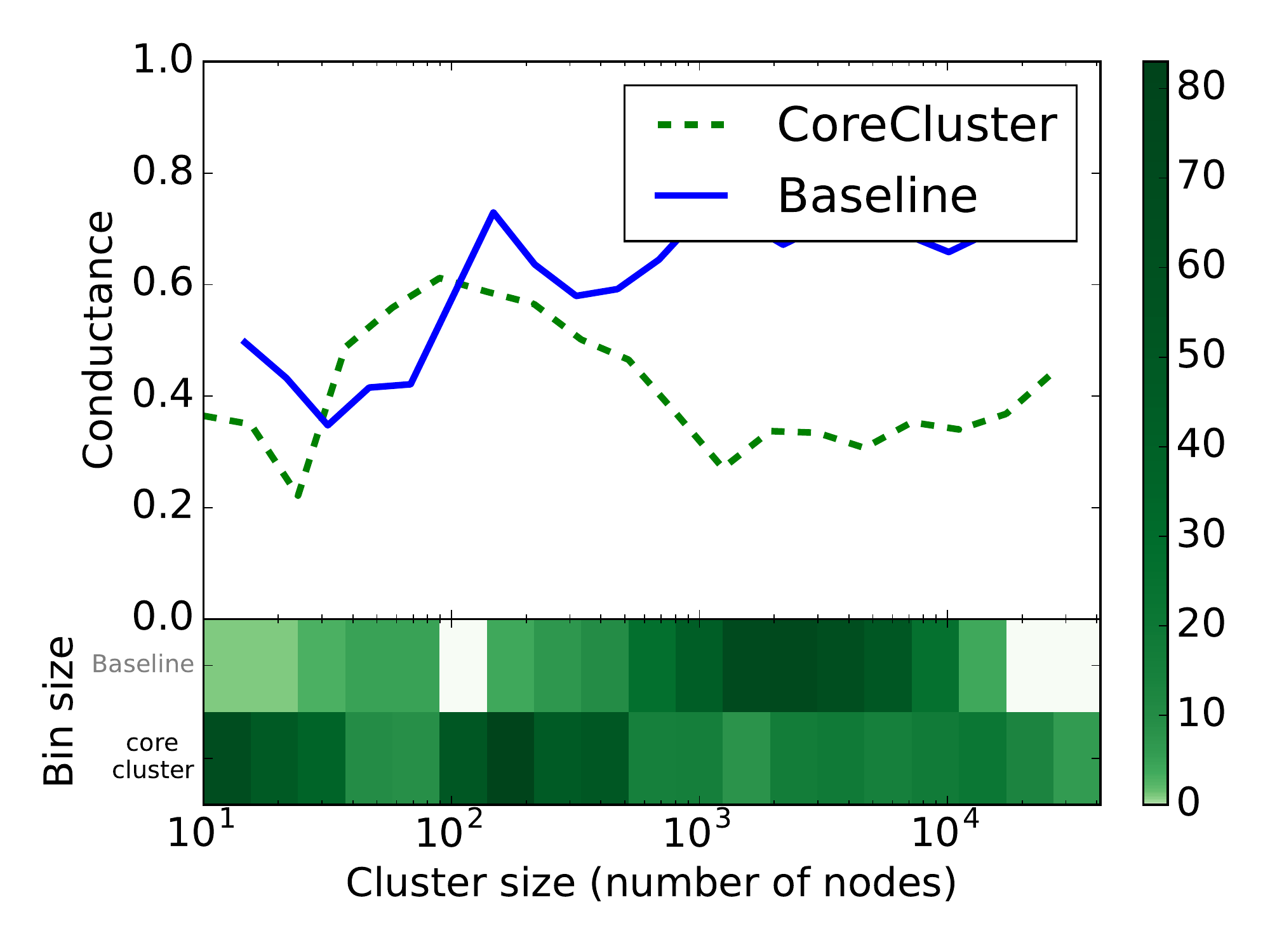}
		\caption{Spectral Clustering}
	\end{subfigure} 
	
	\begin{subfigure}{0.5\textwidth}
		\centering
		\includegraphics[scale=0.35]{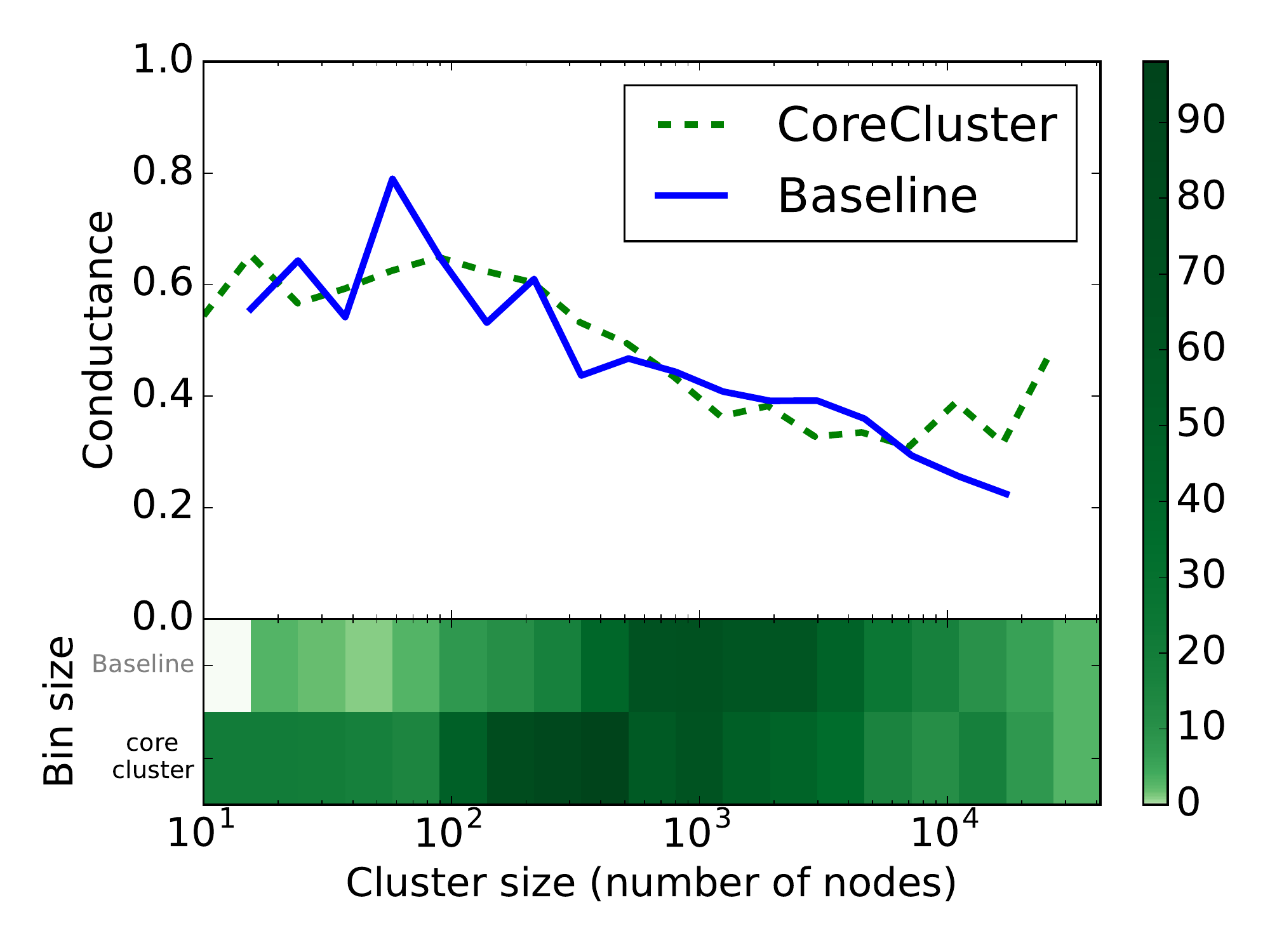}
		\caption{Leading Eigenvector}
	\end{subfigure}
	\begin{subfigure}{0.5\textwidth}
		\centering
		\includegraphics[scale=0.35]{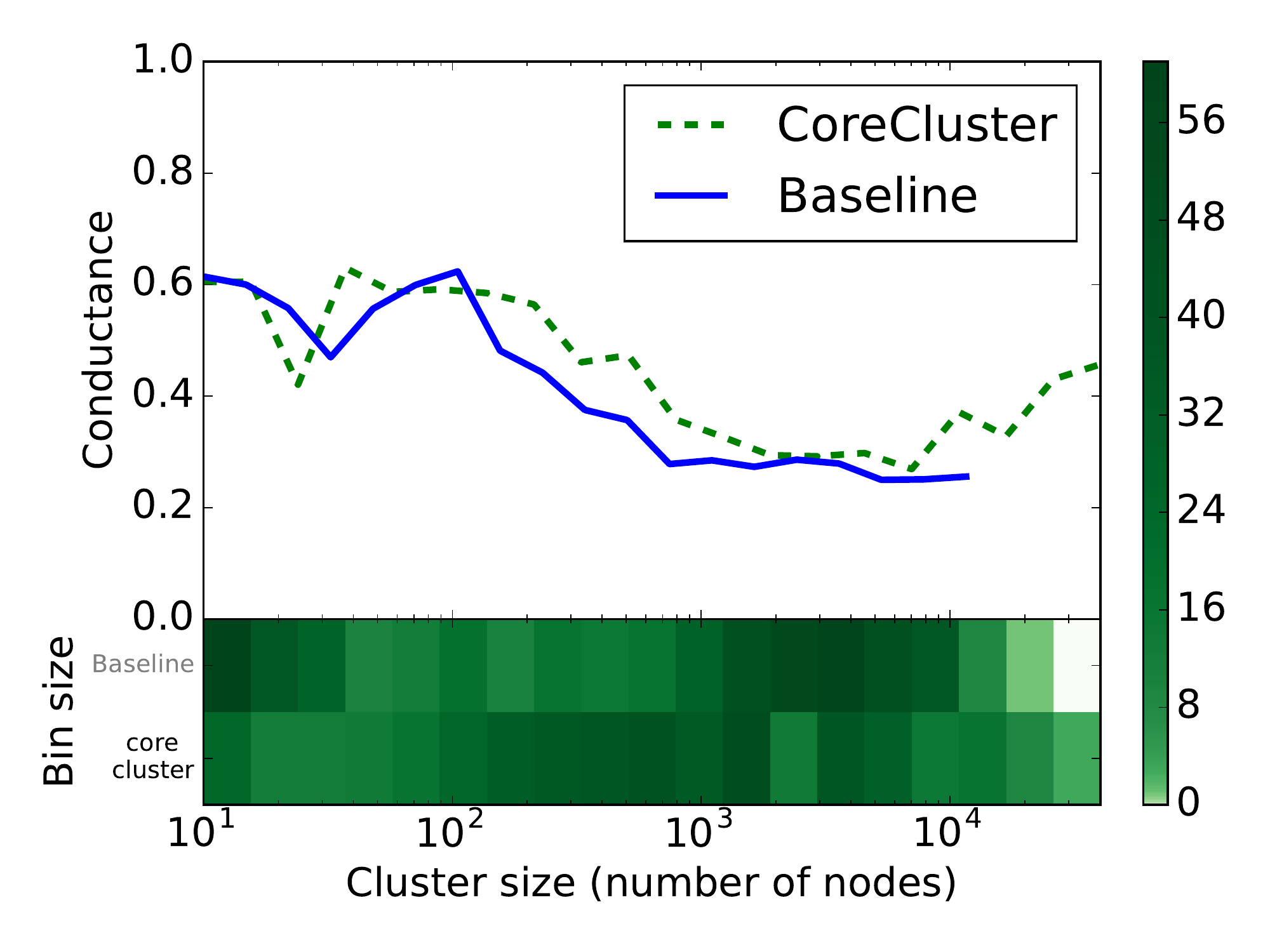}
		\caption{Fast Greedy}
	\end{subfigure}
	
	\begin{subfigure}{0.5\textwidth}
		\centering
		\includegraphics[scale=0.35]{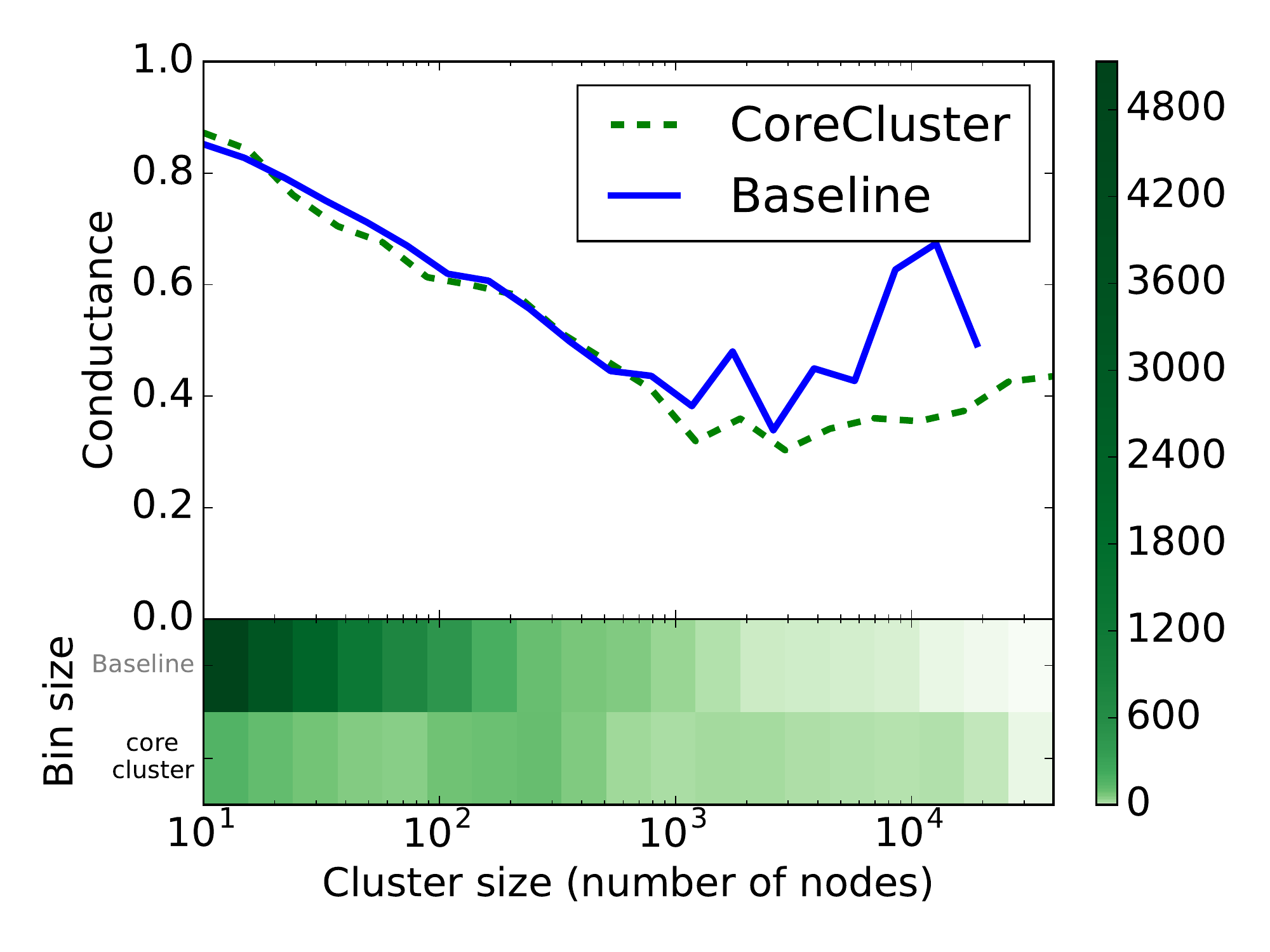}
		\caption{MCL}
	\end{subfigure}
	\begin{subfigure}{0.5\textwidth}
		\centering
		\includegraphics[scale=0.35]{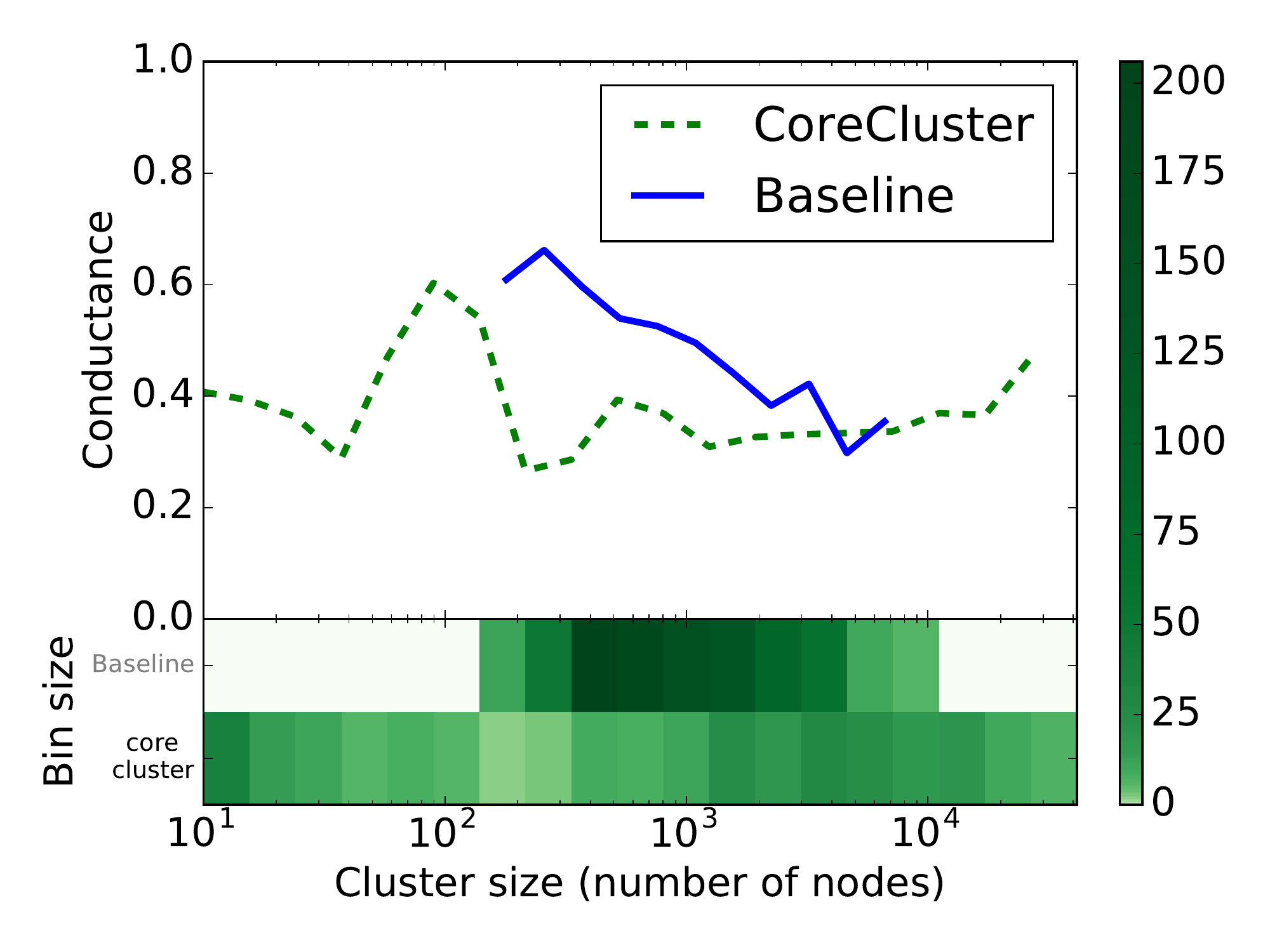}
		\caption{Metis}
	\end{subfigure}
	\caption{Facebook conductance (Part I).}
	\label{fig:facecond}
\end{figure}

\begin{figure}[t]
	\ContinuedFloat 
	\begin{subfigure}{0.5\textwidth}
		\centering
		\includegraphics[scale=0.35]{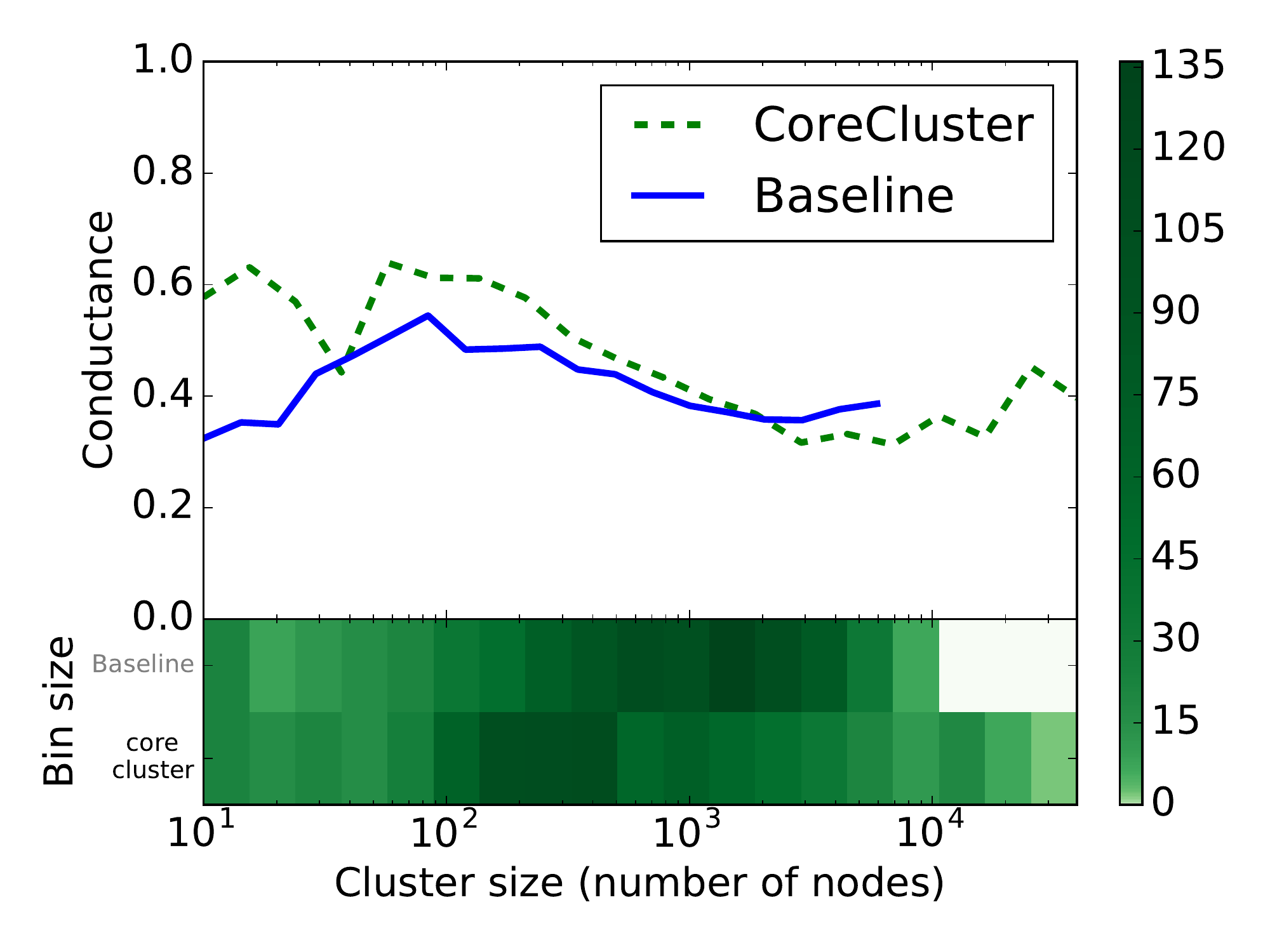}
		\caption{MultiLevel}
	\end{subfigure}
	\begin{subfigure}{0.5\textwidth}
		\centering
		\includegraphics[scale=0.35]{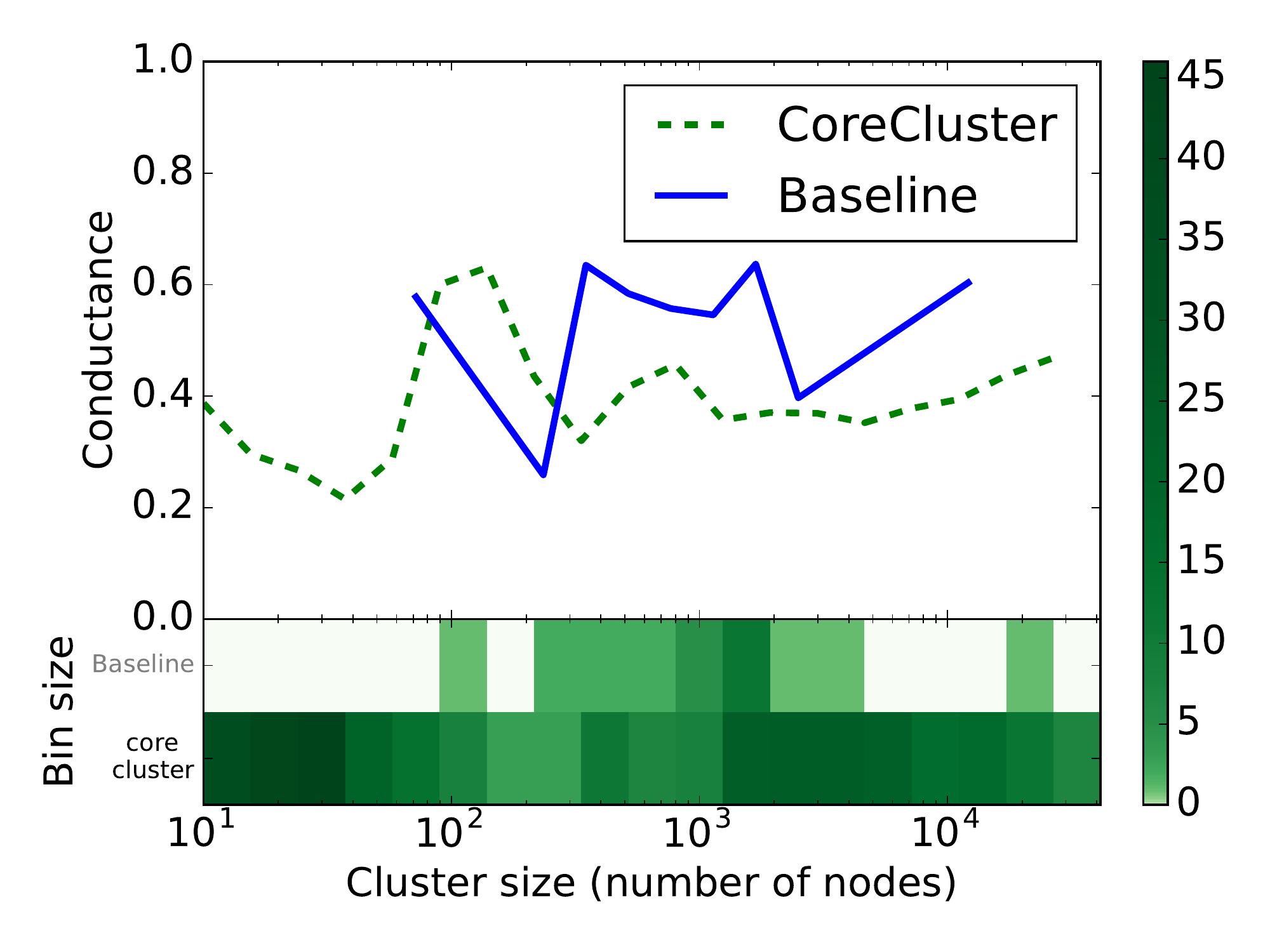}
		\caption{SpinGlass}
	\end{subfigure}
	
	\begin{subfigure}{0.5\textwidth}
		\centering
		\includegraphics[scale=0.35]{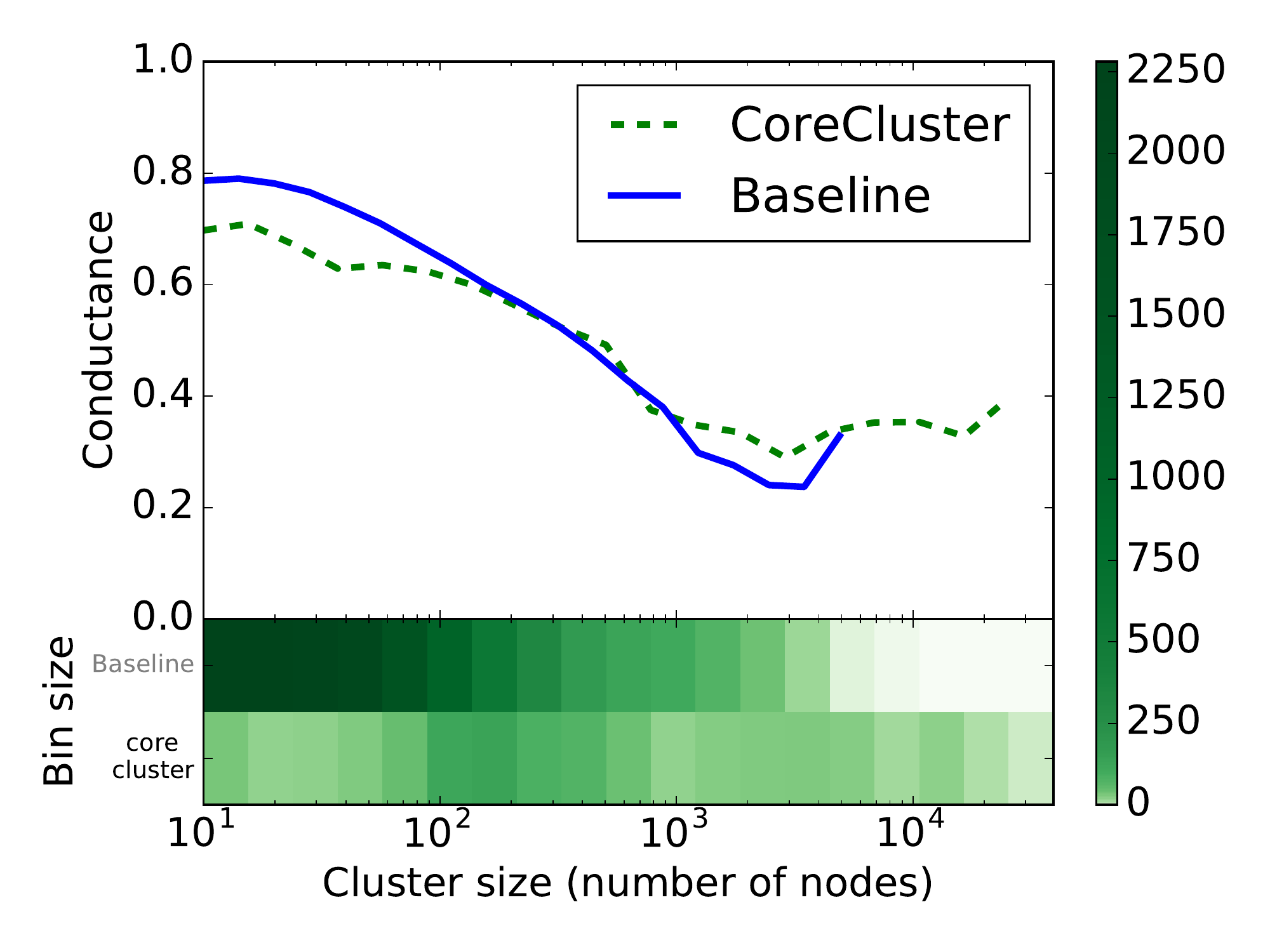}
		\caption{InfoMap}
	\end{subfigure}
	\begin{subfigure}{0.5\textwidth}
		\centering
		\includegraphics[scale=0.35]{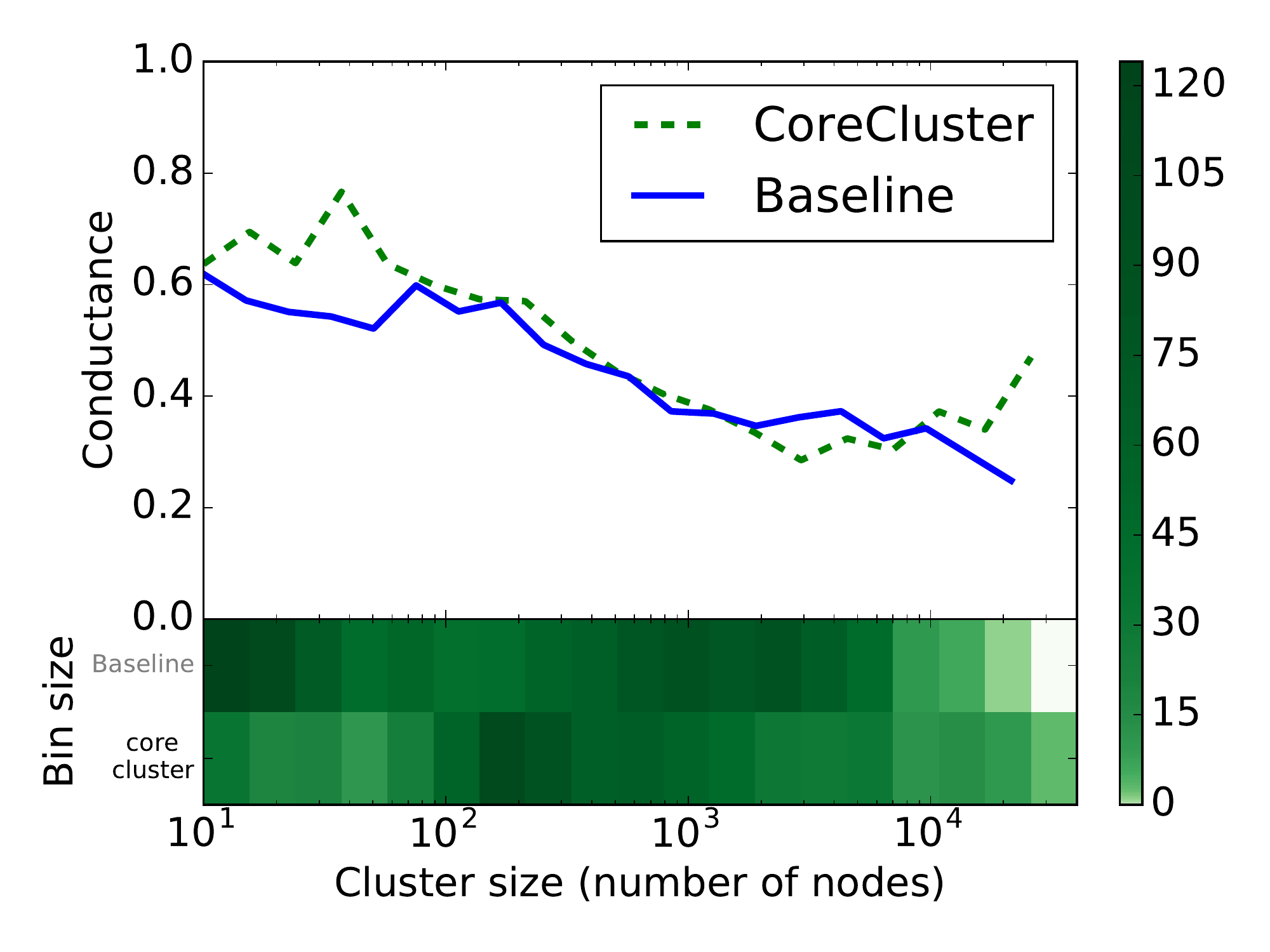}
		\caption{Walktrap}
	\end{subfigure}
	\caption{Facebook conductance (Part II).}
\end{figure}

The previous sections offered a great detail on the inner workings of our framework and gave an easy set of techniques to determine how appropriate is the application of our framework without having to perform exhaustive evaluations. Given that the artificial data range over a wide variety of properties, the results  on the performance (NMI) might not always reflect the performance on real data. Simply put, the artificial data were intentionally created to stress test our framework in even extreme cases. 

In this section, we apply \textsc{CoreCluster} into a real dataset along with all of the aforementioned algorithms, and we evaluate its performance with the well know {\em conductance} metric. 
In this manner, we provide a complete picture of how well our framework performs and we justify its' usefulness with real and concrete examples. 

In Figure~\ref{fig:facecond}, the comparison between the aforementioned ``Baselines'' (Section ~\ref{sec:Algorithms}) and the corresponding \textsc{CoreCluster} ``combination'' is presented. 
Given that the {\em Facebook} dataset is composed by 100 different graphs, we summarize the clustering performance in the following two aspects:
\begin{enumerate}
	\item {\itshape Conductance per cluster size}: The upper ``parts'' of the sub-figures in Figure~\ref{fig:facecond} represents the comparison of the clusters found (along all the graphs) in terms of conductance (lower values are better).  While conductance reflects in both a practical and intuitive manner the quality of a cluster, the size of the clusters found plays an important role (e.g., we can't consider a clustering ``good'' if there is one ``huge'' cluster with high conductance and many ``small'' ones with low). On an additional note, we excluded from the comparison small non connected components as those are trivial to be detected by any algorithm and don't provide additional information for the comparison.
	\item {\itshape Frequency of the bin (bin size):} For presentation purposes, the aforementioned comparison is done by binning the cluster sizes in logarithmic bins. This is to avoid noise in the plot and to provide a better picture on the average behaviour. In this point, we evaluate the frequency of the bins to further investigate the clustering ``tendencies'' of the Baselines and their combinations.  On the plots we refer to the frequency of the bin as {\itshape Bin size}. 
\end{enumerate} 

With the exception of ``MultiLevel'', we see that \textsc{CoreCluster} either outperforms the corresponding Baselines  or has an almost identical quality of results. Regarding the sizes of the clusters, we see from the frequency of the bins that some of the Baselines ({\em Spinglass, InfoMap, Metis, MCL,}) are more prone to a specific range of sizes. On the other hand, \textsc{CoreCluster} framework maintains a good clustering quality while being able to detect cluster in all possible ranges indiscriminately of the ``input'' algorithm. Evidently, this is important as it is not an attribute that all algorithms can demonstrate.

\subsubsection{Execution Time}\label{sec:time}

\begin{figure}
	\begin{subfigure}{1.0\textwidth}
		\centering
		\includegraphics[scale=0.5]{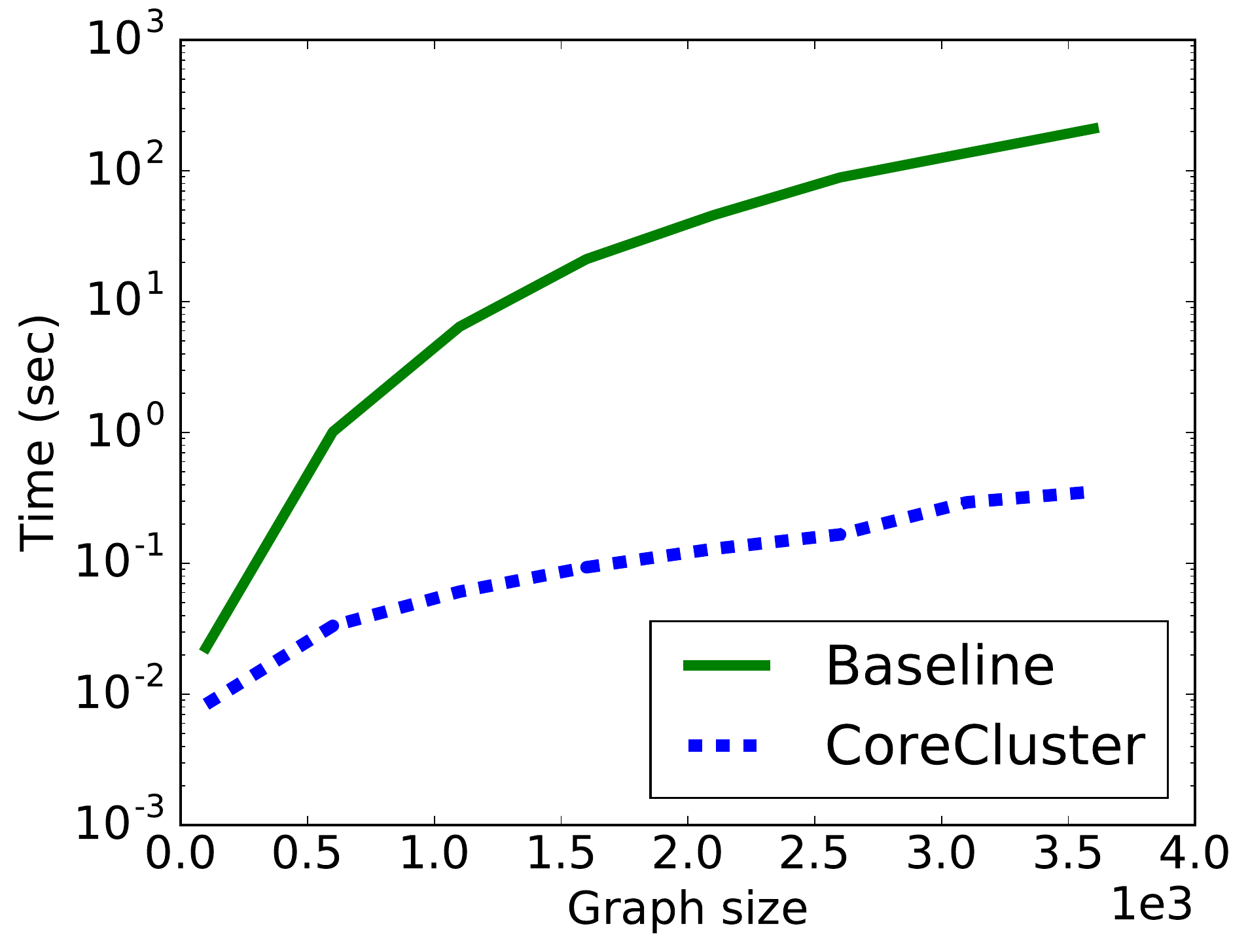}
		\caption{Spectral Clustering}
	\end{subfigure}
		
	\begin{subfigure}{0.5\textwidth}
		\centering
		\includegraphics[scale=0.35]{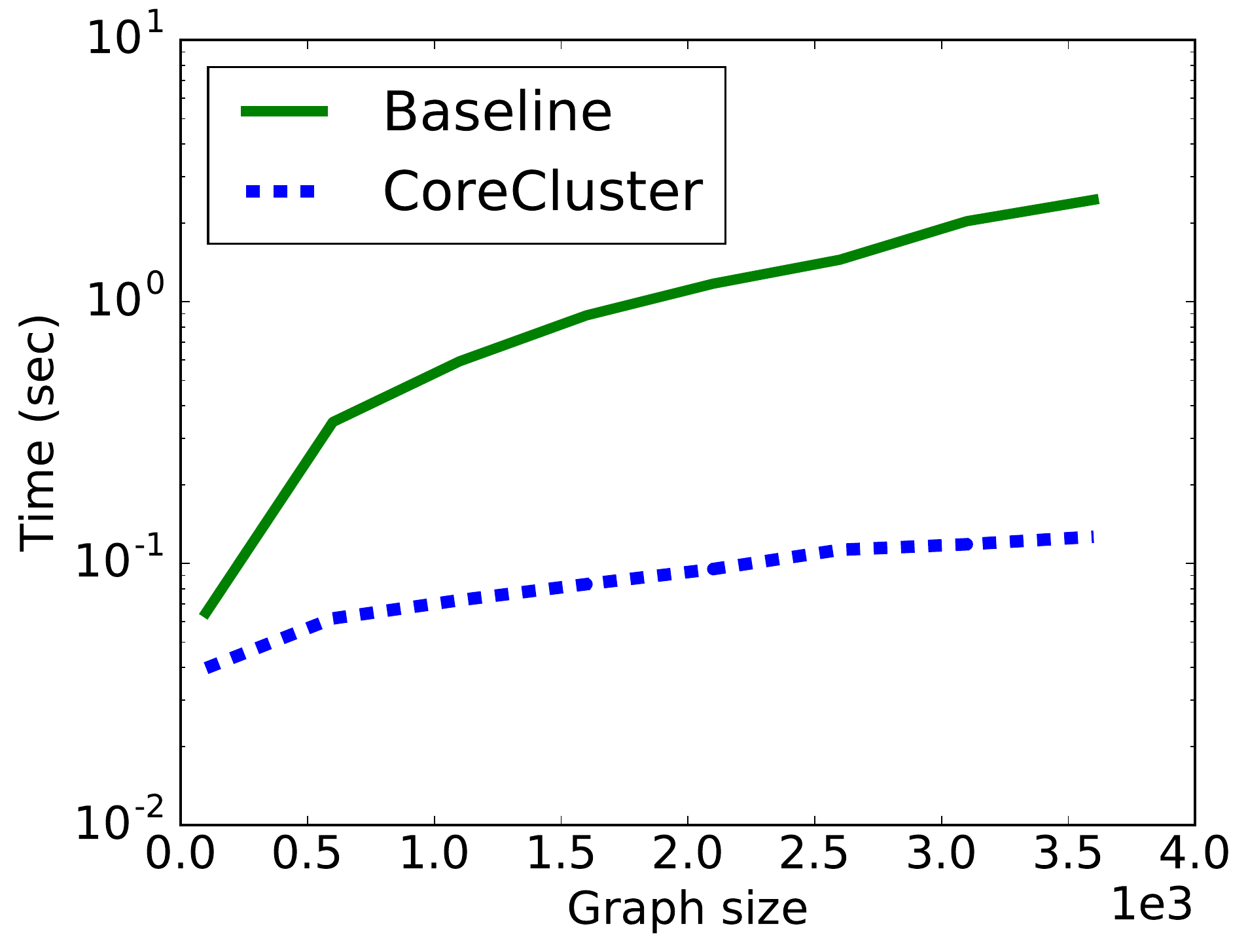}
		\caption{Leading Eigenvector}
	\end{subfigure}
	\begin{subfigure}{0.5\textwidth}
		\centering
		\includegraphics[scale=0.35]{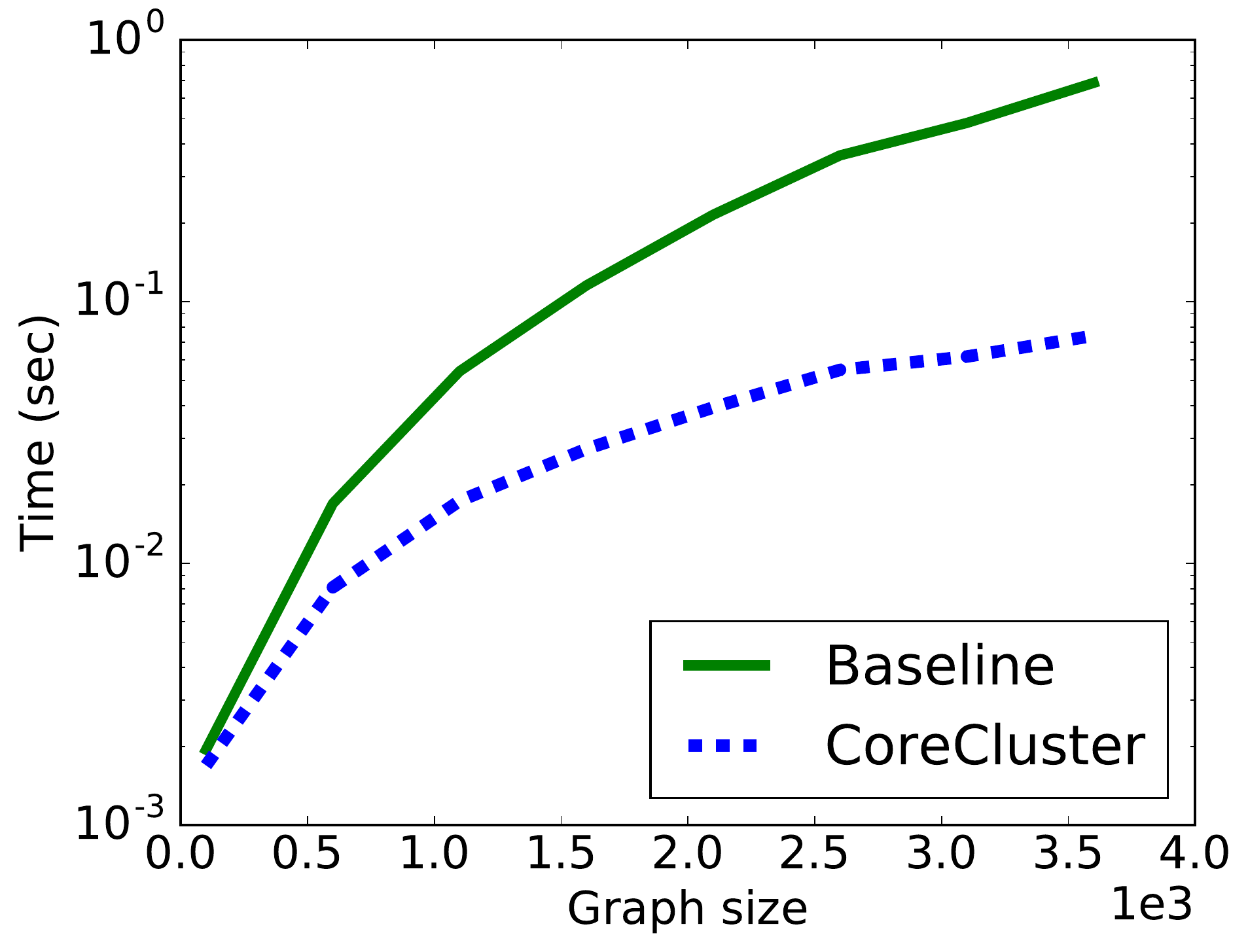}
		\caption{Fast Greedy}
	\end{subfigure}
	
	\begin{subfigure}{0.5\textwidth}
		\centering
		\includegraphics[scale=0.35]{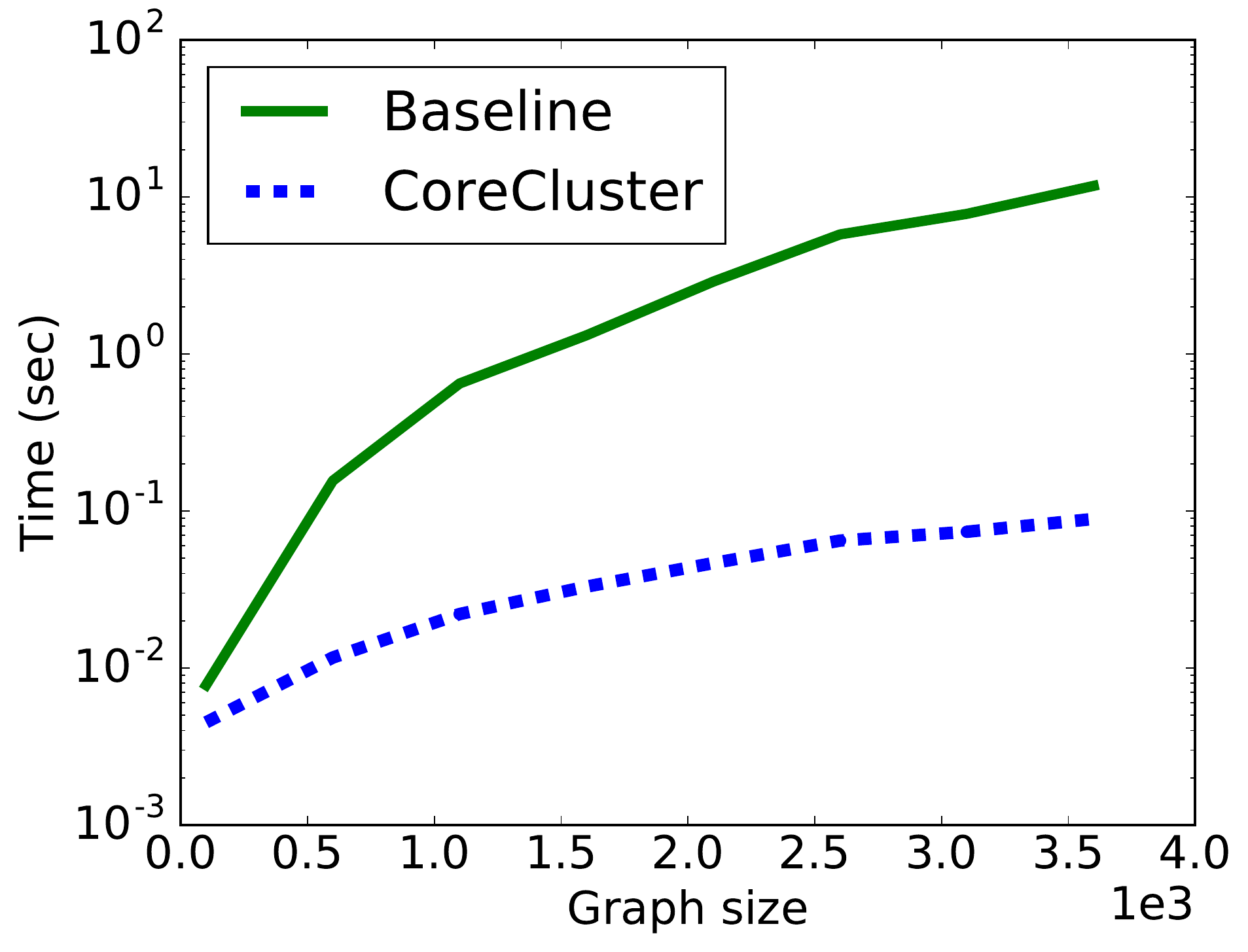}
		\caption{MCL}
	\end{subfigure}
	\begin{subfigure}{0.5\textwidth}
		\centering
		\includegraphics[scale=0.35]{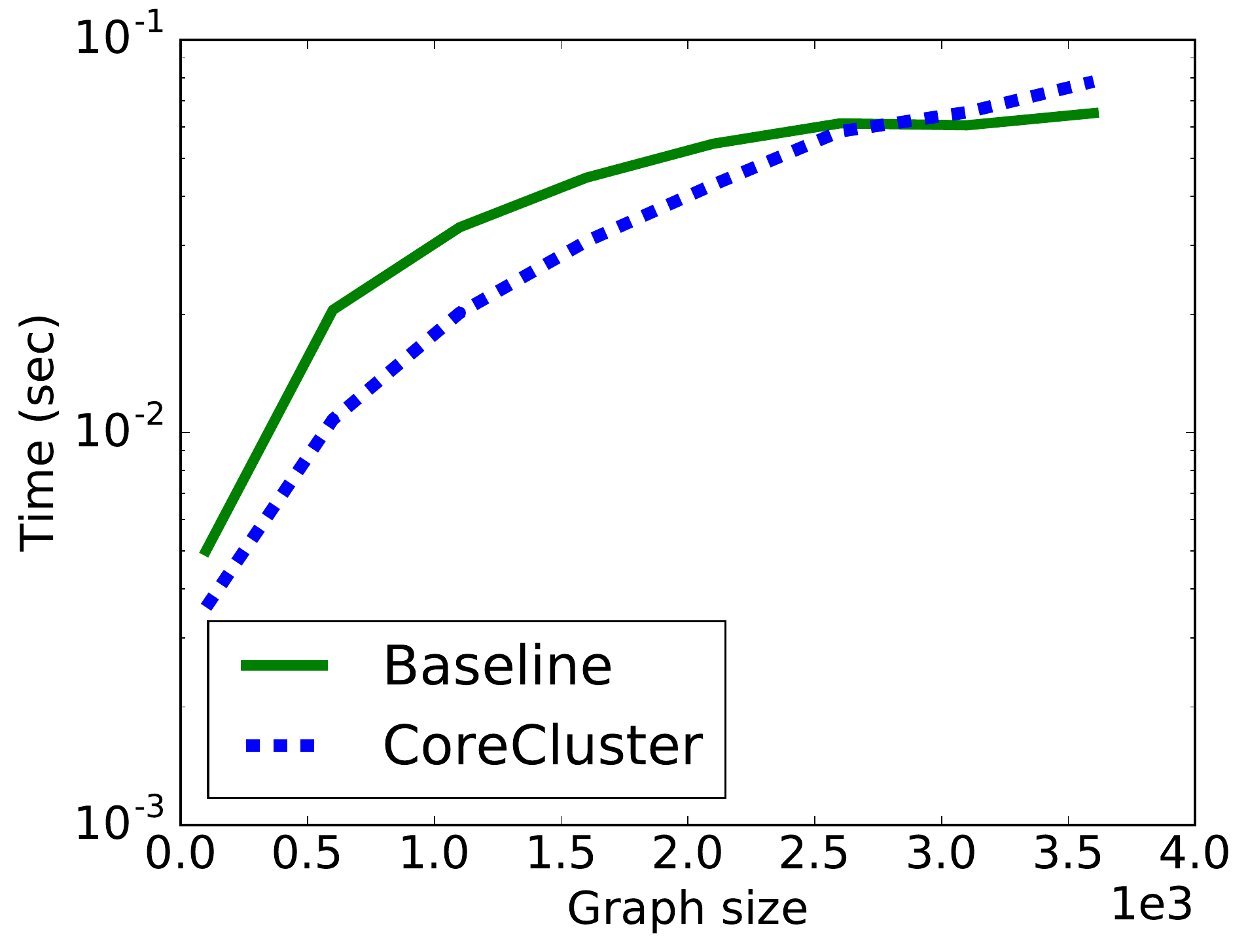}
		\caption{Metis}
	\end{subfigure} 
	\caption{Artificial Graphs execution time (Part I).}
	\label{fig:timeART}
\end{figure}

\begin{figure}[t]
	\ContinuedFloat 
	\begin{subfigure}{0.5\textwidth}
		\centering
		\includegraphics[scale=0.35]{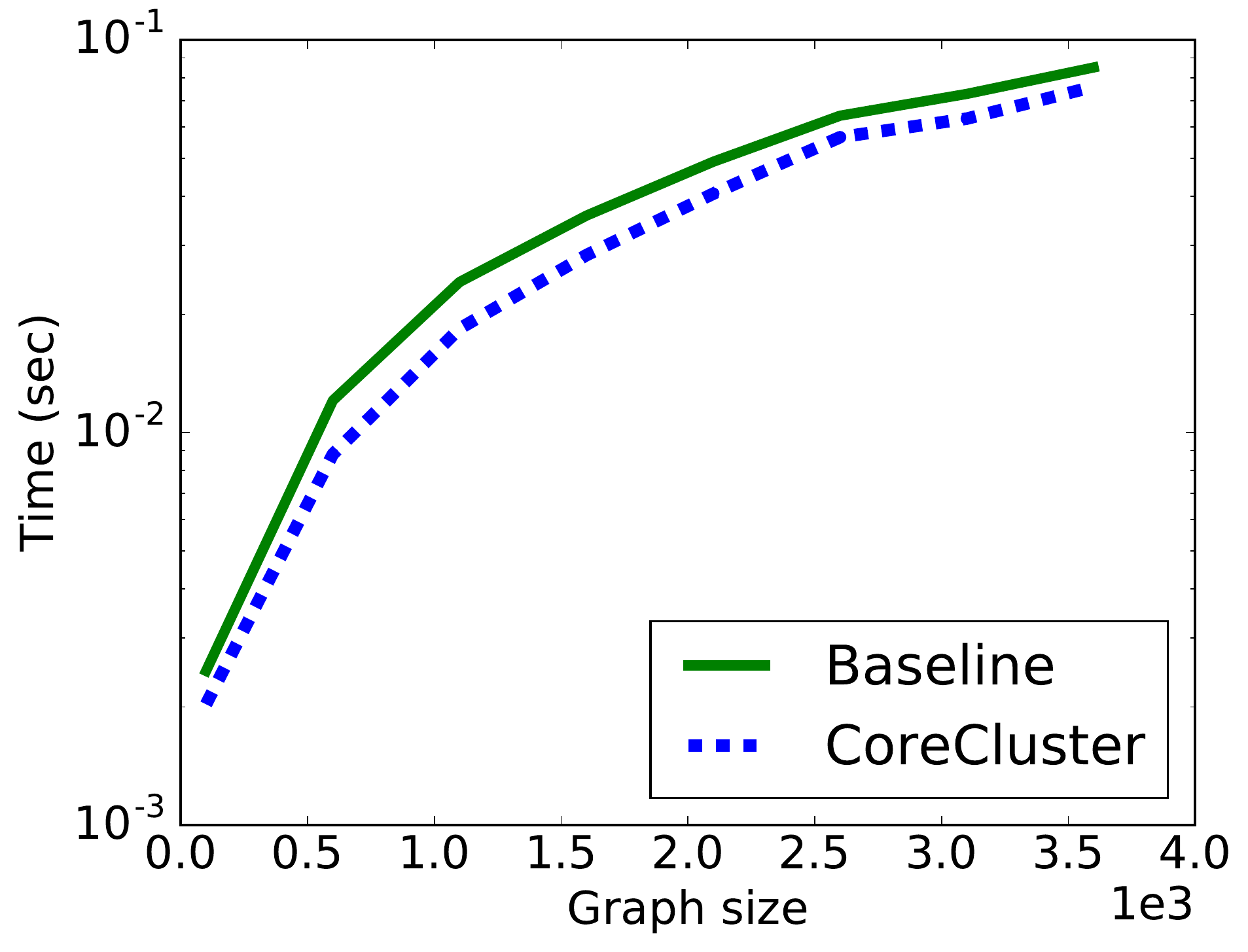}
		\caption{MultiLevel}
	\end{subfigure}
	\begin{subfigure}{0.5\textwidth}
		\centering
		\includegraphics[scale=0.35]{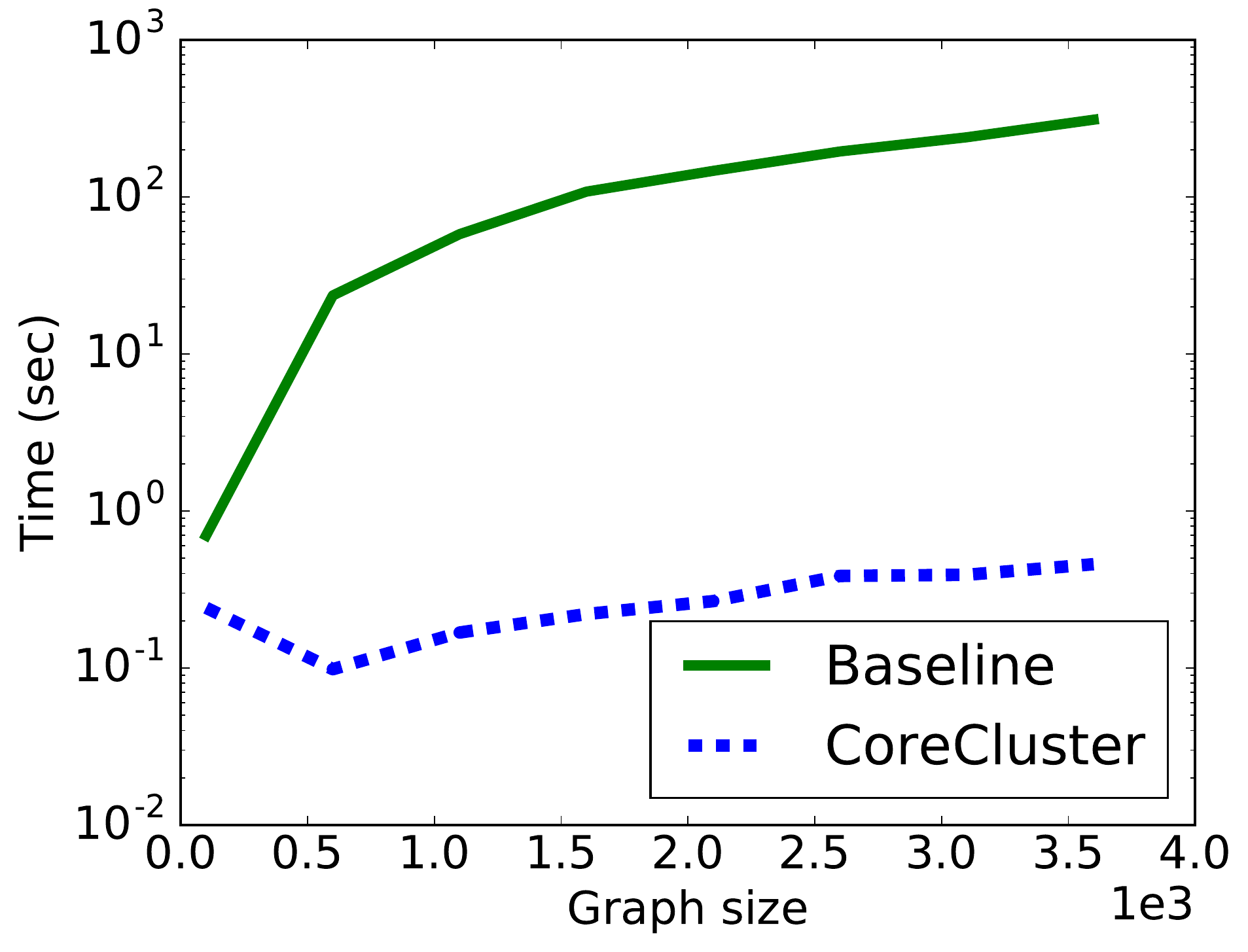}
		\caption{SpinGlass}
	\end{subfigure}
	
	\begin{subfigure}{0.5\textwidth}
		\centering
		\includegraphics[scale=0.35]{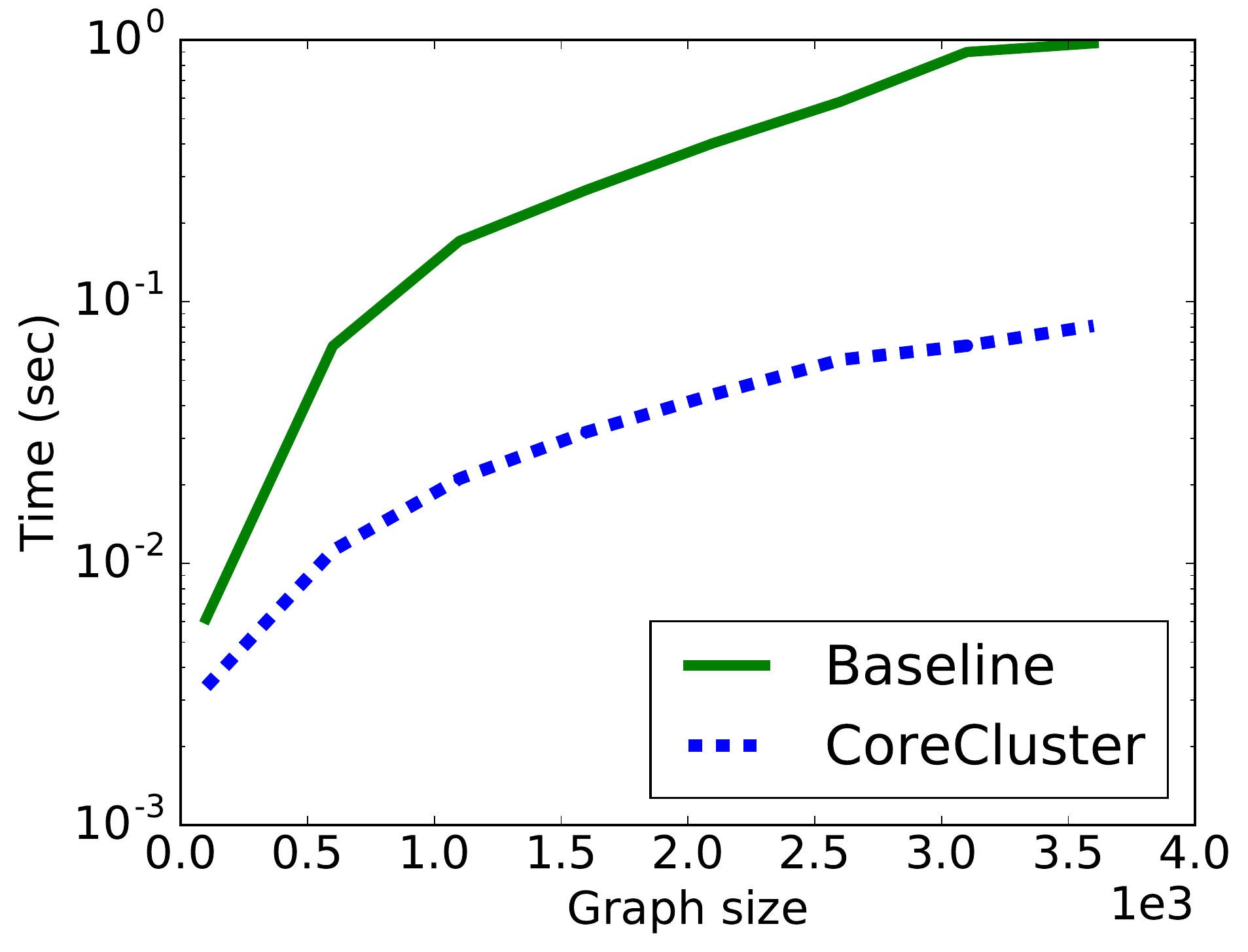}
		\caption{Info Map}
	\end{subfigure}
	\begin{subfigure}{0.5\textwidth}
		\centering
		\includegraphics[scale=0.35]{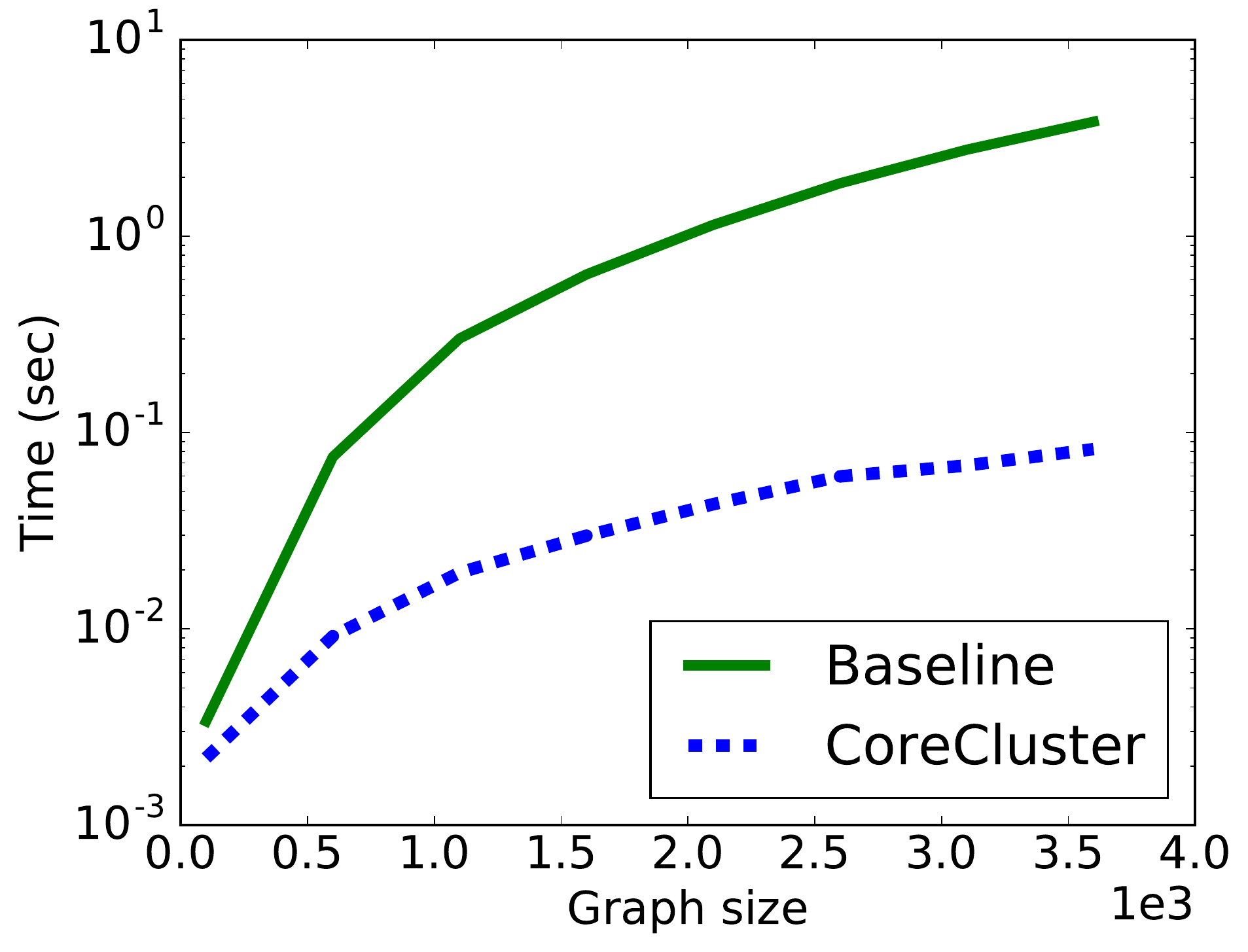}
		\caption{Walktrap}
	\end{subfigure}
	\caption{Artificial Graphs execution time (Part II).}
\end{figure}

\begin{figure}
	\begin{subfigure}{1.0\textwidth}
		\centering
		\includegraphics[scale=0.5]{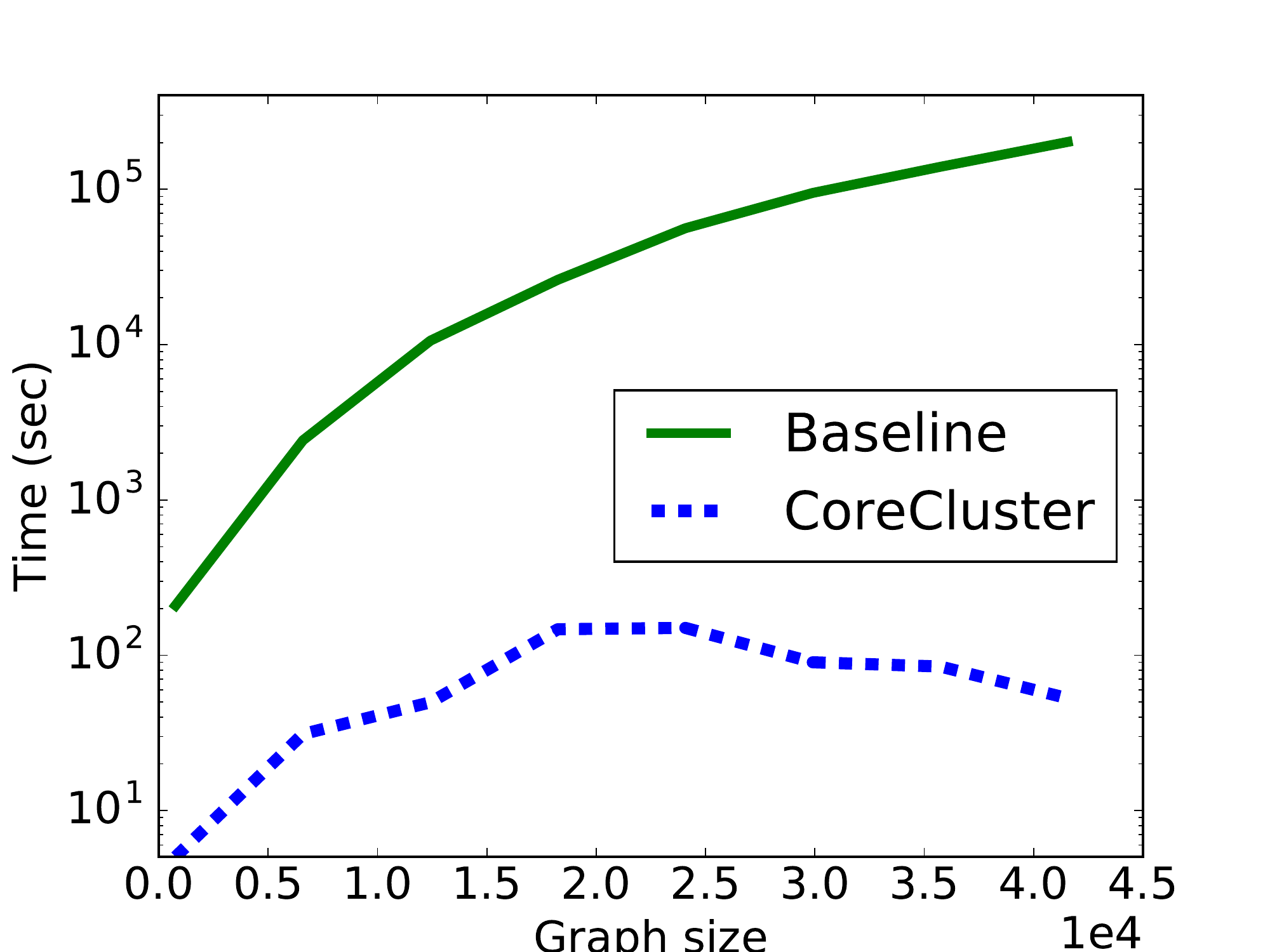}
		\caption{Spectral Clustering}
	\end{subfigure}
	
	\begin{subfigure}{0.5\textwidth}
		\centering
		\includegraphics[scale=0.35]{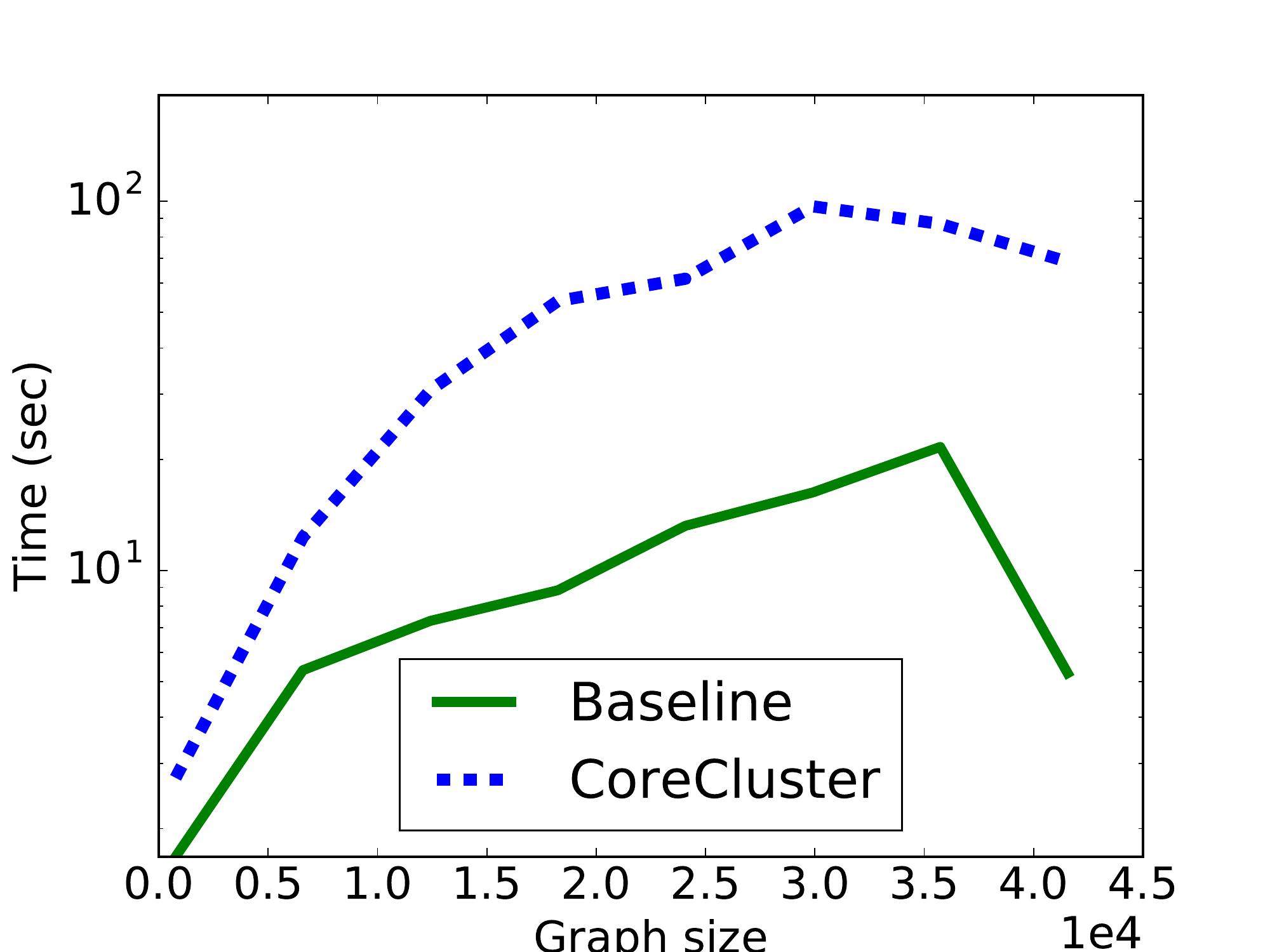}
		\caption{Leading Eigenvector}
	\end{subfigure}
	\begin{subfigure}{0.5\textwidth}
		\centering
		\includegraphics[scale=0.35]{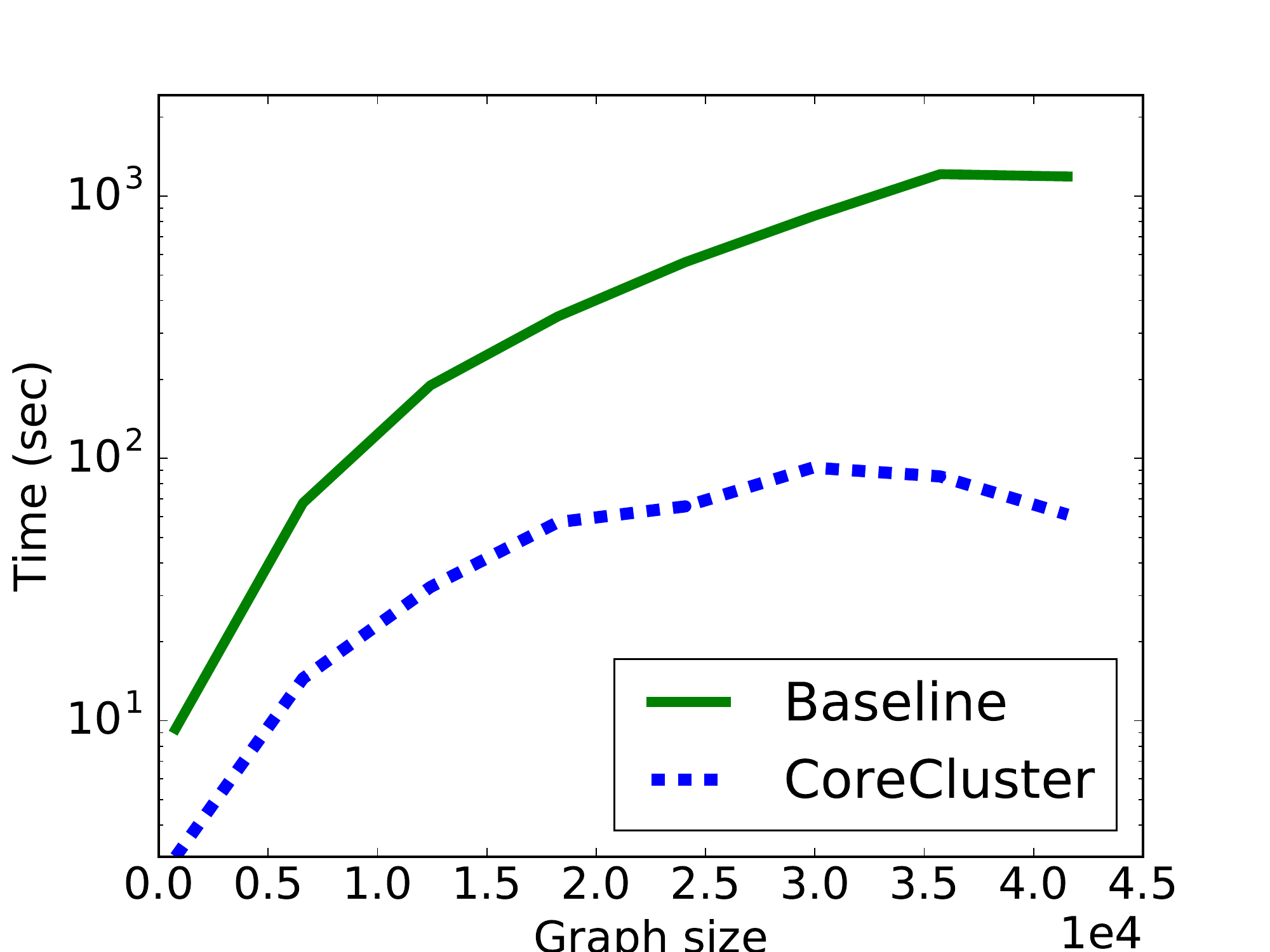}
		\caption{Fast Greedy}
	\end{subfigure}
	
	\begin{subfigure}{0.5\textwidth}
		\centering
		\includegraphics[scale=0.35]{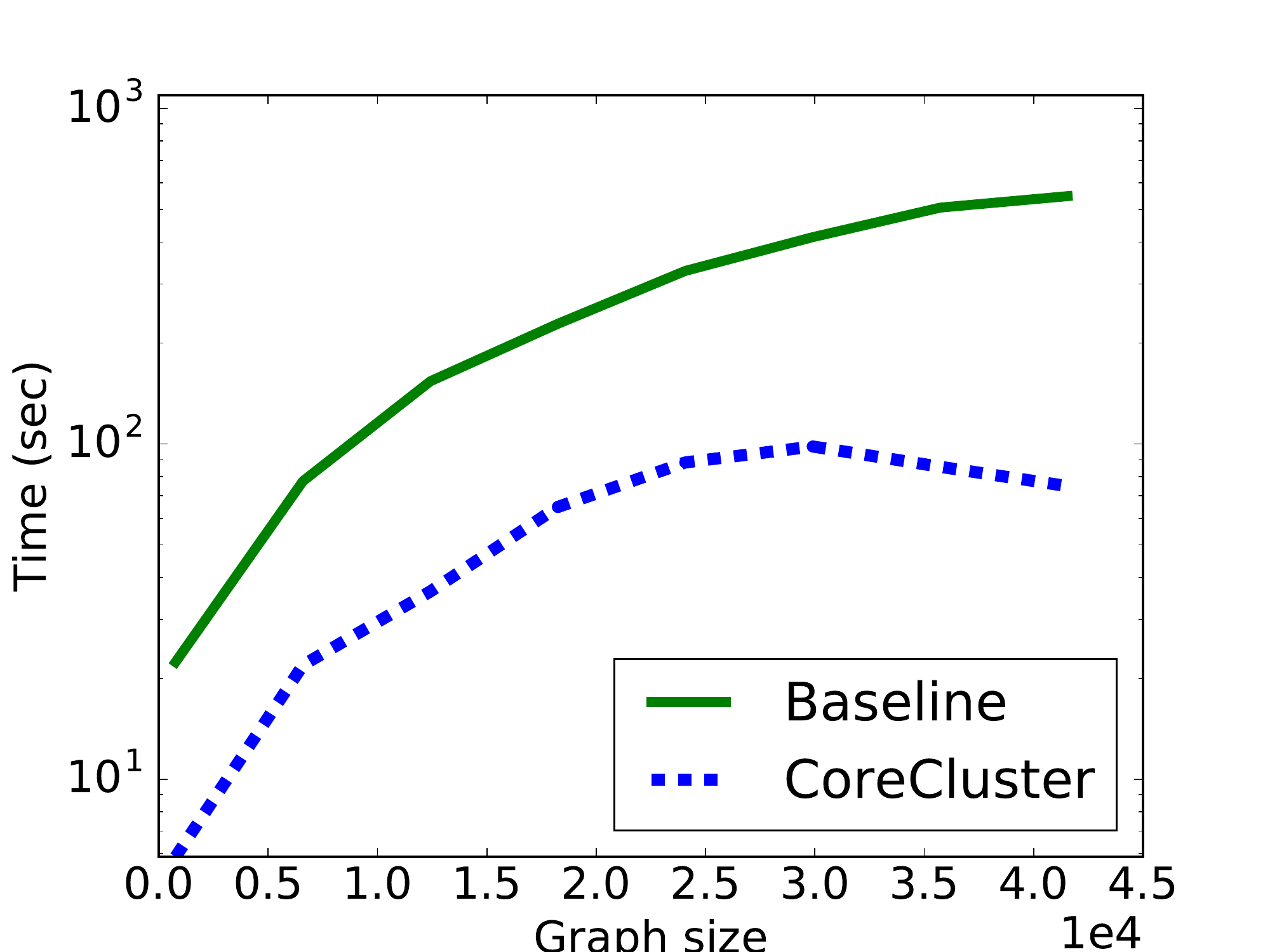}
		\caption{MCL}
	\end{subfigure}
	\begin{subfigure}{0.5\textwidth}
		\centering
		\includegraphics[scale=0.35]{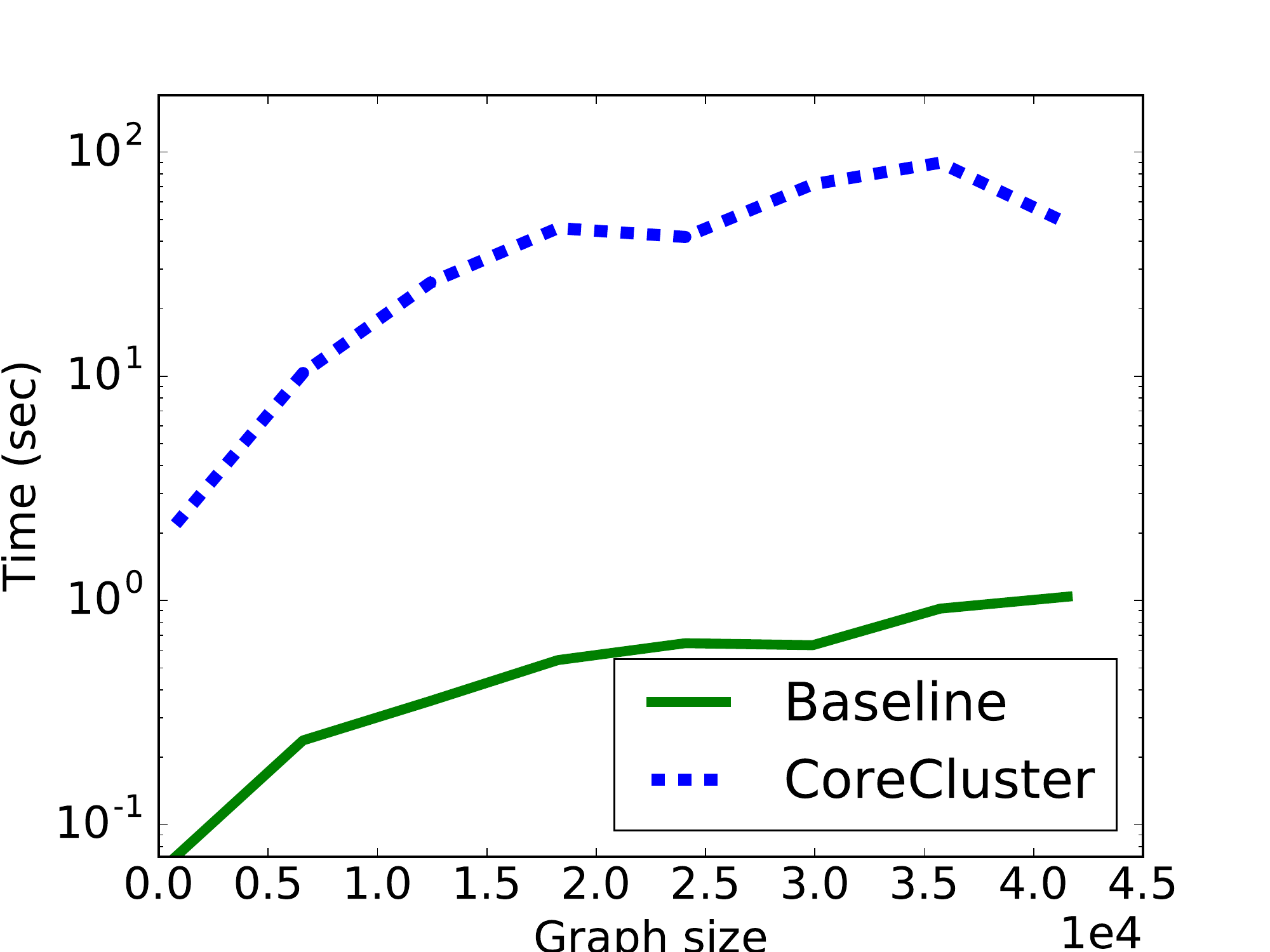}
		\caption{Metis}
	\end{subfigure} 
	\caption{Facebook execution time (Part I).}
\end{figure}

\begin{figure}[h!]
	\ContinuedFloat 
	\begin{subfigure}{0.5\textwidth}
		\centering
		\includegraphics[scale=0.35]{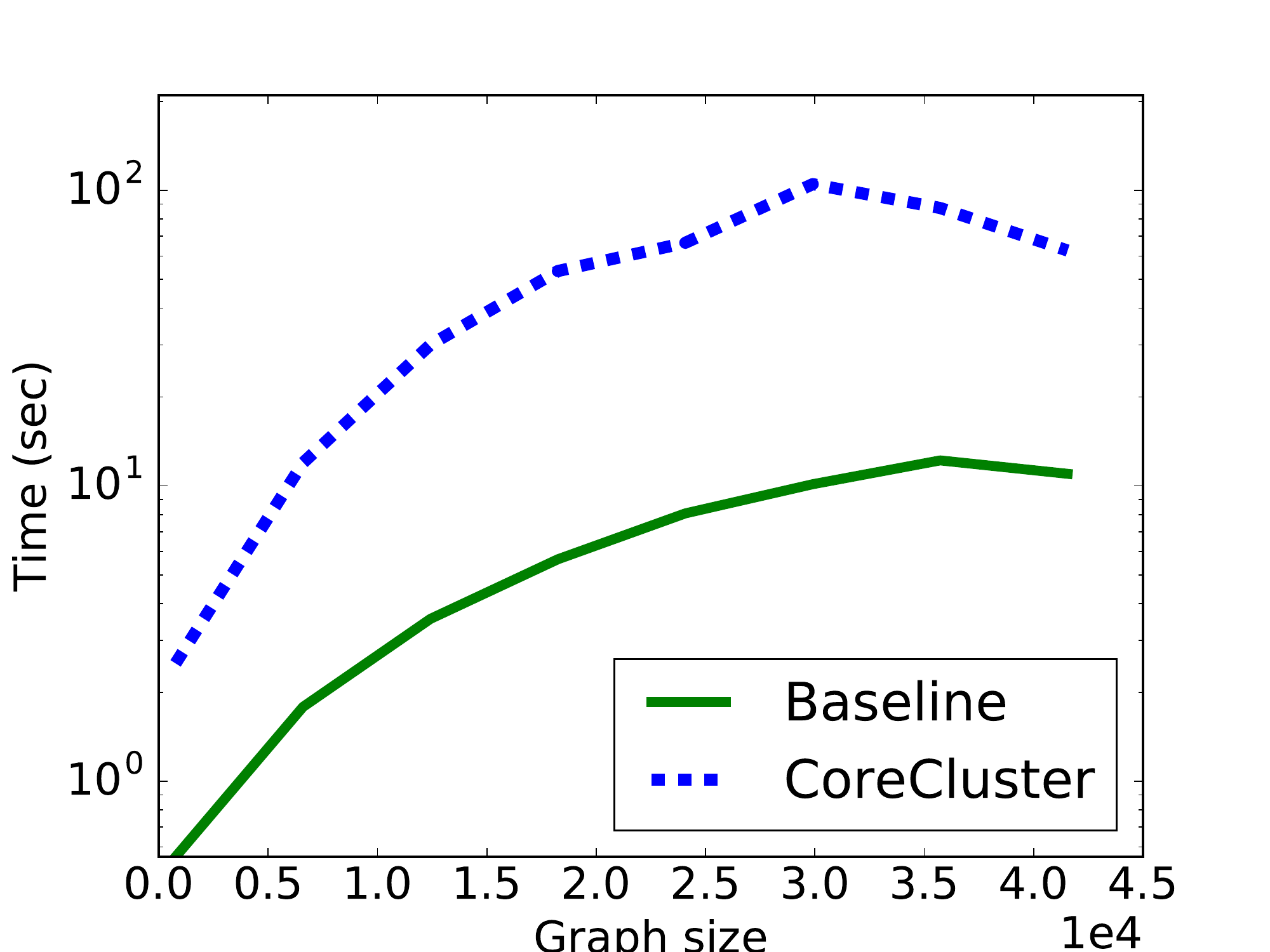}
		\caption{MultiLevel}
	\end{subfigure}
	\begin{subfigure}{0.5\textwidth}
		\centering
		\includegraphics[scale=0.35]{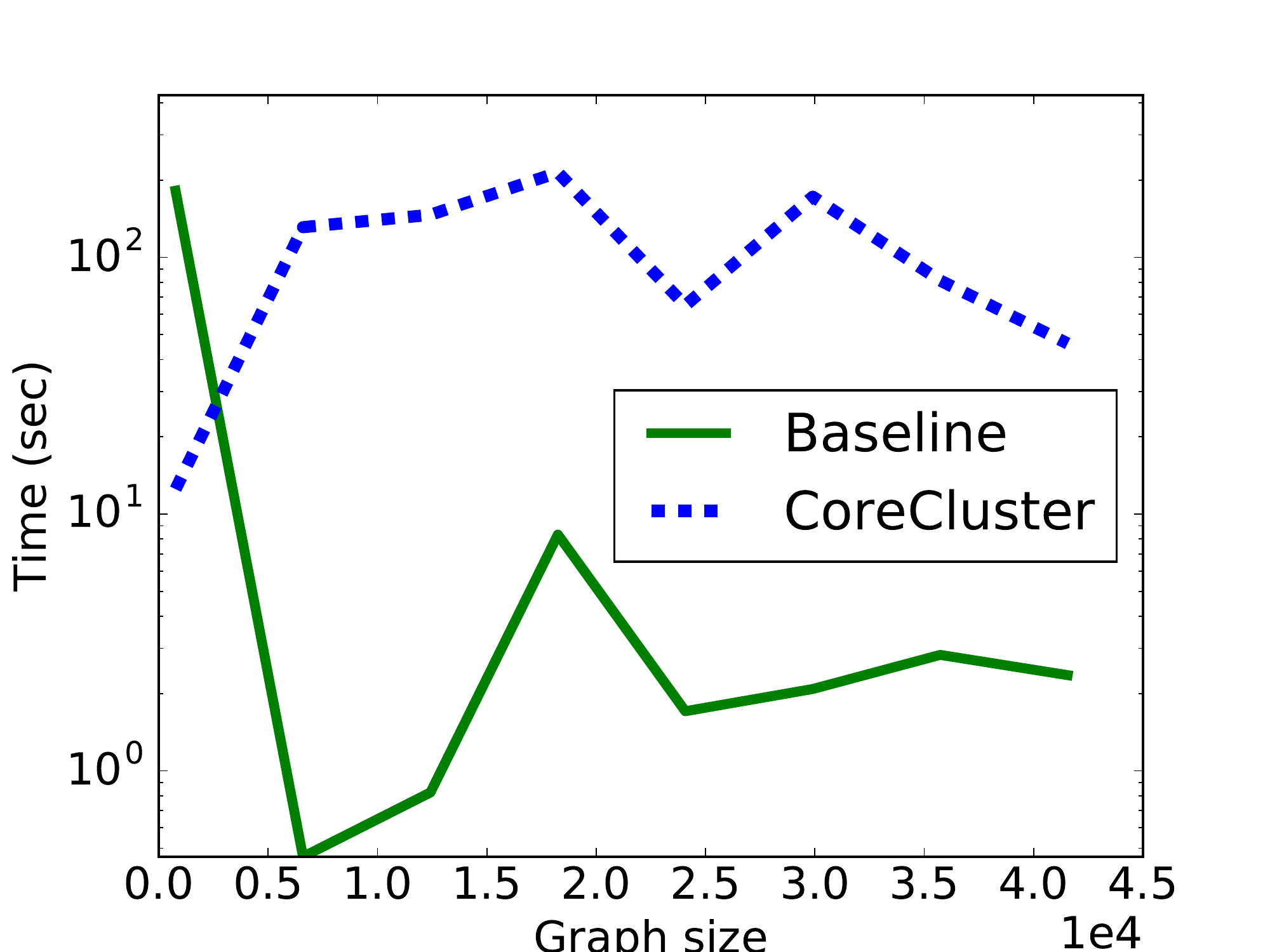}
		\caption{SpinGlass}
	\end{subfigure}
	
	\begin{subfigure}{0.5\textwidth}
		\centering
		\includegraphics[scale=0.35]{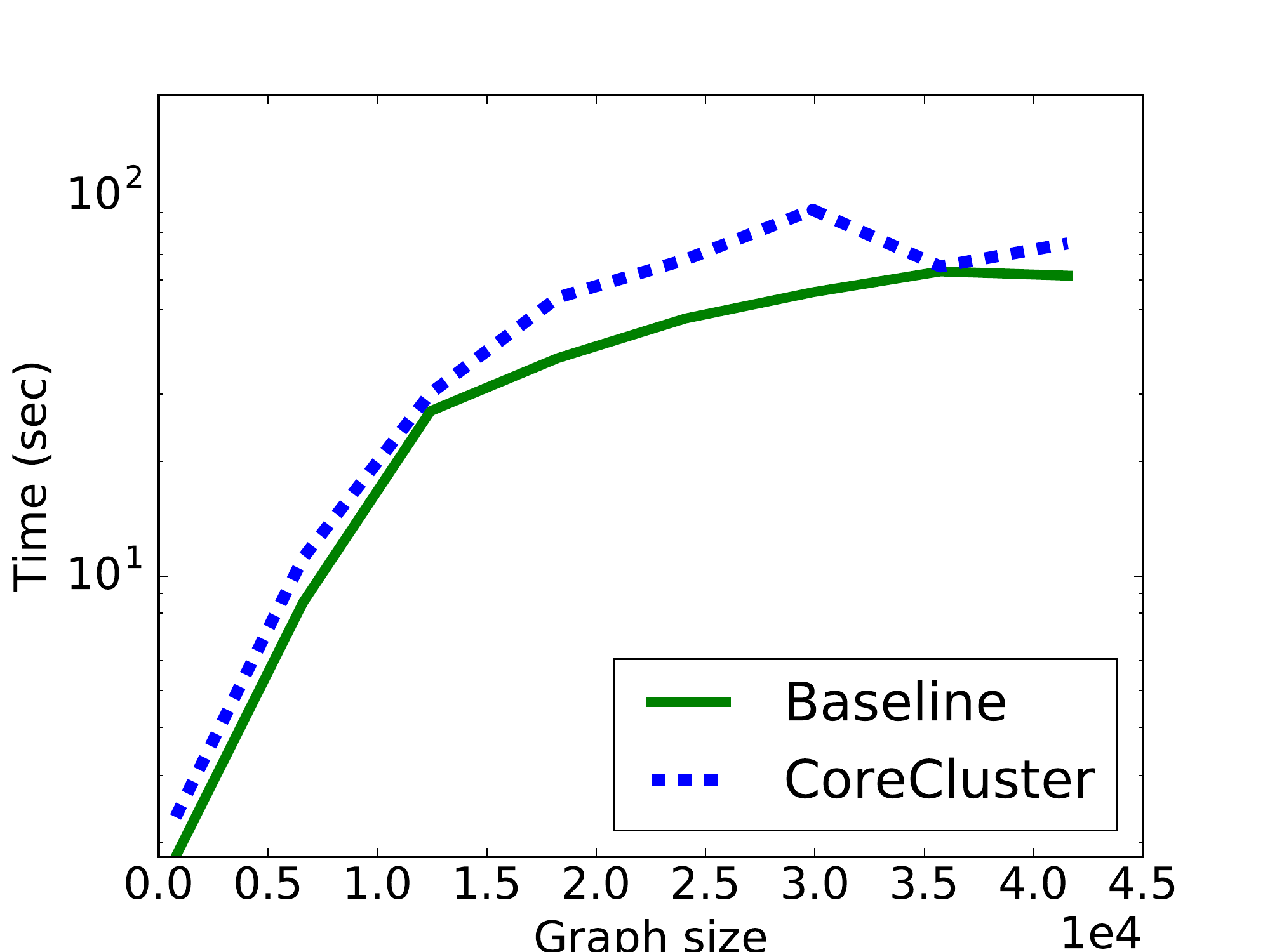}
		\caption{Info Map}
	\end{subfigure}
	\begin{subfigure}{0.5\textwidth}
		\centering
		\includegraphics[scale=0.35]{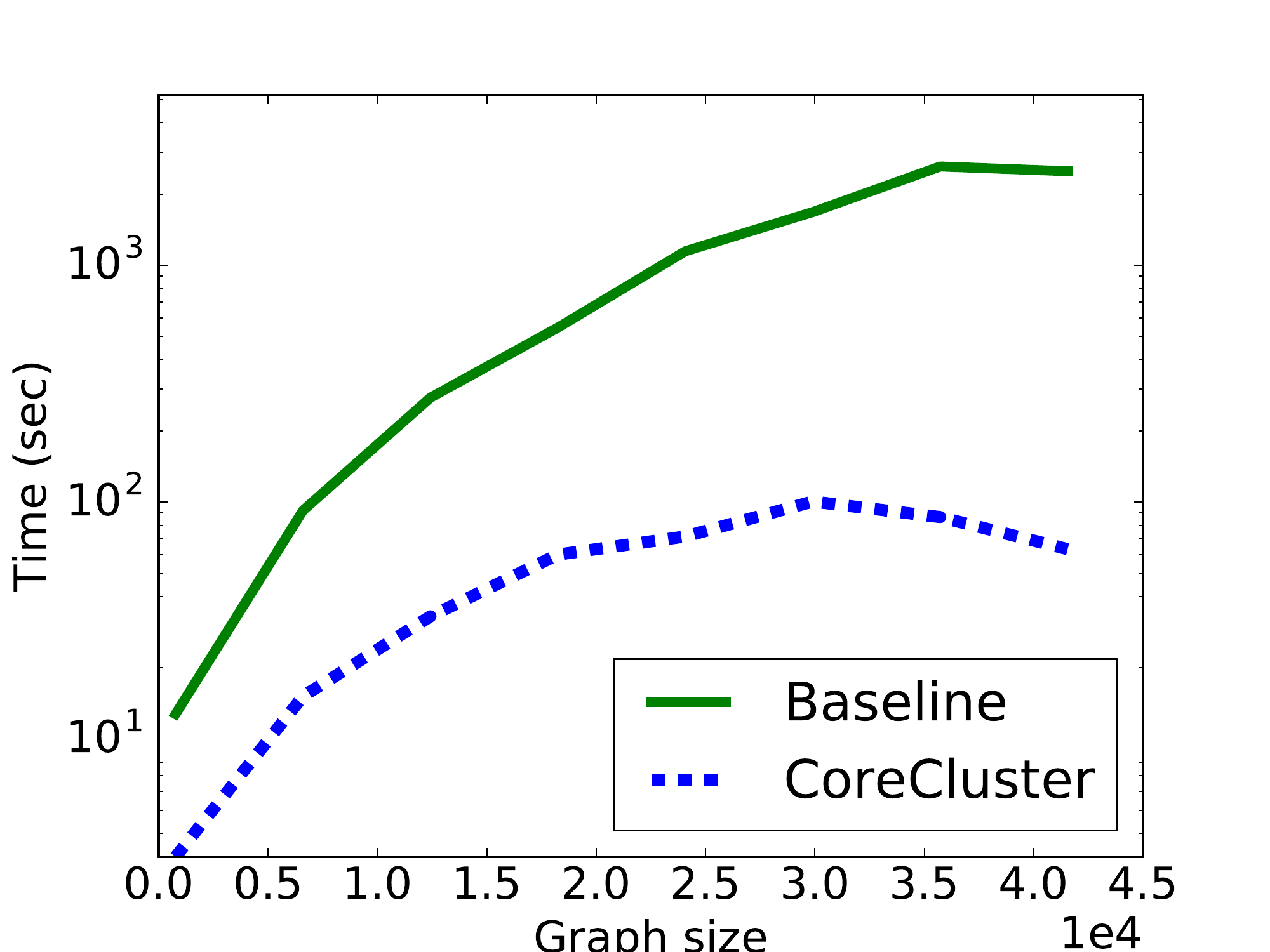}
		\caption{Walktrap}
	\end{subfigure}
	\caption{Facebook execution time (Part II).}
	\label{fig:timeRL}
\end{figure}
As a final part of our evaluation, we present the average execution time of the Baselines in comparison to the corresponding one when \textsc{CoreCluster} is applied. We once more clarify that not all of the Baseline algorithms are appropriate for ``acceleration'' with this framework. 
As mentioned at Section~\ref{sec:method}, \textsc{CoreCluster} is designed to accelerate a computationally expensive algorithm. 
For the cases of ``complex'' algorithms, we can see in Figures~\ref{fig:timeART}~and~\ref{fig:timeRL} the examples of \textit{Spectral Clustering, SpinGlass, Walktrap }and \textit{ MCL} algorithms. In these examples, we see orders of magnitude in the difference between the Baseline and  \textsc{CoreCluster} that increases exponentially for greater graph sizes. 

In the case of the algorithms that are not as computationally expensive there is the added overhead to compute the core structure with the $k$-core algorithm. This overhead makes the \textsc{CoreCluster} slower but we see that the difference is great as the $k$-core algorithm can be linear to the number of edges \citep{batagelj-2003}. This means that the additional cost will not increase immensely for greater graph sizes but, as we also observe in the plots, it will remain within a practically logical range.

Additionally, we note that \textsc{CoreCluster} does not require the full graph in memory as it processes parts of the graph incrementally. Added with the fact that \textsc{CoreCluster} can outperform the Baseline (e.g., Metis for conductance at Figure~\ref{fig:facecond} ) we can envision \textsc{CoreCluster}  being applicable with ``fast'' algorithms as well.

\section{Conclusions}
\label{sec:conclusions}

In this paper, we have presented in full detail a framework for optimizing the efficiency of graph clustering. This framework, is capitalizing on the structures produced by the $k$-core decomposition algorithm. The intuition is that the extreme $k$-cores maintain the most crucial parts of the clustering structure  of the original graph. Our main contribution is the  \textsc{CoreCluster}  framework that processes the graph in an incremental manner and in parts of small size (compared to the entire graph).

We have described how this framework scales-up from an analytical view and why the $k$-core structure offers an appropriate partition with increased ``importance'' (to the clustering structure) of vertices. We displayed both of these in an experimental manner as well with a multitude of evaluations. In more detail, we have presented an exhaustive set of experiments on real and artificial data which:
\begin{itemize}
\item Displayed the acceleration of computationally expensive algorithms while maintaining high cluster quality.
\item Through thorough analysis, described a set of properties -easy to identify and detect- for the proper usage of our framework.
\end{itemize}
Moreover, we provided an indirect review of the used algorithms and offered insights on their individual performance.

As future work, we plan to further extend and improve our framework in the following aspects:
\begin{itemize}
	\item Optimize the heuristics and clustering assignments our framework uses when not utilizing the provided clustering algorithm.
	\item Merging adjacent cores. On a few hand picked occasions, we noticed that the framework would have been more effective if adjacent cores were merged as one. This is mainly for cores that were only marginally different. It could potentially increase the efficiency of execution and the quality of the results if a second pass over the cores merged them appropriately. 
	\item Extend it for directed graphs. For the application of the same model in a directed network, one would have to take into account that degeneracy creates partitions in two different dimensions (incoming and outgoing vertices). Exploring all possible partitions in this case becomes a computation issue by itself.
	
\end{itemize}
Finally, we provide our implementation of \textsc{CoreCluster} for the use of further experiments and its' application to real world scenarios. The full implementation along with instructions on how to utilize the algorithms described in this paper can be found at: \url{removed  will be in the published version}

\appendix
\section*{Appendix A. Empirical Analysis of the CoreCluster's Quality}
\label{sec:AppendixA}
As described in the main part, the intuition behind the \textsc{CoreCluster} framework is that 
 the core expansion sequence
$V_{k},V_{k-1},\ldots,V_{0}$
gives a good sense of direction on how to do clustering in an incremental
way. Moreover, the observation is that the early $i$-cores  (i.e., $i$-cores where $i$ is close to $k$) are already dense, and therefore \textit{sufficiently coherent}, to provide a good starting clustering that will expand well because of the selection criterion.

\begin{figure}[th!]
\begin{center}
\includegraphics[scale=0.4]{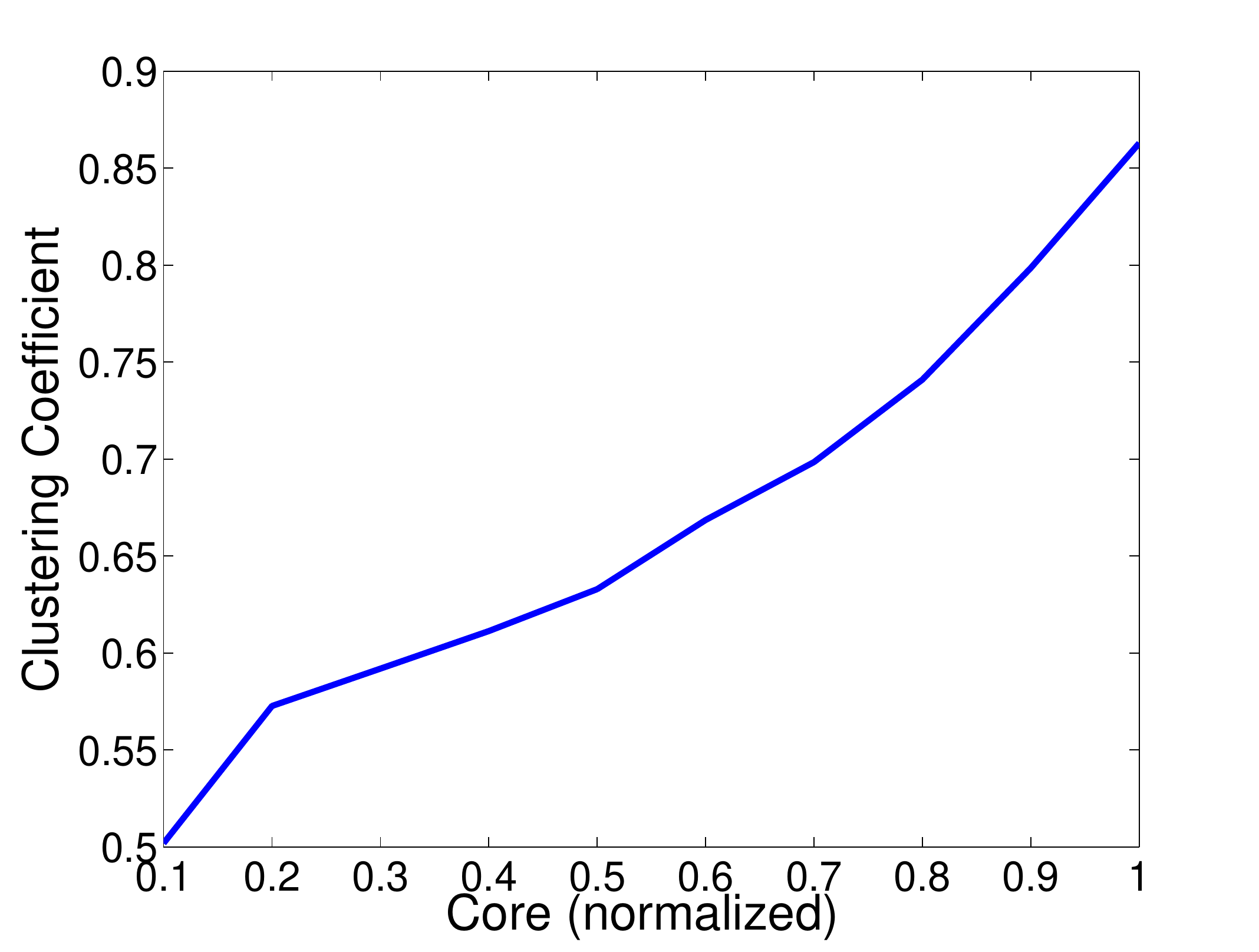}
\end{center}
\caption{Average clustering coefficient compared to $k$-core index
normalized by the maximum value.}
\label{fig:clco}
\end{figure}

Figure~\ref{fig:clco} depicts the above intuition about the datasets used in our experiments (as they are described in the main document).
More specifically, it represents the average clustering coefficient of a $k$-core subgraph with regards to the core index value $k$ (normalized by the maximum $k$ of each graph).
This indicates that, overall, a clearer clustering can be given at the maximum $k$-core.
As a reminder, the reader may see the properties of the artificial data presented on Table ~\ref{table:first}.

The particular results, along with the theoretical justification (described in the main
document, Section~\ref{sec:TheoreticalAnalysis}), provide the conclusion that the highest ranked $k$-core will most
likely  have the nodes of the graph that belong to the largest clusters and at
the same the  clustering structure will be the best under the measure of the
clustering coefficient.
As the \textsc{CoreCluster} framework starts from that subgraph (the maximal $k$-core), it will perform the best into separating the clusters of this subgraph and then it will move one the "next best" subset of nodes and edges as they are given from the core expansion sequence.

\vskip 0.2in

\bibliography{corecluster}

\end{document}